\documentclass{article}
\usepackage{amssymb,amsmath,amsthm}
\usepackage{mathtools} 
\usepackage{xfrac} 
\usepackage{a4wide}
\usepackage{pgf,tikz}
\usepackage{geometry}
\usepackage[utf8]{inputenc}
\usepackage[english]{babel}
\usepackage{geometry}
\usepackage{array}
\usepackage{breqn}
\usepackage{physics}
\usetikzlibrary{arrows}
\usepackage{epsfig}
\usepackage{color}

\usepackage{hyperref}

\usepackage{authblk}

\usepackage{subcaption}

\renewcommand{\theequation}{\arabic{section}.\arabic{equation}}

\newcommand{\beq}{\begin{equation}}  
\newcommand{\eeq}{\end{equation}}  
\newcommand{\bea}{\begin{eqnarray}} %% already in Fordy macros 
\newcommand{\eea}{\end{eqnarray}}   %% ditto 
\newcommand{\bear}{\begin{array}}  
\newcommand{\eear}{\end{array}} 

\newtheorem{thm}{Theorem}[section] 

\newtheorem{propn}[thm]{Proposition}

\newtheorem{lem}[thm]{Lemma}

\newenvironment{prf}{\trivlist \item [\hskip 
\labelsep {\bf Proof:}]\ignorespaces}{\qed \endtrivlist}

\theoremstyle{definition}

\newtheorem{remark}[thm]{Remark}%[section]

\newcommand{\Q}{{\mathbb Q}}
\newcommand{\Z}{{\mathbb Z}}
\newcommand{\C}{{\mathbb C}}
\newcommand{\R}{{\mathbb R}}
\newcommand{\Proj}{{\mathbb P}}

\newcommand{\rN}{\mathrm{N}}
\newcommand{\rA}{\mathrm{A}}
\newcommand{\ra}{\mathrm{a}}
\newcommand{\rB}{\mathrm{B}}

\newcommand{\rD}{\mathrm{D}}
\newcommand{\rx}{\mathrm{x}}
\newcommand{\ry}{\mathrm{y}}

\newcommand{\rd}{\mathrm{d}}
\newcommand{\ri}{\mathrm{i}}
\newcommand{\rg}{\mathrm{g}}
\newcommand\la{{\lambda}}

\newcommand\ka{{\kappa}}
\newcommand\al{{\alpha}}
\newcommand\be{{\beta}}

\newcommand\si{{\sigma}}

\newcommand\om{{\omega}}
\newcommand\Om{{\Omega}}

\DeclareMathOperator{\Jac}{Jac}
\DeclareMathOperator{\Pic}{Pic}

\newcommand\htau{\hat{\tau}}
\newcommand\tmu{\tilde{\mu}}

\newcommand\bv{{\bf v}}
\newcommand\bx{{\bf x}}
\newcommand\by{{\bf y}}
\newcommand\bu{{\bf u}}

\newcommand\tx{\tilde{x}}
\newcommand\im{\mathrm{im}\,} 
\newcommand\hphi{\hat{\varphi}}
\newcommand\hatom{\hat{\omega}}
\newcommand\mon{{\mathfrak{m}}}

\title{New cluster algebras from old: integrability beyond Zamolodchikov periodicity}

\author[1]{Andrew N. W. Hone\footnote{Corresponding author e-mail: A.N.W.Hone@kent.ac.uk}} 
\author[2]{Wookyung Kim}
\author[3]{Takafumi Mase}

\affil[1]{School of Mathematics, Statistics \&  Actuarial Science \protect\\ 
University of Kent,
Canterbury CT2 7FS, U.K.%~\\ e-mail: A.N.W.Hone@kent.ac.uk
}

\affil[2]{Department of Mathematics and Statistics  \protect\\ %,
Lancaster University,
Lancaster
LA1 4YF, U.K. 
}

\affil[3]{Graduate School of Mathematical Sciences  \protect\\ %,
University of Tokyo, 3-8-1 Komaba, Tokyo 153-8914, Japan.
}
\date{}

\begin{document} 

\maketitle
 
\begin{abstract} 
We consider discrete dynamical systems obtained as deformations 
of mutations in cluster algebras  associated with finite-dimensional simple Lie algebras. 
The original (undeformed) dynamical systems provide the simplest examples of 
Zamolodchikov periodicity: they are affine birational maps for which every orbit is periodic with the 
same period. Following on from preliminary work % results obtained 
by one of us with Kouloukas,  
here we present integrable maps obtained from deformations of cluster mutations related to the %following 
simple root systems %: 
$\rA_3$,  $\rB_2$, $\rB_3$ and $\rD_4$. 
We further show how new cluster algebras arise, by considering Laurentification, that is, a lifting to a higher-dimensional map expressed in a set  of new variables (tau functions), for which the dynamics exhibits the Laurent property. For the integrable map obtained by deformation of type $\rA_3$, which already appeared 
in our previous work, we show that there is a commuting map of Quispel-Roberts-Thompson (QRT) type which is built from a composition of mutations and a permutation applied to 
the same cluster algebra of rank 6, with an additional 2 frozen variables. Furthermore, both the deformed $\rA_3$ map and the QRT map correspond to translation by a generator in the Mordell-Weil group of a rational elliptic surface of rank two, and the underlying cluster algebra comes from a quiver that is mutation equivalent to  the $q$-Painlev\'e III quiver found by Okubo. 
The deformed integrable maps of types     $\rB_2$, $\rB_3$ and $\rD_4$ are also related to elliptic surfaces. 

From a dynamical systems viewpoint, 
the message of the paper is that special families of birational maps %whose iteration yields 
with completely periodic dynamics under  iteration  admit natural deformations that are aperiodic yet completely integrable.    
\end{abstract} 

\section{Introduction} 

\setcounter{equation}{0}  

The recurrence relation of second order given by  
\beq\label{lynessper}
x_{n+2} x_{n}=x_{n+1}+1 
\eeq 
is commonly referred  to by the name Lyness \cite{lyness}, after the British schoolteacher who observed that any pair 
of initial values $x_1,x_2$ produces the cycle of values 
$$ 
x_1,x_2,\frac{x_2+1}{x_1},  \frac{x_1+x_2+1}{x_1x_2}, \frac{x_1+1}{x_2}, 
$$ 
repeating with period 5. The Lyness 5-cycle has many avatars, including the associahedron $K_4$ \cite{fzassoc} and Abel's pentagon identity \cite{nakanishi}, 
and it also arises as one of the frieze patterns found by Coxeter \cite{coxeter}, who revealed a much earlier connection with the results of Gauss on the {\it pentagramma mirificum} 
(see \cite{director} for an interesting but somewhat idiosyncratic historical account of the latter). More recently,  (\ref{lynessper}) was found by Zamolodchikov in the 
context of  integrable quantum field theory, as one among many  functional relations (Y-systems) that were observed to display 
periodic behaviour; and a general axiomatic framework for describing such relations soon appeared in the shape of coefficient mutations 
in Fomin and Zelevinsky's theory of cluster algebras \cite{fz1, fz2}. Due to its relevance to Yangians, quantum affine algebras and solvable lattice models, 
the theory of Y-systems and other associated relations (T-systems and Q-systems) has now been extended considerably, and cluster algebras and other techniques 
have been applied extensively to resolve Zamolodchikov's periodicity conjecture for Y-systems \cite{inoue, kun}.  

The starting point of the recent work \cite{hk} was the observation that (\ref{lynessper}) admits a 2-parameter deformation, 
\beq\label{lynessgen}
x_{n+2} x_{n}=ax_{n+1}+b,  
\eeq 
with parameters $a,b$, which is 
also often referred to as the Lyness map \cite{tran}. 
This deformed map no longer has periodic orbits, except when the parameters are constrained to the case $b=a^2$ (which gives (\ref{lynessper}) when $a=1$).  
Also, while (\ref{lynessper}) corresponds to a mutation in a cluster algebra, and the 5-cycle corresponds to  a sequence of seeds (in the cluster algebra of finite type $\rA_2$), 
the deformed map does not have this property.   
However, (\ref{lynessgen}) preserves the same log-canonical symplectic structure 
$$ 
\om = \frac{\rd x_1\wedge \rd x_2}{x_1x_2} 
$$ 
as (\ref{lynessper}), and has a rational first integral for any $a,b$, so it is an integrable map in the Liouville sense \cite{bruschi, maeda, veselov}. 
Moreover, the general Lyness map admits a lift from the plane to  7-dimensional affine space, defined by the transformation 
\beq\label{lynesstau} 
x_n = \frac{\tau_n\tau_{n+5}}{\tau_{n+2}\tau_{n+3}},  
\eeq 
where the tau function $\tau_n$ satisfies the bilinear recurrence 
\beq\label{s7rec} 
\tau_{n+7}\tau_n = a\, \tau_{n+6}\tau_{n+1} + b\, \tau_{n+4}\tau_{n+3}, 
\eeq 
and (as pointed out in \cite{fordymarsh}) this particular Somos-7 relation is generated by mutations in a cluster algebra of rank 7, with the parameters $a,b$ regarded as frozen variables. 

In \cite{hk}, we derived the most general deformation of a cluster mutation that preserves the same symplectic (or more generally, presymplectic) structure,  
and showed how other Liouville integrable maps arise as deformations of cluster maps that exhibit Zamolodchikov periodicity.  In particular, in addition to (\ref{lynessgen}), which is the deformation 
of the type $\rA_2$ cluster map, we found integrable maps from deformations of types $\rA_3$ and $\rA_4$. The purpose of this article is twofold: 
firstly, we aim to get a better understanding of the deformed $\rA_3$ map, together with another commuting map of the type considered by 
Quispel, Roberts and Thompson (QRT) \cite{qrt}; and secondly,   we begin to investigate the 
result of applying an 
analogous deformation process to cluster algebras associated with  other Dynkin types (including examples of the non-simply laced case). 
We are conducting a separate, parallel, investigation that  is concerned with  analysing 
the deformations of higher rank cluster algebras of type $\rA$ \cite{grab}.

\subsection{Zamolodchikov periodicity and cluster mutations}

%Within Fomin and Zelevinsky's theory of cluster algebras \cite{fz1, fz2}, 

Some of the simplest examples of cluster algebras 
are provided by starting from the  Cartan matrix $C$ of a finite-dimensional Lie algebra 
%(more generally, a generalized Cartan matrix, in the case of an affine or Kac-Moody Lie algebra), 
and then constructing 
an associated companion matrix $B$, called the exchange matrix, which is skew-symmetrizable (i.e.\ there is a 
diagonal integer matrix $D$ such that $BD$ is skew-symmetric). The exchange matrix $B$ is the raw combinatorial data 
that is needed to define a cluster algebra. 

%The complete and the most 
Although the general definition of a cluster algebra is rather intricate, and appears somewhat complicated at first  sight, 
one of the original motivations behind %for introducing 
this definition was the remarkable %and apparently simple %periodicity 
phenomenon called   Zamolodchikov periodicity. It was observed by Zamolodchikov in \cite{zam} that for a certain family of 
integrable quantum field theories, namely deformations of conformal field theories associated with 
%finite-type 
simple Lie algebras, the thermodynamic Bethe ansatz allowed %showed that 
the form factors of correlation functions to be determined from systems of difference equations called Y-systems, and the solutions of these equations 
were conjectured to be periodic with the period being the same for any initial data (namely the Coxeter number plus 2).

A general cluster algebra involves two sets of variables: cluster variables, and coefficients. The generators of the algebra 
(given by clusters) are defined recursively by a process called mutation, which modifies the cluster variables and 
the coefficients at each step. The mutation formula for coefficients is modelled on Y-systems, while 
cluster mutation corresponds to so-called T-systems (see \cite{kun} for full details). One of Fomin and Zelevinsky's 
first important results was to classify cluster algebras of finite type: they showed that               
cluster algebras with finitely many cluster variables correspond precisely to $B$ matrices whose Cartan companions $C$ 
define Lie algebras of finite type \cite{fz2}. So this part of the theory of cluster algebras mirrors Dynkin's classification, 
and  eventually 
this led to a method to  prove   Zamolodchikov's periodicity conjectures for Y-systems, and various generalizations thereof.

Below we will just be concerned with exchange relations for cluster variables without coefficients (coefficient-free cluster algebras). 
However, even without coefficients, T-systems provide another avatar of Zamolodchikov periodicity: sequences 
of mutations in coefficient-free cluster algebras associated with Dynkin diagrams of finite type produce periodic maps. 
To set the scene, a coefficient-free cluster algebra ${\cal A}({\bf x}, B)$ of rank\footnote{The reader unfamiliar with the terminology of cluster algebras should be advised that the word 
\textit{rank} here refers to the number of cluster variables in each seed; in the finite type case, this happens to coincide with the rank of the root system of the associated Dynkin diagram.} 
$N$ is constructed by starting from a seed $({\bf x},B)$, which consists 
of an initial cluster ${\bf x}$, that is an $N$-tuple ${\bf x}=(x_1,\ldots,x_N)$, and an exchange matrix $B=(b_{ij})\in\mathrm{Mat}_N(\Z)$, which 
is required to be skew-symmetrizable. Then the mutation $\mu_k$ in the direction $k$ produces the new seed $ ({\bf x}', B')= \mu_k({\bf x}, B)$, 
where $B'=(b'_{ij})$ is obtained via matrix mutation, as specified by 
\beq\label{matmut}
b_{ij}' =  \begin{cases}
 -b_{ij} &\text{if}  \,\,i=k \,\, \text{or} \,\, j=k , \\
 b_{ij}+\text{sgn}(b_{ik})[b_{ik}b_{kj}]_+ & \text{otherwise},
\end{cases}
\eeq
and the new cluster ${\bf x}'=(x_j')$ is defined by cluster mutation, that is  
\beq \label{clustmut}
x_j'=
 \begin{cases}
x_k^{-1}\, \left( \prod_{i=1}^N x_i^ {[b_{ki}]_+} + \prod_{i=1}^N x_i^ {[-b_{ki}]_+}\right)&\text{for}  \,\,j=k  \\
 \ x_j   &\text{for}  \,\, j \neq k.
\end{cases}
\eeq 
(In the above, $[r]_+ =\max (r,0)$ for $r\in\R$.)  The cluster algebra ${\cal A}({\bf x}, B)$ is the subalgebra of $\Q(\bx)$ generated by the union of all cluster variables obtained 
from arbitrary 
sequences of mutations applied to the initial seed. These cluster variables satisfy the Laurent property: they belong to  $\Z[x_1^{\pm 1}, \ldots, x_N^{\pm 1}]$, the ring of Laurent polynomials
in  the variables from the initial cluster $\bx$.

T-systems are functional relations between variables that can be constructed from compositions of cluster mutations, as in 
(\ref{clustmut}). For Y-systems, one requires the more general setup of cluster algebras with coefficients \cite{fz1}, 
requiring  the introduction of another $N$-tuple of coefficient variables ${\bf y}=(y_1,\ldots,y_N)$, which are subject to their own 
set of rules for coefficient mutation. However, our starting point in what follows will be  periodic relations between cluster variables 
$x_j$ (essentially, T-systems rather than Y-systems), so we will omit any further discussion of coefficient mutation. 
Nevertheless, when we construct deformations we will be led to consider exchange relations between cluster variables 
involving additional frozen variables, which provide a natural way to treat coefficients appearing in these relations that are constant 
(they do not mutate).  
For further background material, 
the reader is referred to \cite{hk} and references.

\subsection{Outline of the paper} 

In the next section we briefly review the construction of the integrable deformation of the $\rA_3$ map from \cite{hk}, 
which reduces to a Liouville integrable map in the plane, depending on two arbitrary parameters, that preserves a pencil of biquadratic curves. As such, there is an associated QRT map 
associated with the same invariant pencil, and these two birational maps in the plane commute with one another. We proceed to show that the two maps admit a simultaneous 
Laurentification, in the sense defined in \cite{hhkq}, meaning   
that they each admit a lift to maps with the Laurent property, acting on the same 6-dimensional space of tau functions. Moreover, the tau functions  correspond to cluster variables in a skew-symmetric cluster algebra of rank 6, extended by an additional 2 frozen variables (associated with the two arbitrary parameters). The lift is  such that  each of the maps is obtained by composing a suitable sequence of cluster mutations with a permutation.  The fact that the two maps commute means that the tau functions take values on the lattice $\Z^2$, and we are able to obtain an explicit formula for the degrees of all tau functions on the lattice, as Laurent 
polynomials in 6 initial data. 

Section 3 concerns the construction of an integrable deformation of the $\rB_2$ cluster map, which is given by a 1-parameter family of symplectic maps in the plane, and we show that this lifts to a cluster algebra of rank 5 extended by a single frozen variable.   Sections 4 and 5 are concerned with integrable deformations of the periodic cluster maps of types $\rB_3$ and $\rD_4$, respectively. The situation is more complicated for these latter two examples, as we find that in each case the periodic map admits more than one inequivalent deformation that is integrable. 
Nevertheless, we are able to find tau functions for each of the different deformations in these cases as well. 

Overall, the message of the paper is the observation that (in all examples studied so far) the simplest cluster algebras of all, namely the finite type cluster algebras, hide within them cluster algebras that are 
much larger  (in the sense that they are both higher rank, and infinite), whose more intricate structure is revealed via deformation. From the point of view of discrete dynamical systems, 
the message is that there are special families of birational maps %whose iteration yields 
with completely periodic dynamics under  iteration, which  admit natural deformations that are aperiodic yet completely integrable.    
 
\section{Integrable deformation of the  $\rA_3$ cluster map} 
\setcounter{equation}{0}

The Cartan matrix for the $\rA_3$ root system is 
$$ 
C=\left(\begin{array}{ccc} 2 & -1 & 0 \\ 
-1 & 2 & -1 \\ 
0 & -1 & 2
\end{array}\right), 
$$ 
which is the companion of the skew-symmetric exchange matrix 
\beq\label{A3B} 
B=\left(\begin{array}{ccc} 0 & 1 & 0 \\ 
-1 & 0 & 1 \\ 
0 & -1 & 0 
\end{array}\right).  
\eeq 
We begin with a seed $( \bx, B)$ and consider sequences of mutations in the associated 
cluster algebra.   

\subsection{Periodic map from the $\rA_3$ cluster algebra} 

Starting from the exchange matrix (\ref{A3B}),  
we consider the following sequence of three mutations acting 
on an initial cluster $\bx = (x_1,x_2,x_3)$: 
\beq \label{A3maps}\begin{array}{rcl}
\mu_1: \quad (x_1,x_2,x_3)\mapsto (x_1',x_2,x_3), \qquad x_1'x_1 & = & x_2+1, \\
 \mu_2: \quad (x_1',x_2,x_3)\mapsto (x_1',x_2',x_3), \qquad x_2'x_2 & = & x_1'x_3+1, \\ 
\mu_3: \quad (x_1',x_2',x_3)\mapsto (x_1',x_2',x_3'), \qquad x_3'x_3 & = & x_2'+1. 
\end{array}
\eeq 
At each step, a prime is affixed only to the cluster variable that is mutated. 
The matrix (\ref{A3B}) is cluster mutation-periodic with respect to this sequence of mutations, in the sense that 
$$ 
\mu_3\mu_2\mu_1(B) = B, 
$$ 
and the corresponding cluster map $\varphi$ given by the composition 
$$
(x_1,x_2,x_3)\xmapsto{\mu_1}
%\overset{\mu_1}{\mapsto}  
(x_1',x_2,x_3)\xmapsto{\mu_2}%\overset{\mu_2}{\mapsto}  
(x_1',x_2',x_3)\xmapsto{\mu_3}%\overset{\mu_3}{\mapsto}  
(x_1',x_2',x_3')
$$ 
is periodic with period $6=4+2$ (two more than the 
Coxeter number of $\rA_3$):  
$$ 
\varphi=\mu_3\mu_2\mu_1, \qquad \varphi^6 (\bx) = \bx. 
$$ 

In order to interpret the above map $\varphi$ and its deformations in terms of Liouville integrability, we need to reduce it to a symplectic map in 2D. To begin with, we  note that, by Theorem 1.3 in \cite{hk}, the log-canonical presymplectic   2-form $\om$ associated with the matrix (\ref{A3B}) is preserved by the action of $\varphi$, 
that is 
\beq\label{omA3} 
\om = \frac{1}{x_1x_2}\rd x_1\wedge \rd x_2 +  \frac{1}{x_2x_3}\rd x_2\wedge \rd x_3, 
\qquad \varphi^*(\om)=\om.
\eeq 
The matrix $B$ has rank two, with the nullspace $\ker B$ spanned by $(1,0,1)^T$, and 
$\im B=(\ker  B)^\perp=<\bv_1,\bv_2>$,  spanned by the vectors 
$$ 
\bv_1=(0,1,0)^T, \qquad \bv_2= (-1,0,1)^T. 
$$ 
Then the  monomial quantities 
\beq\label{yjA3} 
u_1=\bx^{\bv_1} = x_2 , \qquad u_2=\bx^{\bv_2}= \frac{x_3}{x_1}
\eeq 
provide (local) coordinates for the leaves of the null foliation for $\om$, transverse to the flow of the null vector 
field $x_1\partial_{x_1}+x_3\partial_{x_3}$, 
and, from computing $\varphi^*(u_1)$ and $\varphi^*(u_2)$, we find that  the rational map 
\beq\label{pimapA3} 
 \begin{array}{lcrcl}
    \pi&:&\C^3&\to&\C^2\\
    & &\bx=(x_1,x_2,x_3)&\mapsto&\bu=(u_1,u_2) \; 
  \end{array}
\eeq 
intertwines $\varphi$ with the % reduced 
birational sympectic map $\hat\varphi$ given by 
\beq\label{phihatmapA3} 
 \begin{array}{lcrcl}
    \hat{\varphi}&:&\C^2&\to&\C^2\\
    & &\bu=(u_1,u_2)&\mapsto&\left(\frac{u_1u_2+u_2+1}{u_1},\frac{u_2+1}{u_1u_2}\right) \;, 
  \end{array}
\eeq 
that is 
$$ 
\hat{\varphi}\cdot\pi = \pi\cdot\varphi, \qquad \hphi^*(\hatom)=\hatom, 
$$ 
where $\pi^*(\hatom)=\om$ is the pullback of the symplectic form 
\beq\label{canA3}
\hatom=\frac{1}{u_1u_2}\rd u_1\wedge \rd u_2 
\eeq 
under $\pi$. 

It is clear that the reduced map $\hat{\varphi}$ in (\ref{phihatmapA3}) must also be periodic, because it arises from 
the pullback of $\varphi$ on the monomials (\ref{yjA3}), but in fact, its period is 3 (half that of $\varphi$): 
$\hphi^3(\bu)=\bu$.  Thus any symmetric function averaged over the period of an orbit is an invariant (first integral) for 
the map $\hphi$. In particular, the functions $$K_1=\prod_{j=1}^3 (\hphi^*)^{j-1}(u_1)=2 + \sum_{j=1}^3 (\hphi^*)^{j-1}(u_1), \quad %$ and $
K_2 =   \sum_{j=1}^3 (\hphi^*)^{j-1}(u_2)$$ 
provide two independent invariants. Both of the latter are Laurent polynomials in $u_1,u_2$, with the first being given by 
\beq\label{k1A3}
\begin{array}{rcl} 
K_1 & = & u_1 +u_2 +  \frac{u_2}{u_1}+\frac{2}{u_1}+\frac{1}{u_2}+\frac{1}{u_1u_2}+ 2 \\ 
& = & \mon_1+\mon_2+\mon_3+2\,\mon_4+\mon_5+\mon_6+\mathrm{const}, 
\end{array} 
\eeq 
where we have labelled the non-constant Laurent monomials appearing above by  
$$ 
\mon_1=u_1,\quad \mon_2=u_2,\quad \mon_3=\frac{u_2}{u_1}, \quad\mon_4=\frac{1}{u_1}, \quad \mon_5 = 
\frac{1}{u_2}, \quad \mon_6= \frac{1}{u_1u_2}. 
$$ 
An integrable deformation of the map $\hphi$ can be obtained by introducing parameters into the mutations $\mu_1,\mu_2,\mu_3$ and 
finding conditions on the parameters such that an analogue of one of these invariants survives under the deformation.

\subsection{Deformed $\rA_3$ map} 

In the $\rA_3$ case, the particular deformed mutations previously considered in \cite{hk} are of the same form as (\ref{A3maps}), 
but with constant parameters $a_j,b_j$ ($j=1,2,3$)
introduced into the exchange relations as follows: 
\beq \label{dA3maps}\begin{array}{rcl}
\mu_1: \quad %(x_1,x_2,x_3)\mapsto (x_1',x_2,x_3), \qquad 
x_1'x_1 & = & a_1x_2+b_1, \\
 \mu_2: \quad %(x_1',x_2,x_3)\mapsto (x_1',x_2',x_3), \qquad 
x_2'x_2 & = & a_2x_1'x_3+b_2, \\ 
\mu_3: \quad %(x_1',x_2',x_3)\mapsto (x_1',x_2',x_3'), \qquad 
x_3'x_3 & = & a_3x_2'+b_3. 
\end{array}
\eeq 
The deformed mutations above, which we denote by the same symbols $\mu_j$ as in the undeformed case, destroy the Laurent property: they do not generate Laurent polynomials in $x_1,x_2,x_3$ and the parameters $a_j,b_j$. 
Nevertheless, as shown in \cite{hk}, the map $\varphi=\mu_3\mu_2\mu_1$ formed from the composition of the exchange 
relations (\ref{dA3maps}) still preserves the same 
presymplectic form, that is $\varphi^*(\om)=\om$ with $\om$ as in (\ref{omA3}). Moreover, the reduction to the leaves of the null foliation 
defined by $\om$ produces a birational map in 2D, which can be written in terms of the same coordinates $u_1,u_2$ as in (\ref{yjA3}). 

Before considering the deformed 2D map,  there is a further simplification to be made, by considering the freedom to rescale the cluster 
variables, so that $x_j\to \la_jx_j$, $x_j'\to \la_jx_j'$ with arbitrary $\la_j\neq 0$ for $j=1,2,3$. Exploiting this freedom means that, given generic 
non-zero parameters $a_j,b_j$ in (\ref{dA3maps}), we can always make a choice of coordinates so that 3 of the 6 parameters in the deformed map 
$\varphi$ can be removed.  Following \cite{hk}, we scale the parameters in (\ref{dA3maps}) so that 
$$   
b_2 \to c, \qquad a_2\to d, \qquad b_3\to e,  \qquad a_1,b_1,a_3 \to 1.
$$ 
With this choice of scale for the parameters of the deformation, we can reduce the composition 
$\varphi=\mu_3\mu_2\mu_1$ to a 2D map on the leaves of the null foliation for $\om$, given by 
\beq\label{dphihatmapA3} 
 \begin{array}{lcrcl}
    \hat{\varphi}&:&\C^2&\to&\C^2\\
    & &\bu=(u_1,u_2)&\mapsto&\left(\frac{du_1u_2+du_2+c}{u_1},\frac{du_2+c}{u_1u_2}+\frac{e-c}{u_2(u_1+1)}\right) \;. 
  \end{array}
\eeq 
According to Theorem 2.1 in \cite{hk}, the above map 
preserves the same rational symplectic form on $\C^2$ as before, that is  $\hphi^*(\hatom)=\hatom$,  
with 
\beq\label{rat2formA3} 
\hatom=\rd \log u_1\wedge \rd \log u_2 .
\eeq 

For the deformed map to be integrable in the Liouville sense, we require that at least one of the invariants $K_1,K_2$ 
identified for the periodic map must survive the deformation. Thus we begin by  considering a deformed version of $K_1$, 
given by taking arbitrary linear combinations of the Laurent monomials $\mon_j$ appearing in (\ref{k1A3}), so 
that 
\beq\label{Klincomb}
K_1  =
 \ka_1\mon_1+\ka_2\mon_2+\ka_3\mon_3+\ka_4\mon_4+\ka_5\mon_5+\ka_6\mon_6+\mathrm{const}, 
\eeq 
where $\ka_j$ are coefficients. Without loss of generality, we can fix the leading coefficient 
$\ka_1=1$. Then a direct calculation shows that this rational function is invariant if and only if 
the following conditions hold:
$$ 
c=e, \qquad \ka_2=\ka_3=\ka_5= d\ka_1, \qquad \ka_4 = (c+d^2)\ka_1, 
\qquad \ka_6=cd\ka_1.  
$$ 
If we impose these conditions, then the reduced map becomes 
\beq\label{phihatA3} 
\hphi: \qquad 
\left(\begin{array}{c}u_1 \\ 
u_2 \end{array}\right) \mapsto 
\left(\begin{array}{c}u_1^{-1}\,\big(c+d(u_1+1)u_2\big) \\ 
(u_1u_2)^{-1}\,\big(c+du_2\big) \end{array}\right), 
\eeq
and by fixing $\ka_1=1$ and adding the constant $c+1$ to the given linear combination 
of $\mon_j$, we find that $\hphi^*(K_1)=K_1$, with the invariant $K_1$ taking the 
factorized form 
\beq\label{hphiK1A3} 
K_1= \frac{\Big(u_1+du_2+c\Big)\Big((u_1+1)u_2+d\Big)}{u_1u_2} .  
\eeq  

The conclusion of this calculation with the rational function $K_1$ is stated in Theorem 2.1 in \cite{hk}: when $a_1=a_3=b_1=1$ and $b_2=b_3=c$, $a_2=d$ 
(for arbitrary $c,d$) the deformed $\rA_3$ map   (\ref{dA3maps}) reduces to the map (\ref{phihatA3}) in the plane, which is 
integrable in the Liouville sense.

\subsection{Compatible QRT map} 

The family of level sets of the invariant $K_1$, given by fixing the value 
$K_1=\ka$,  defines a pencil of biquadratic curves, of which a generic member has genus 1, that is 
\beq\label{biqk1} 
\Big(u_1+du_2+c\Big)\Big((u_1+1)u_2+d\Big) =\ka \, u_1u_2. 
\eeq 
Each curve in such a pencil admits the pair of  involutions 
\beq\label{invol}
\iota_h: \, \, (u_1,u_2)\mapsto (u_1^\dagger, u_2), \qquad 
\iota_v: \, \, (u_1,u_2)\mapsto (u_1,u_2^\dagger),
\eeq 
referred to as the horizontal switch and the vertical switch \cite{duistermaat}, defined 
by mapping each point on the curve to the other intersection point with a 
horizontal/vertical line, respectively. Using the Vieta formula for the product of the roots of a quadratic, it is clear that each of these involutions is a  birational map, and is given by a 
formula that is independent of the parameter value $\ka$. Then their composition 
$$\hat\psi=\iota_v\cdot \iota_h$$ 
can be written as 
\beq\label{psihatmapA3}  % cf. {dphihatmapA3} 
 \begin{array}{lcrcl}
    \hat{\psi}&:&\C^2&\to&\C^2\\
    & &\bu=(u_1,u_2)&\mapsto& (\bar{u}_1,\bar{u}_2) = 
\left(\frac{(du_2+c)(u_2+d)}{u_1u_2},\frac{\bar{u}_1+c}{u_2(\bar{u}_1+1)}\right) \;.  
  \end{array}
\eeq  

By construction, the map (\ref{psihatmapA3}) preserves the same first integral (\ref{hphiK1A3}) as   (\ref{phihatA3}) does 
(so $\hat{\psi}^*(K_1)=K_1$),  as well as leaving the same symplectic form 
(\ref{rat2formA3}) invariant (so $\hat{\psi}^*(\hatom)=\hatom$).  Hence $\hat\psi$ is a Liouville integrable map in the plane.  From standard arguments about QRT maps (see e.g.\ \cite{duistermaat}), we can infer a lot more: on each generic fibre  of the pencil, %curve, 
either of the two maps      $\hat\psi$ and $\hphi$ acts as an automorphism without fixed points, and since 
a generic fibre has genus 1, this implies that both maps must act as translation by a point in the abelian group law of the corresponding elliptic curve; hence it follows that the two maps commute, that is 
\beq\label{commuting} 
\hat{\psi}\cdot \hphi = \hphi\cdot \hat{\psi}, 
\eeq 
a fact which is readily verified by direct calculation. An interesting question is whether this commutativity is trivial, as would be the case if the two maps were both iterated powers of the same (possibly simpler) birational map. Neither map can be a power of the other, because $\hat\psi$ becomes an involution (period 2) when $c=d=1$, while in that case $\hphi$ is the reduction of the $\rA_3$ cluster map, hence has period 3. However, we will see below that in fact the maps     $\hat\psi$ and $\hphi$ are independent of one another, in the sense that 
(for generic values of the parameters $c,d$) the Mordell-Weil group of the corresponding elliptic surface has rank 2, and they correspond to independent translations in this group.

\subsection{Laurentification and cluster structure of commuting maps} 

To make contact with the notation used in \cite{hk}, and to avoid excessive use of indices, we will use $(y,w)$ to denote the coordinates of 
$\bu\in\C^2$, so that 
$$ \bu =(u_1,u_2)=(y,w).$$ 
It will further be convenient to adopt the convention that iterates of $\hphi$ are denoted by a lower index $n$, so that the coordinates along an orbit are labelled thus:   
$$\hphi^n (\bu) =(y_n,w_n). $$ 
One of the main results on the deformed $\rA_3$ map  in our previous work was that it admits Laurentification, in the sense that we can lift 
$\hphi$ to a map in 6 dimensions that has the Laurent property, related via the monomial map %rational transformation 
\beq\label{tauf} 
\tilde{\pi}: \qquad 
y_n=\frac{ \tau_{n-1} \tau_{n+2} }{ \tau_{n} \tau_{n+1} }, \quad 
w_n = \frac{\sigma_{n+1} \tau_n }{\sigma_n \tau_{n+1}}, 
\eeq  
where $\tilde{\pi}: \C^6 \to \C^2$, and the sequence of tau functions $\sigma_n,\tau_n$ satisfy the pair of bilinear equations 
\beq\label{tausys} 
\begin{array}{lcl}
\sigma_{n+2}\,\tau_{n-1}& =& d\,\sigma_{n+1}\tau_{n}+c\, \sigma_n\tau_{n+1}, \\ 
\tau_{n+3} \,\sigma_n  & = & \tau_{n+1}\sigma_{n+2}+ d\, \tau_{n+2}\sigma_{n+1}.  
\end{array} 
\eeq 
(Note that the index on $\tau_n$ has been shifted compared with  \cite{hk}.) 

\begin{figure}
 \centering
\epsfig{file=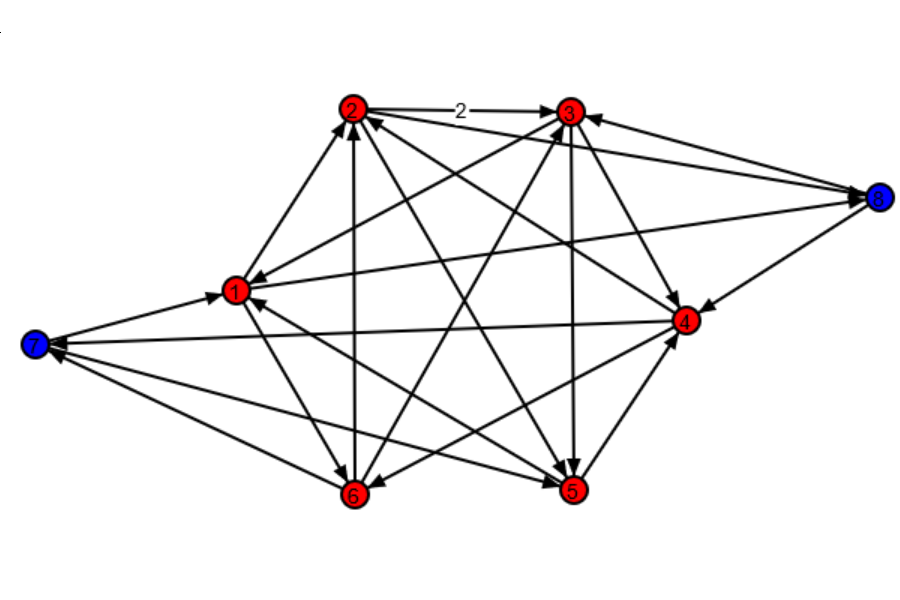, height=3in, width=4.8in}
\caption{The initial quiver $\tilde{Q}$ associated with the extended exchange matrix (\ref{BtauA3ext}).} 
\label{quiverA3}
\end{figure} 

A set of initial data for (\ref{tausys}) is provided by a set of six tau functions, namely $${\bx}^\dagger=(\tau_{-1},\tau_{0},\tau_{1},\tau_{2},\sigma_0,\sigma_1)
=(\tilde{x}_j)_{1\leq j\leq 6}.$$ It is possible to prove directly that under iteration, %The fact 
the bilinear system (\ref{tausys}) generates Laurent polynomials in these initial data, belonging to the ring 
$\Z[c,d,\tau_{-1}^{\pm 1}, \tau_0^{\pm 1},\tau_{1}^{\pm 1}, \tau_2^{\pm 1}]$. However, this is more easily seen as a 
consequence of the fact that each of the  bilinear equations correspond to mutations in a cluster algebra, whose coefficient-free part is given by an exchange matrix ${B}^\dagger$  obtained 
by pulling  back the symplectic form (\ref{rat2formA3}) by $\tilde\pi$, as in (\ref{tauf}), to find 
\beq\label{ompullback} 
\tilde{\om}=\tilde{\pi}^* (\hatom) = \sum_{i<j} \tilde{b}_{ij} \frac{\rd \tilde{x}_i \wedge \rd \tilde{x}_j}{\tilde{x}_i\tilde{x}_j}, 
\eeq 
where all the indices above run from 1 to 6. The matrix $B^\dagger=(\tilde{b}_{ij})_{1\leq i,j\leq 6}$ is skew-symmetric, and is the $6\times 6$ square submatrix appearing at the top of the extended $8\times 6$ 
exchange matrix $\tilde{B}$ given by 
\beq\label{BtauA3ext}
\tilde{B}=\left(\begin{array}{cccccc}
0 & 1& -1 & 0 & -1 & 1 \\ 
-1 & 0 & 2 & -1 & 1 & -1 \\ 
1 & -2 & 0 & 1 & 1 & -1 \\ 
0 & 1 & -1 & 0 & -1 & 1 \\ 
1 & -1 & -1 & 1 & 0 & 0 \\ 
-1 & 1 & 1 & -1 & 0 & 0 \\ 
1 & 0 & 0 & -1 & 1 & -1 \\ 
-1 & -1 & 1 & 1 & 0 & 0 
\end{array}
\right). 
\eeq 
The quiver $\tilde{Q}$ associated with $\tilde{B}=(\tilde{b}_{ij})$ is shown in Fig.\ref{quiverA3}: the convention is that $|b_{ij}|$ is the number of arrows between node $i$ and node $j$, 
with the  sign fixed according to whether $|b_{ij}|$ arrows run $i\to j$ or vice versa.  

\begin{remark}\label{qpiii} 
Note that the subquiver of  $\tilde{Q}$ in Fig.\ref{quiverA3} containing the 6 unfrozen nodes is mutation equivalent to the quiver that was shown by Okubo to produce the 
$q$-Painlev\'{e} III equation via an appropriate sequence of coefficient mutations (cf. Figure 21 in \cite{okubo}).  
\end{remark}

To specify the cluster algebra that contains the appropriate sequence of tau functions, we proceed to extend the initial data, in order to get an extended cluster 
$$\tilde{\bx}=(\tau_{-1},\tau_{0},\tau_{1},\tau_{2},\sigma_0,\sigma_1,c,d)
=(\tilde{x}_j)_{1\leq j\leq 8},$$
which includes the coeffients $c=\tilde{x}_7$ and $d=\tilde{x}_8$ as additional frozen variables (which are not allowed to be mutated),    
Then we see that the pair of successive mutations 
\beq \label{tmu15maps}\begin{array}{rcl}
\tilde{\mu}_1: \quad  
\tx_1' \tx_1 & = & \tx_8 \tx_2 \tx_6+\tx_7\tx_3\tx_5 , \\
\tilde{\mu}_5: \quad 
\tx_5' \tx_5 & = & \tx_8 \tx_4\tx_6+ \tx_1'\tx_3 
\end{array}
\eeq 
together 
correspond %precisely 
to iterations of (\ref{tausys}), provided that these two mutations are composed with the inverse of the cyclic permutation 
$\rho =(123456)$ (which only acts on the non-frozen variables). More precisely, we have 
\beq\label{phildef} 
\tilde{\mu}_5 \tilde{\mu}_1 (\tilde{B}) = \rho (\tilde{B}), \qquad \tilde{\pi}\cdot \tilde{\varphi} = \hphi \cdot  \tilde{\pi}, 
\eeq 
where the lifted map $\tilde{\varphi}$ acts as the shift $n\to n+1$ on the tau functions, that is, it acts as 
\beq \label{philift}
 \tilde{\varphi}=\rho^{-1} \tilde{\mu}_5 \tilde{\mu}_1 : \quad 
(\tau_{n-1},\tau_{n},\tau_{n+1},\tau_{n+2},\sigma_n,\sigma_{n+1},c,d) \mapsto 
(\tau_{n},\tau_{n+1},\tau_{n+2},\tau_{n+3},\sigma_{n+1},\sigma_{n+2},c,d)
 \eeq 
for all $n$, but leaves $\tilde{B}$ invariant, since $\tilde{\varphi}(\tilde{B}) = \rho^{-1} \tilde{\mu}_5 \tilde{\mu}_1(\tilde{B})=\tilde{B}$ 
from (\ref{phildef}).

The preceding observations about the Laurentification of $\hphi$, the reduction to the plane of the integrable deformation of the $\rA_3$ cluster map,  were summarized in Theorem 2.3 in \cite{hk}. To go beyond the latter, we now explain how 
the same ideas can be  extended to the QRT map  (\ref{psihatmapA3}) that commutes with   (\ref{phihatA3}). To do this, it will initially be convenient to abuse notation by adding an index $m$ to the pair of coordinates $(y,w)$, and write the sequence of points on an orbit as  
$$\hat{\psi}^m (\bu) =(y_m,w_m). $$ 
To avoid creating any confusion, we henceforth take the convention that the letters $m,n$ are used exclusively to label iterates of $\hat\psi$, $\hphi$, respectively.  
Then a  lift of $\hat\psi$  
%%$\hphi$ 
to a map in 6 dimensions is defined by %  that has the Laurent property, via the rational transformation 
\beq\label{taupsi} 
\bar{\pi}: \qquad 
y_m=\frac{ \eta_{m} \chi_{m} }{ \xi_{m} \theta_{m} }, \quad 
w_m = \frac{\xi_{m+1} \theta_m }{\xi_m \theta_{m+1}}, 
\eeq  
where the tau functions $\eta_m,\chi_m,\xi_m,\theta_m$ satisfy the bilinear system 
\beq\label{thetasys} 
\begin{array}{lcl}
\eta_{m+1}\,\chi_{m}& =& d\,\xi_{m+1}\theta_{m}+c\, \xi_m\theta_{m+1}, \\ 
\chi_{m+1}\,\eta_{m}& =& \xi_{m+1}\theta_{m}+d\, \xi_m\theta_{m+1}, \\ 
\xi_{m+2}\,\theta_{m}& =& c\,\xi_{m+1}\theta_{m+1}+\chi_{m+1}\eta_{m+1}, \\ 
\theta_{m+2} \,\xi_m  & = & \xi_{m+1}\theta_{m+1}+\chi_{m+1}\eta_{m+1}.
\end{array} 
\eeq 
This above bilinear system for the QRT map has the Laurent property, as summarized in the following result (which is a direct analogue of  Theorem 2.3 in \cite{hk}).  

\begin{thm}\label{QRTtaus}
The tau function sequences $(\eta_m)$, $(\chi_m)$, $(\xi_m)$, $(\theta_m)$ for the integrable map  (\ref{psihatmapA3}) 
consist of  elements of the ring of Laurent polynomials with positive coefficients, lying in 
$\Z_{>0}[c,d,\chi_{0}^{\pm 1},\theta_{0}^{\pm 1},\xi_{0}^{\pm 1},\eta_{0}^{\pm 1},\theta_{1}^{\pm 1},\xi_{1}^{\pm 1}]$,  being generated 
by the action of a permutation composed with a sequence of mutations in the cluster algebra defined by the quiver in Fig.\ref{quiverA3}.  %with the addition of two frozen nodes. 
\end{thm}
\begin{prf} Setting $m=0$ in (\ref{taupsi}), 
the pullback of the symplectic form  (\ref{rat2formA3}) is 
%\beq\label{ompullback} 
$$
\tilde{\om}=\tilde{\pi}^* \left( \rd \log \left(\frac{ \eta_{0} \chi_{0} }{ \xi_{0} \theta_{0} }\right) \wedge \rd \log  
\left( \frac{\xi_{1} \theta_0 }{\xi_0 \theta_{1}}\right) \right) = \sum_{1\leq i<j\leq 6} \tilde{b}_{ij} \frac{\rd \tilde{x}_i \wedge \rd \tilde{x}_j}{\tilde{x}_i\tilde{x}_j}, 
$$  
which coincides precisely with   (\ref{ompullback}), for the same $6\times 6$ skew-symmetric submatrix $B^\dagger= (  \tilde{b}_{ij})$, if we identify the unfrozen variables as  
$${\bx}^\dagger=(\chi_{0},\theta_{0},\xi_{0},\eta_{0},\theta_{1},\xi_{1})
=(\tilde{x}_j)_{1\leq j\leq 6}.$$ Extending this to the full matrix $\tilde{B}$, as in (\ref{BtauA3ext}), with the same frozen variables as before, namely $c=\tilde{x}_7$ and $d=\tilde{x}_8$, 
we see that repeatedly applying the sequence of four successive mutations 
\beq \label{tmu1423maps}\begin{array}{rcl}
\tilde{\mu}_1: \quad  
\tx_1' \tx_1 & = & \tx_8 \tx_2 \tx_6+\tx_7\tx_3\tx_5 , \\
\tilde{\mu}_4: \quad 
\tx_4' \tx_4 & = & \tx_2 \tx_6+\tx_8\tx_3\tx_5 , \\
\tilde{\mu}_2: \quad  
\tx_2' \tx_2 & = & \tx_7 \tx_5 \tx_6+\tx_1'\tx_4' , \\
\tilde{\mu}_3: \quad 
\tx_3' \tx_3 & = & \tx_5 \tx_6+\tx_1'\tx_4' 
% \tx_8 \tx_4\tx_6+ \tx_1'\tx_3 
\end{array}
\eeq 
is equivalent to iterating  the bilinear system  (\ref{thetasys}), provided that these four mutations are composed with the inverse of the cyclic permutation 
$\bar{\rho} =(14)(2536)$ (which only affects the non-frozen variables). To be precise we have 
\beq\label{psildef} 
\tilde{\mu}_3 \tilde{\mu}_2\tilde{\mu}_4\tilde{\mu}_1 (\tilde{B}) = \bar{\rho} (\tilde{B}), \qquad \bar{\pi}\cdot \tilde{\psi} = \hat{\psi} \cdot  \bar{\pi}, 
\eeq 
where the lifted map $\tilde{\psi}$ acts as the shift $m\to m+1$ on the tau functions, that is, it acts as %sends 
\beq \label{psilift}
 \tilde{\psi}=\bar{\rho}^{-1} \tilde{\mu}_3\tilde{\mu}_2\tilde{\mu}_4 \tilde{\mu}_1 : \quad 
(\chi_{m},\theta_{m},\xi_{m},\eta_{m},\theta_{m+1},\xi_{m+1},c,d) \mapsto 
(\chi_{m+1},\theta_{m+1},\xi_{m+1},\eta_{m+1},\theta_{m+2},\xi_{m+2},c,d)
 \eeq 
for all $m$, but leaves $\tilde{B}$ invariant, since $\tilde{\psi}(\tilde{B}) = \bar{\rho}^{-1}   \tilde{\mu}_3\tilde{\mu}_2\tilde{\mu}_4\tilde{\mu}_1(\tilde{B})=\tilde{B}$ 
from (\ref{psildef}). The fact that the sequences of tau functions generated by (\ref{thetasys}) consist of Laurent polynomials in the initial cluster variables, with integer coefficients, is just the Laurent 
phenomenon  for the cluster algebra \cite{fz1}, while the coefficients being in $\Z_{>0}$ is a consequence of the positivity conjecture, proved for skew-symmetric exchange matrices in \cite{ls}. 
\end{prf} 

\subsection{Somos relations for tau functions} 

A sequence $(\rx_n)$ generated by a quadratic recurrence relation of the form   
\begin{equation} \label{somosk} 
    \rx_{n+k}\rx_{n} = \sum_{j=1}^{\lfloor k/2\rfloor} \alpha_{j}\, \rx_{n+j}\rx_{n+k-j}
\end{equation}
is known as a Somos-$k$ sequence. There are  particular cases of  Somos-type recurrences that fit into the framework of cluster algebras, namely those that 
have a sum of two monomials on the right-hand side of \eqref{somosk}.\footnote{Somos recurrences with %a sum of 
three monomials on the right can be constructed in the 
more general framework of LP algebras \cite{lp}.} 
One such case is that of Somos-5 sequences, which 
satisfy a bilinear relation  given in the form  
\begin{align*}
    \rx_{n+5}\rx_{n} = \tilde{\al}\,\rx_{n+4}\rx_{n+1} +  \tilde{\be}\, \rx_{n+3}\rx_{n+2} . 
\end{align*}
It turns out that the Somos-5 recurrence can be reduced to a certain QRT map, equivalent to the recurrence 
\begin{equation}\label{somos5QRTB2}
    u_{n+1}u_{n}u_{n-1} = \tilde{\al}\, u_{n} + \tilde{\be} ,  
\end{equation}
where $u_n$ is given by a certain ratio composed of tau functions $\tau_n$ (see  \cite{hones5}, for instance). 
We refer to the above iteration 
as the Somos-5 QRT map.

The Laurentification of the map  (\ref{phihatA3}) in the plane produces two sequences of tau functions $\sigma_n,\tau_n$. The form of the substitution 
for $y_n$ in \eqref{tauf} is the same as was used for Somos-5 in \cite{hones5}, which suggests a connection with Somos sequences. 
It turns out that, along each orbit of the cluster map $\tilde{\varphi}$, the tau functions   $\sigma_n$ and $\tau_n$ both satisfy the same Somos-5 relation. 

\begin{thm}\label{A3S5thm} 
Along each orbit of the cluster map $\tilde{\varphi}$, given by iteration of the bilinear system \eqref{tausys}, the sequence of tau functions $(\tau_n)$ 
satisfies the Somos-5 relation 
\beq\label{A3S5eq} 
\tau_{n+5}\tau_n = \tilde{\al}\, \tau_{n+4}\tau_{n+1} + \tilde{\be}\, \tau_{n+3}\tau_{n+2}, 
\eeq 
with constant coefficients given by  
$$ 
 \tilde{\al} =d^2 - c  
\qquad
 \tilde{\be} = cK_1+(c+1)(d^2-c)
,
$$   	
where $K_1$ is the corresponding value of the first integral for the map 
(\ref{phihatA3}), obtained from the initial tau functions by 
setting 
\beq\label{K1tausubs} 
u_1 =
\frac{ \tau_{-1} \tau_{2} }{ \tau_{0} \tau_{1} }, \quad 
u_2=  \frac{\sigma_{1} \tau_0 }{\sigma_0 \tau_{1}},
\eeq 
in the formula \eqref{hphiK1A3}. Similarly, on each orbit of   $\tilde{\varphi}$,  the sequence of tau functions $(\sigma_n)$  
satisfies the Somos-5 relation 
\eqref{A3S5eq} with the same  coefficients $\tilde{\al},\tilde{\be}$.  
\end{thm}
\begin{prf}
As a recursion relation, the first component of the map $\hat\varphi$ given by  (\ref{phihatA3}) is equivalent to 
$$ 
y_{n+1}= \frac{c+d(y_n+1)w_n}{y_n}, 
$$ 
while the first component of the inverse map  $\hat{\varphi}^{-1}$ gives 
$$
y_{n-1}= \frac{cd+cw_n+dy_n}{w_ny_n}.  
$$
Then a direct calculation shows that 
$$ 
y_{n+1}y_ny_{n-1} = (d^2-c)\, y_n + cK_1+(c+1)(d^2-c), 
$$ 
where $K_1$ is the first integral for the map $\hat\varphi$ with $u_1=y_n$, $u_2=w_n$. Note that 
as $K_1$ is invariant under $\hat\varphi$, from \eqref{phildef}  we have 
$\tilde{\varphi}^*(\tilde{\pi}^*(K_1))=\tilde{\pi}^*(\hat{\varphi}^*(K_1))=\tilde{\pi}^*(K_1)$, 
so $K_1$ pulls back to a first integral for the cluster map $\tilde{\varphi}$, obtained by 
making the substitutions \eqref{K1tausubs} in \eqref{hphiK1A3}. Thus we have shown that 
$u_n=y_n$ satisfies the Somos-5 QRT map, in the form of the recurrence \eqref{somos5QRTB2}
%\beq\label{A3S5rec}y_{n+1}y_ny_{n-1} = \tilde{\al}\, y_n +\tilde{\be} \eeq 
with the required values of the coefficients $\tilde{\al},\tilde{\be}$, and 
this implies immediately that  $\tau_n$ is a solution of the Somos-5 relation 
\eqref{A3S5eq}, by substituting for $y_n$ 
 as in  \eqref{tauf}. 
Now from the other sequence of tau functions  $(\sigma_n)$ we can define the ratio 
$$ 
y_n^* :=\frac{\si_{n-1}\si_{n+2}}{\si_n\si_{n+1}}=\frac{y_nw_{n+1}}{w_{n-1}}, 
$$ 
where the final expression on the right-hand side above is obtained from \eqref{tauf}. Using 
$$ 
w_{n+1} = \frac{c+dw_n}{w_ny_n}, \quad w_{n-1}= \frac{y_n}{d+w_n} 
$$ 
from the second components of $\hat\varphi$ and  $\hat{\varphi}^{-1}$, we can 
rewrite $y_n^*$ in terms of $y_n,w_n$ alone, and then calculate analogous expressions for 
$y_{n+1}^*$ and $y_{n-1}^*$. Hence we find that $u_n=y_n^*$ is another solution of the  
Somos-5 QRT map  \eqref{somos5QRTB2}, with the same coefficients $\tilde{\al},\tilde{\be}$, and as a 
direct consequence 
the sequence  $(\sigma_n)$ also satisfies  \eqref{A3S5eq}. 
\end{prf}

Both the preceding observation and its proof were based on recognizing the specific ratio of tau functions $\tau_n$ for  $y_n$ appearing in \eqref{tauf}. However, a more systematic approach to deriving and proving Somos-type relations, which does not require any \textit{a priori} information,  
is to regard them as linear relations between degree 2 products of tau functions. We will illustrate this approach by considering the tau functions $\chi_m,\theta_m,\xi_m,\eta_m$ generated by  iterating  the bilinear system  (\ref{thetasys}). 

Let us suppose that the sequence $(\xi_m)$ generated by iteration of  (\ref{thetasys}) satisfies a Somos-$k$ recurrence relation for some $k$. The simplest non-trivial case 
to try is $k=4$. In that case, we write down two adjacent iterations of a Somos-4 recurrence for $\xi_m$, in the form 
\beq\label{s4linsys} 
\begin{array}{rcl}
\xi_{m+4}\xi_m & = &  \al\, \xi_{m+3}\xi_{m+1}+\be\, \xi_{m+2}^2, \\ 
\xi_{m+5}\xi_{m+1} & = &  \al\, \xi_{m+4}\xi_{m+2}+\be\, \xi_{m+3}^2 .
\end{array} \eeq 
A direct way to check whether such a relation is valid is to write down one more iteration of the recurrence, and verify that a corresponding $3\times 3 $ determinant vanishes; and this method can 
be extended to check  a Somos-$k$ relation of arbitrary order $k$ (cf.\ the proof of Theorem \ref{thmforsomos5} in the next section). 
In any case, the equations \eqref{s4linsys} provide a linear system for the coefficients $\al,\be$, with solution 
$$ 
\left(\begin{array}{c} \al \\ \be \end{array}\right) = 
\left(\begin{array}{cc}  \xi_{m+3}\xi_{m+1} & \xi_{m+2}^2 \\ 
 \xi_{m+4}\xi_{m+2} & \xi_{m+3}^2
\end{array}\right)^{-1}  
\left(\begin{array}{c}  \xi_{m+4}\xi_m \\ \xi_{m+5}\xi_{m+1}  \end{array}\right), 
$$ 
so the Somos-4 relation is valid if and only if the components of the above vector are constant (that is, independent of the index $m$). In the particular case at hand, we have initial data given by the cluster 
${\bx}^\dagger=(\chi_{0},\theta_{0},\xi_{0},\eta_{0},\theta_{1},\xi_{1})$, 
and to determine all the terms appearing in  \eqref{s4linsys} with $m=0$ we need to perform 4 iterations of the bilinear system (\ref{thetasys}). Then from solving the pair of linear equations for the coefficients, we find that $\al=(c-1)^2d$, which only depends on the parameters $c,d$, so is clearly constant, 
while 
$\be = \be (c,d, {\bx}^\dagger )$ 
is a rather complicated rational function (in fact, a Laurent polynomial) of 
$c,d$ and the initial data 
${\bx}^\dagger$. 
Nevertheless, under  the action of the cluster map 
$\tilde{\psi}$ 
it can be verified directly that 
$\tilde{\psi}^*(\be)=\be$, 
so that this coefficient is constant along each orbit. This then suggests that $\be$ can be rewritten as a 
polynomial function of $c,d$ and $K_1$, the first integral for the map  (\ref{psihatmapA3}), which indeed turns out to be the case. 
We can apply exactly the same method to seek 
Somos-type relations for the other tau functions $\chi_m,\theta_m,\eta_m$, and find that they all satisfy the same Somos-4 recurrence. The final result is summarized as follows. 

\begin{thm}
\label{psimapS4} %%{A3S5eq} 
Along each orbit of the cluster map $\tilde{\psi}$, 
given by iteration of the bilinear system (\ref{thetasys}), the sequence of tau functions $(\xi_n)$ satisfies the Somos-4 relation 
\beq\label{A3S4eq} \xi_{m+4}\xi_m  =  \al\, \xi_{m+3}\xi_{m+1}+\be\, \xi_{m+2}^2, \eeq 
with constant coefficients given by  
$$ {\al} =(c-1)^2 d,    \qquad {\be} = c K_1^2 + (c + 1) d^2 K_1 + d^4 + (c - 1)^2 d^2, $$   	
where $K_1$ is the corresponding value of the first integral \eqref{hphiK1A3} for (\ref{psihatmapA3}), obtained in terms of the initial tau functions by 
pulling it back via the map $\bar{\pi}$, for  
$u_1 =y_0$, $u_2=w_0$  given by \eqref{taupsi}   with $m=0$. Similarly, on each orbit of  $\tilde{\psi}$,  the sequences 
$(\chi_m),(\theta_m),(\eta_m)$ all satisfy the Somos-4 relation 
\eqref{A3S4eq} with the same  coefficients ${\al},{\be}$. 
\end{thm}  
 
\subsection{Tau functions on the $\Z^2$ lattice} 

The preceding results show that the action of the commuting integrable birational maps (\ref{psihatmapA3}) and   (\ref{phihatA3}) in the plane lifts to a  pair of commuting cluster maps $\tilde{\psi}$ 
and $\tilde{\varphi}$,  
which act on seeds in the same cluster algebra of rank 6 (with 2 additional frozen variables). Thus far we have used the two letters $m,n$ to index iterations of  the two  different maps and/or their lifts, so now it makes sense to combine them into a pair $(m,n)\in\Z^2$, and write 
$$ 
\hat{\psi}^m\hat{\varphi}^n (\bu) = (y_{m,n},w_{m,n}), \qquad (m,n)\in\Z^2, 
$$ 
as well as  introducing  a tau function $T_{m,n}$ on the $\Z^2$ lattice, such that 
\beq\label{ywmntau} 
y_{m,n} = \frac{T_{m,n-1}T_{m,n+2}}{T_{m,n}T_{m,n+1}}, \qquad w_{m,n} =  \frac{T_{m+1,n+1}T_{m,n}}{T_{m+1,n}T_{m,n+1}}. 
\eeq 
Then the initial seed in the associated cluster algebra is $(\tilde{\bx}, \tilde{B})$, with the extended exchange matrix $\tilde{B}$ as in  (\ref{BtauA3ext}), and the initial cluster 
being  specified by 
\beq\label{initclusterA3} 
\tilde{\bx}= (\tx_j)_{1\leq j\leq 8} = (T_{0,-1},T_{0,0},T_{0,1},T_{0,2},T_{1,0},T_{1,1},c,d). 
\eeq 

Under the combined actions of the cluster maps $\tilde{\psi}$ 
and $\tilde{\varphi}$, which are equivalent to iterating the bilinear equations given by the systems (\ref{thetasys}) and (\ref{tausys}), respectively, 
a generic set of initial values results in a complete set of tau functions defined at each point in the lattice, that is $(T_{m,n})_{(m,n)\in\Z^2}$. More precisely, 
due to the Laurent phenomenon, if each of the initial $T_{i,j}$ appearing in (\ref{initclusterA3}) is non-zero, then all other $T_{m,n}$ 
are obtained by evaluating suitable Laurent polynomials at these initial values.  In fact, these functions on the lattice 
are completely characterized as the %This evolution is completely described by the following 
solution of a system of four bilinear lattice equations. 

\begin{thm}\label{latmn} 
The tau functions on $\Z^2$,  associated via (\ref{ywmntau}) with combined iteration of the commuting integrable maps  (\ref{psihatmapA3}) and   (\ref{phihatA3}), 
satisfy the following system of bilinear equations: 
\beq\label{bilattice} 
\begin{array}{rcl}
T_{m+1,n+2}T_{m,n-1} & = & 
d\,  T_{m+1,n+1}T_{m,n}+ c\, T_{m+1,n}T_{m,n+1}, \\ 
T_{m,n+2}T_{m+1,n-1} & = & 
d\, T_{m,n+1}T_{m+1,n}+ T_{m+1,n+1}T_{m,n}, \\ 
T_{m+2,n+1}T_{m,n} & = & 
c\, T_{m+1,n+1}T_{m+1,n}+ T_{m+1,n+2}T_{m+1,n-1}, \\ 
T_{m+2,n}T_{m,n+1} & = & 
 T_{m+1,n+1}T_{m+1,n}+ T_{m+1,n+2}T_{m+1,n-1}. 
\end{array}  
\eeq 
Conversely, any solution of this system of bilinear lattice equations produces a simultaneous solution $(y_{m,n},w_{m,n})$ of the pair of 
iterated maps (\ref{psihatmapA3}) and   (\ref{phihatA3}). 
\end{thm}
\begin{prf} 
Essentially this is just a matter of rewriting the results of Theorem \ref{QRTtaus} with appropriate indices in $\Z^2$, and similarly for the preceding statements  about the 
bilinear system for the tau functions of the map (\ref{phihatA3}), and checking that they are compatible. First of all, 
for the tau functions of the QRT map (\ref{psihatmapA3}), we can identify the non-frozen part of any cluster via 
$$ 
(T_{m,n-1},T_{m,n},T_{m,n+1},T_{m,n+2},T_{m+1,n},T_{m+1,n+1}) \equiv  (\chi_{m},\theta_{m},\xi_{m},\eta_{m},\theta_{m+1},\xi_{m+1}), 
$$ 
where the equivalence means that, going from left to right, we simplify suppress the dependence on the second index $n$. The action of a single iteration of the 
map $\tilde{\psi}$ on any such cluster, corresponding to the shift $m\to m+1$, is obtained by solving each of the four equations (\ref{bilattice}) in turn, to obtain the transformation  
\small
$$
\tilde{\psi}: \quad (T_{m,n-1},T_{m,n},T_{m,n+1},T_{m,n+2},T_{m+1,n},T_{m+1,n+1}) \mapsto 
(T_{m+1,n-1},T_{m+1,n},T_{m+1,n+1},T_{m+1,n+2},T_{m+2,n},T_{m+2,n+1}), 
$$
\normalsize  
which is equivalent to one iteration of the four bilinear equations (\ref{thetasys}), or to  the composition 
of the four successive cluster mutations (\ref{tmu1423maps}) together with a permutation. 
Similarly, for the tau functions of the map (\ref{phihatA3}), we can write the ``forgetful'' equivalence 
$$ 
(T_{m,n-1},T_{m,n},T_{m,n+1},T_{m,n+2},T_{m+1,n},T_{m+1,n+1}) \equiv  (\tau_{n-1},\tau_{n},\tau_{n+1},\tau_{n+2},\sigma_n,\sigma_{n+1}), 
$$ 
where the dependence on the first index $m$ is suppressed upon moving from left to right. Then an iteration of the map $\tilde{\varphi}$ acting  on such a cluster  corresponds to the shift 
$n\to n+1$, which is achieved by solving the first equation in (\ref{bilattice}) to find  $T_{m+1,n+2}$, then using the second equation in the shifted form 
$$ T_{m,n+3}T_{m+1,n}  =  
d\, T_{m,n+2}T_{m+1,n+1}+ T_{m+1,n+2}T_{m,n+1}
$$ 
to find $T_{m,n+3}$, so that overall one has 
\small 
$$
\tilde{\varphi}: \quad (T_{m,n-1},T_{m,n},T_{m,n+1},T_{m,n+2},T_{m+1,n},T_{m+1,n+1}) \mapsto 
(T_{m,n},T_{m,n+1},T_{m,n+2},T_{m,n+3},T_{m+1,n+1},T_{m+1,n+2}), 
$$ 
\normalsize
and clearly this is equivalent to performing one iteration of the pair of bilinear equations (\ref{tausys}), or to  the composition 
of the two cluster mutations (\ref{tmu15maps}) together with a permutation. 
The converse statement follows immediately, because whenever the 
double sequence $(y_{m,n},w_{m,n})_{(m,n)\in\Z^2} $ is specified in terms of a solution of the system (\ref{bilattice}) by the formulae 
(\ref{ywmntau}), the bilinear equations imply that 
$\hat{\psi}\big((y_{m,n},w_{m,n})\big) = 
(y_{m+1,n},w_{m+1,n})$ and $\hat{\varphi}\big((y_{m,n},w_{m,n})\big) = 
(y_{m,n+1},w_{m,n+1})$ hold for all $m,n$. 
%The complete solution of the above equations on $\Z^2$ is generated by specifying initial data on a stencil of 6 points, which can be chosen as $,\tau_{0,-1},\tau_{0,0},\tau_{0,1},\tau_{0,2}\tau_{1,0},\tau_{1,1}$. 
%All of these initial values should be non-zero, and then (due to the Laurent property) it follows that the values $T_{m,n}$ can be determined on the whole lattice.  
\end{prf} 

\begin{remark}\label{somosA3T} 
By Theorem \ref{A3S5thm} and Theorem \ref{psimapS4}, whenever $T_{m,n}$ is a solution of the bilinear lattice system \eqref{bilattice}, it also satisfies a Somos-5 relation in $n$, and a Somos-4 relation in $m$. Moreover, the coefficients of both of these Somos relations are constant (that is, independent of both $m$ and $n$). 
\end{remark}

\subsection{Tropical dynamics and degree growth} %rank 2 elliptic surface} 

Following \cite{fz4}, it is constructive to consider the structure of the Laurent polynomials in a cluster algebra in terms of the so-called d-vectors, which permit the degree growth of 
the cluster variables to be determined from the tropical analogues of the exchange relations. In this case, we can write each cluster of tau functions related by 
iteration of the lattice  system (\ref{bilattice}) as 
\beq\label{clusterTmn} 
\tilde{\bx}_{m,n} :=  (T_{m,n-1},T_{m,n},T_{m,n+1},T_{m,n+2},T_{m+1,n},T_{m+1,n+1}) = \big(\tx_j(m,n)\big)_{1\leq j\leq 6}, 
\eeq 
where, in terms of the initial cluster $\tilde{\bx}$ given by (\ref{initclusterA3}), the Laurent property means that we may write  
\beq\label{lpolyA3}
 \tx_j(m,n)=\frac{\rN_j(m,n;\tilde{\bx})}{\tilde{\bx}^{{\bf d}_j(m,n)}}, \qquad j=1,\ldots,6, 
\eeq  
with the numerators $\rN_j(m,n;\tilde{\bx})\in\Z[\tilde{\bx}]$ being polynomials in the variables of the initial cluster $\tilde{\bx}$ which are not divisible by 
any of $\tx_1,\ldots,\tx_6$, and the denominators being monomials whose 
exponents are encoded in the integer d-vectors  ${\bf d}_j(m,n)\in\Z^6$. Then it is convenient to combine the six d-vectors in each cluster into a 
$6\times 6$ matrix $D_{m,n}$, that is 
\beq\label{bigdenergy} 
D_{m,n} :=\Big({\bf d}_1(m,n)\,\,{\bf d}_2(m,n)\,\,{\bf d}_3(m,n)\,\,{\bf d}_4(m,n)\,\,{\bf d}_5(m,n)\,\,{\bf d}_6(m,n)\Big).
\eeq 
While the degrees of the denominators grow, so that for large enough $m,n$ all the d-vectors belong to    $\Z_{>0}^6$, the initial conditions require that 
$$ \rN_j(0,0;\tilde{\bx})=1 \quad\mathrm{for}\quad j=1,\ldots,6$$
and 
$$ 
D_{0,0}=-I, 
$$ 
where $I$ denotes the  $6\times 6$ identity matrix. Due to homogeneity of the cluster variables (equivalently, the fact that the tau functions satisfy bilinear equations, so they all have 
the same weight), the d-vectors encode everything about the degrees of the Laurent polynomials: indeed, homogeneity requires that the total degree of each numerator 
(regarding $c,d$ as fixed constants) is 
$$ 
\deg_{\tilde{\bx}}\left(\rN_j(m,n;\tilde{\bx})\right) = {\bf e}^T {\bf d}_j(m,n) +1, \qquad {\bf e}=(1,1,1,1,1,1)^T, 
$$ 
that is, one more than the total degree of the monomial denominators.

Using standard arguments from \cite{fz4}, it can be shown that under the action of cluster mutation, the components of the d-vectors satisfy the $(\max,+)$  tropical version  of   the 
exchange relations. To be precise, the action of the cluster map $\tilde\varphi=\rho^{-1}\tmu_5\tmu_1$ on an initial cluster of d-vectors (a tropical seed) takes the form 
$$ 
\tilde\varphi: \quad 
\big({\bf d}_1,{\bf d}_2,{\bf d}_3,{\bf d}_4,{\bf d}_5,{\bf d}_6\big) \mapsto 
\big({\bf d}_2,{\bf d}_3,{\bf d}_4,{\bf d}_5',{\bf d}_6,{\bf d}_1'\big) , 
$$ 
which is composed of the combination of 
the $(\max,+)$ analogues of the mutations $\tmu_1,\tmu_5$, namely 
%tropical version of 
\beq \label{trop15maps}\begin{array}{rcl}
\tilde{\mu}_1: \quad  
{\bf d}_1'+{\bf d}_1 & = & \max({\bf d}_2 +{\bf d}_6,{\bf d}_3+{\bf d}_5 ), \\
\tilde{\mu}_5: \quad 
{\bf d}_5'+{\bf d}_5 & = & \max ({\bf d}_4+{\bf d}_6, {\bf d}_1' +{\bf d}_3 ), 
\end{array}
\eeq 
followed by the inverse of the cyclic permutation 
$\rho =(123456)$. (Note that the terms corresponding to the frozen variables $\tx_7,\tx_8$ are absent from 
\eqref{trop15maps}, since the components of the d-vectors only measure degrees of the non-frozen variables appearing in the monomial denominators
of cluster variables.) The action of $\tilde\varphi$ on a tropical seed corresponds to the shift $n\to n+1$ which transforms \eqref{bigdenergy} to a new matrix 
$D_{m,n+1}$.  Similarly, the shift $m\to m+1$ which transforms \eqref{bigdenergy} to a new d-vector matrix  $D_{m+1,n}$, 
is achieved via the action of $\tilde\psi=\bar{\rho}^{-1}\tmu_3\tmu_2\tmu_4\tmu_1$ on a tropical seed, taking the form 
$$ 
\tilde\psi: \quad 
\big({\bf d}_1,{\bf d}_2,{\bf d}_3,{\bf d}_4,{\bf d}_5,{\bf d}_6\big) \mapsto 
\big({\bf d}_4',{\bf d}_5,{\bf d}_6,{\bf d}_1',{\bf d}_3',{\bf d}_2'\big) , 
$$ 
which is the composition of the $(\max,+)$ versions of four mutations, given by 
\beq \label{trop1423maps}\begin{array}{rcl}
\tilde{\mu}_1: \quad  
{\bf d}_1'+{\bf d}_1 & = & \max({\bf d}_2 +{\bf d}_6,{\bf d}_3+{\bf d}_5 ), \\
\tilde{\mu}_4: \quad  
{\bf d}_4'+{\bf d}_4 & = & \max({\bf d}_2 +{\bf d}_6,{\bf d}_3+{\bf d}_5 ), \\
\tilde{\mu}_2: \quad  
{\bf d}_2'+{\bf d}_2 & = & \max({\bf d}_5 +{\bf d}_6,{\bf d}_1'+{\bf d}_4' ), \\
\tilde{\mu}_3: \quad  
{\bf d}_3'+{\bf d}_3 & = & \max({\bf d}_5 +{\bf d}_6,{\bf d}_1'+{\bf d}_4' ), 
\end{array}
\eeq 
followed by the inverse of the cyclic permutation 
$\bar{\rho} =(14)(2536)$.   (Note that, due to the absence of frozen variables, in \eqref{trop1423maps} the right-hand sides of 
$\tmu_1$ and $\tmu_4$ are identical, and the same is true for $\tmu_2$ and $\tmu_3$.) 

It was noted in the last subsection  that we can determine all the seeds obtained via iteration of $\tilde\varphi$ and $\tilde\psi$ from a single cluster variable 
indexed by $(m,n)\in\Z^2$, that is 
$$ 
T_{m,n} = \frac{\rN(m,n;\tilde{\bx})}{\tilde{\bx}^{{\bf d}(m,n)}},
$$
where we identify $\rN_2(m,n; \tilde{\bx}) = \rN (m,n; \tilde{\bx})$ 
and 
$
 {\bf d}_2(m,n) ={\bf d}(m,n)
$ 
for all $m,n$. This simplifies the analysis of the tropical dynamics considerably, 
as we see from \eqref{clusterTmn} that the d-vector matrix of any cluster is specified by the single $\Z^2$-indexed d-vector 
${\bf d}(m,n)$, according to 
$$ 
D_{m,n} =\Big({\bf d}(m,n-1)\,\,{\bf d}(m,n)\,\,{\bf d}(m,n+1)\,\,{\bf d}(m,n+2)\,\,{\bf d}(m+1,n)\,\,{\bf d}(m+1,n+1)\Big). 
$$
Then, by Theorem \ref{latmn}, it follows that all the components of ${\bf d}(m,n)$, and hence all components of the 
d-vector matrix, satisfy the same $(\max,+)$ difference equations on the lattice, 
which immediately leads to the following result.
\begin{propn}\label{dmatrix}
The matrix of d-vectors satisfies the tropical analogue of the system (\ref{bilattice}), that is  
\beq\label{dlattice} 
\begin{array}{rcl}
D_{m+1,n+2}+D_{m,n-1} & = & 
\max ( D_{m+1,n+1}+D_{m,n}, D_{m+1,n}+D_{m,n+1}), \\ 
D_{m,n+2}+D_{m+1,n-1} & = & 
\max (D_{m,n+1}+D_{m+1,n}, D_{m+1,n+1}+D_{m,n}), \\ 
D_{m+2,n+1}+D_{m,n} & = & 
\max (D_{m+1,n+1}+D_{m+1,n}, D_{m+1,n+2}+D_{m+1,n-1}), \\ 
D_{m+2,n}+D_{m,n+1} & = & 
 \max (D_{m+1,n+1}+D_{m+1,n}, D_{m+1,n+2}+D_{m+1,n-1}). 
\end{array}  
\eeq 
\end{propn} 

Manipulation of the above tropical equations reveals that in both the $m$ and $n$ directions, the d-vectors of the lattice system satisfy linear difference equations, which allow the degree growth to be calculated exactly. A key step in deriving the linear relations satisfied by the d-vectors is the consideration of the tropical analogues of the symplectic coordinates $(y,w)$, which are defined by the $(\max,+)$ versions of the formulae \eqref{ywmntau}, namely 
\beq\label{tropyw}
\begin{array}{rcl} 
{\bf Y}_{m,n} & = & {\bf d}(m,n-1)-{\bf d}(m,n)-{\bf d}(m,n+1)+{\bf d}(m,n+2),  \\ 
%\quad 
{\bf W}_{m,n} & = & {\bf d}(m+1,n+1)-{\bf d}(m+1,n)-{\bf d}(m,n+1)+{\bf d}(m,n) . 
\end{array} 
\eeq  
Each component of the pair of vectors $({\bf Y}_{m,n},{\bf W}_{m,n})$ satisfies the same set of coupled difference equations corresponding to iterations of the shifts $m\to m+1$ and 
$n\to n+1$, and the dynamics of these two tropical maps turns out to be completely periodic in both lattice directions, with periods that are inherited from the original undeformed dynamical systems (with $c=d=1$).

\begin{lem}\label{ywperiod23} %% {tropyw}
Each component of the vectors (\ref{tropyw}) satisfies the tropical analogue of the map \eqref{phihatA3}, namely 
\beq\label{tropphi}
\hat{\varphi}_{trop}: \, (Y_{m,n},W_{m,n}) \mapsto  (Y_{m,n+1},W_{m,n+1}), 
\eeq 
where 
$$ 
Y_{m,n+1}+Y_{m,n} = [W_{m,n}+[Y_{m,n}]_+ ]_+, \quad 
W_{m,n+1}+W_{m,n} = [W_{m,n}]_+ - Y_{m,n},  
$$ 
as well as the tropical analogue of \eqref{psihatmapA3}, given by
\beq\label{troppsi}
\hat{\psi}_{trop}: \, (Y_{m,n},W_{m,n}) \mapsto  
(|W_{m,n}|-Y_{m,n} , -W_{m,n}). 
\eeq 
For arbitrary initial data $(Y_{0,0},W_{0,0})=(Y,W)\in\R^2$, the orbit of 
$\hat{\varphi}_{trop}$ is periodic with period 3, and the orbit of 
$\hat{\psi}_{trop}$ is periodic with period 2.
\end{lem}  
\begin{prf} The calculation of the components of the maps $\hat{\varphi}_{trop}$ 
and $\hat{\psi}_{trop}$ is achieved  directly by taking the appropriate combinations of 
d-vectors as in \eqref{tropyw}, and transforming them under the actions of 
$\tilde\varphi=\rho^{-1}\tmu_5\tmu_1$ and  
$\tilde\psi=\bar{\rho}^{-1}\tmu_3\tmu_2\tmu_4\tmu_1$, respectively, 
according to the formulae in \eqref{trop15maps} and \eqref{trop1423maps}.  
(As in the definition of matrix mutation \eqref{matmut}, in order to write the tropical maps concisely we have found it convenient to use the notation $[r]_+=\max (r,0)$ for real numbers $r$.) 
Proving that any real orbit of the map \eqref{tropphi} has period 3 can be checked directly 
via a tedious case-by-case analysis, by considering  the action on pairs of values $(Y,W)\in\R^2$ lying in different sectors of the plane; we leave the details as an exercise for 
the reader. 
It can also be proved by adapting known results about dynamics on tropical elliptic curves \cite{nobe}. For the map  \eqref{troppsi} 
the analysis is more straightforward: the second component gives the relation 
$$ 
({\cal S}+1)W_{m,n} = W_{m+1,n}+W_{m,n}=0, 
$$  
where ${\cal S}$ denotes the shift operator corresponding to $m\to m+1$, 
and hence 
$$ 
W_{m,n}=(-1)^m\, W_{0,n}
$$ 
for all $m,n$, which oscillates with period 2 in $m$. Together with the first component of 
\eqref{troppsi} this also implies that 
$$
Y_{m+1,n}+Y_{m,n} = |W_{m,n}| = |W_{0,n}| \implies 
({\cal S}^2-1)Y_{m,n} = ({\cal S}-1)|W_{0,n}|=0,   
$$ 
so $Y_{m+2,n}=Y_{m,n}$ as required. 
\end{prf}  
\begin{remark}\label{tropseedrem} 
The preceding lemma is actually much stronger than what is needed to calculate the exact 
growth of the d-vectors  appearing in the matrices \eqref{bigdenergy} that are generated 
by tropical mutations applied to the specific initial seed $D_{0,0}=-I$, which is all that is 
required to calculate the degree growth of clusters generated by the lattice system 
(\ref{bilattice}). 
Indeed, all we really require is that a particular 
set of initial conditions for the maps  $\hat{\varphi}_{trop}$ 
and $\hat{\psi}_{trop}$  should have periodic orbits of length 3 and 2, respectively. The 
exact  periodicity of these particular orbits  follows directly from the known Zamolodchikov periods of the original undeformed maps with $c=d=1$, since these specific initial conditions 
correspond precisely to the d-vectors obtained from seeds of the $\rA_3$ cluster algebra, 
as generated by the cluster mutations \eqref{A3maps}.   
\end{remark} 

We can now determine a system of linear difference equations satisfied by the d-vectors. 
\begin{lem}\label{linlem} 
Let ${\cal S}$, ${\cal T}$ denote the shift operators corresponding to 
$m\to m+1$ and $n\to n+1$, respectively. 
Then any solution of the tropical lattice system (\ref{dlattice}) 
satisfies a linear ordinary difference equation of order 4 in the $m$ direction, namely 
\beq\label{order4S}  
\big({\cal S}^4 -2{\cal S}^3+2{\cal S}-1)\, D_{m,n}=0
\eeq 
and a linear ordinary difference equation of order 6 in the $n$ direction, that is 
\beq\label{order6T}  
\big({\cal T}^6 -  {\cal T}^5-{\cal T}^4+{\cal T}^2+{\cal T}-1)\, D_{m,n}=0,  
\eeq 
together with the mixed linear relations 
\beq\label{mixedST}
({\cal S}-1)({\cal T}^3-1)\, D_{m,n}=0, \qquad 
 ({\cal S}^2-1)({\cal T}-1)\, D_{m,n}=0. 
\eeq 
\end{lem}  
\begin{prf} We begin by noting that, from the definitions  (\ref{tropyw}), each 
component of  ${\bf Y}_{m,n}$ and ${\bf W}_{m,n}$ can be rewritten in the form 
$$ 
Y_{m,n} = ({\cal T}^3 -  {\cal T}^2 - {\cal T} +1) \, d_{m,n-1}, \qquad
W_{m,n} = ({\cal S}-1)({\cal T}-1)\,d_{m,n}, 
$$ 
respectively, 
where the scalar $d_{m,n}$ represents any component of a d-vector.   
The ordinary difference equation of order 6 in the $n$ direction follows immediately from 
the period 3 behaviour of the map $\hat{\varphi}_{trop}$ noted above, as we have 
$$  
0=  ({\cal T}^3 - 1)\, Y_{m,n} = 
 ({\cal T}^3 - 1) ({\cal T}^3 -  {\cal T}^2 - {\cal T} +1) \, d_{m,n-1},
$$ 
and since this  holds for each element of $D_{m,n}$, this produces the relation 
\eqref{order6T}. Similarly, considering shifts in $m$,    
from the proof of the previous lemma we have 
$$ 
({\cal S}+1)W_{m,n}=0 \implies ({\cal S}^2-1)({\cal T}-1)\,d_{m,n}=0 , 
$$
which immediately yields the second mixed relation in \eqref{mixedST}. 
On the other hand, the first linear relation   in \eqref{mixedST} is a direct consequence of the lattice system for $D_{m,n}$: it follows by subtracting the second equation in \eqref{dlattice} from the first, since the right-hand sides of these two equations are identical. 
The most involved part of the proof is  to derive the ordinary difference equation of order 4 in the $m$ direction. To show this, we introduce the 4th order difference operator that appears, which 
is 
$$
{\cal L}:= ({\cal S}-1)^3({\cal S}+1) = {\cal S}^4 -2{\cal S}^3+2{\cal S}-1. 
$$   
Now for convenience we revert to using the previous notation from \eqref{bigdenergy} for the first four components in a cluster of d-vectors, 
writing 
${\bf d}(m,n-1), {\bf d}(m,n), {\bf d}(m,n+1), {\bf d}(m,n+2)$ as 
${\bf d}_1,{\bf d}_2,{\bf d}_3,{\bf d}_4$, and note that, in this notation, the first relation 
  in \eqref{mixedST} is equivalent to 
$$ 
 ({\cal S}-1)\, ({\bf d}_4 - {\bf d}_1)=0 \implies {\cal L} \, ({\bf d}_4 - {\bf d}_1)=0, 
$$ 
while the right-hand sides of the third and fourth equations in \eqref{dlattice} are the same, 
so that subtracting them implies the relation 
$$ 
  ({\cal S}^2-1)\, ({\bf d}_3 - {\bf d}_2)=0 \implies {\cal L} \, ({\bf d}_3 - {\bf d}_2)=0, 
$$ 
and the definition of ${\bf Y}_{m,n}$ in \eqref{tropyw} and the 2-periodicity in $m$, as in 
Lemma \ref{ywperiod23}, yields the relation 
$$ 
  ({\cal S}^2-1)\, ({\bf d}_4 - {\bf d}_3-{\bf d}_2 + {\bf d}_1)=0 \implies {\cal L} \, 
({\bf d}_4 - {\bf d}_3-{\bf d}_2 + {\bf d}_1)=0.   
$$ 
Hence we see that three independent linear combinations of the quantities ${\bf d}_1,{\bf d}_2,{\bf d}_3,{\bf d}_4$ are annihilated by the operator ${\cal L}$, so it suffices to obtain 
one more independent combination lying in the kernel. If we add the third and fourth equations in  \eqref{dlattice}, and subtract twice the first entry in the max from both sides, 
followed by using ${\bf Y}_{m+2,n}={\bf Y}_{m,n}$ once again, 
then we find 
$$ 
  ({\cal S}-1)^2\, ({\bf d}_2 + {\bf d}_3)=2 [{\bf Y}_{m+1,n}]_+ 
\implies 
 {\cal L} \, ({\bf d}_2 + {\bf d}_3)=0, 
$$ 
which is the desired fourth linearly independent relation. Hence each component of the 
quadruple ${\bf d}_1,{\bf d}_2,{\bf d}_3,{\bf d}_4$ lies in the kernel of ${\cal L}$, and the 
required result \eqref{order4S} follows. 
\end{prf} 
These linear relations mean that it is a fairly straightforward matter to obtain the explicit solution for the double sequence of d-vector matrices $D_{m,n}$ subject to specifying the initial 
cluster of d-vectors via $D_{0,0}=-I$. 
\begin{thm} \label{exactdegs} The matrix of d-vectors for the tau function solutions of the system of  bilinear lattice equations (\ref{bilattice}), generated by the initial cluster 
of d-vectors 
$$ 
D_{0,0} = 
\Big({\bf d}(0,-1)\,\,{\bf d}(0,0)\,\,{\bf d}(0,1)\,\,{\bf d}(0,2)\,\,{\bf d}(1,0)\,\,{\bf d}(1,1)\Big)
= -I,  
$$  
%degree growth
is given by the exact  formula 
\beq\label{degd}
\begin{array}{rcl} 
D_{m,n} & = & \left(\frac{m^2}{4}  + \frac{n^2}{12}\right)  {\bf e}{\bf e}^T  %+ O(m) + O(n).   
+\frac{m}{2}(\hat{{\bf e}}{\bf e}^T - {\bf e}\hat{{\bf e}}^T)+\frac{n}{6}({\bf e}{\bf f}^T - {\bf f}{\bf e}^T) +\mathrm{E}  
-\frac{(-1)^m}{24}({\bf e}-2\hat{{\bf e}})({\bf e}^T-2\hat{{\bf e}}^T)   
-\frac{(-1)^n}{8} {\bf g}{\bf g}^T
\\
&& 
+
\Big(\mathrm{F}+(-1)^m \mathrm{G}\Big) \, e^{\sfrac{2\pi\ri n}{3}} + 
 \Big(\mathrm{F}^*+(-1)^m \mathrm{G}^*\Big) \, e^{\sfrac{-2\pi\ri n}{3}}
\end{array} 
\eeq 
for $(m,n)\in\Z^2$, 
where star denotes the complex conjugate, and 
$$ 
\mathrm{E}=-\frac{1}{36} \left(\begin{array}{cccccc} 
14 & 11 & 2 & -13 & 2 & -7 \\ 
11 & 14 & 11 & 2 & 5 & 2 \\ 
2 & 11 & 14 & 11 & 2 & 5 \\ 
-13 & 2 & 11 & 14 & -7 & 2 \\ 
2 & 5    & 2   & -7 & 14 & 11 \\ 
-7 & 2   & 5   & 2  & 11 & 14  
\end{array} 
\right), \qquad 
\mathrm{F}=-\frac{1}{18} {\bf h}{\bf h}^\dagger, 
\qquad \mathrm{G} =  -\frac{1}{6} {\bf k}{\bf k}^\dagger, 
$$ 
with the constant vectors 
$$
\begin{array}{l}  
{\bf e}^T=(1,1,1,1,1,1) , \qquad \hat{{\bf e}}^T=(0,0,0,0,1,1), \qquad  
{\bf f}^T = (0,1,2,3,1,2), \qquad  
{\bf g}^T = (1,-1,1,-1,-1,1), \\
{\bf h}^T = (1,e^{-\sfrac{2\pi\ri}{3}},e^{\sfrac{2\pi\ri}{3}},1,e^{-\sfrac{2\pi\ri}{3}},e^{\sfrac{2\pi\ri}{3}}), \qquad 
{\bf k}^T = (1,e^{-\sfrac{2\pi\ri}{3}},e^{\sfrac{2\pi\ri}{3}},1,e^{\sfrac{\pi\ri}{3}},e^{-\sfrac{\pi\ri}{3}})
\end{array} 
$$
(and the dagger means Hermitian conjugate). 
\end{thm}  
\begin{prf} This is mostly a straightforward exercise in solving linear difference 
equations with matrix coefficents. 
The action of the maps $\tilde{\varphi}$ and $\tilde{\psi}$ 
on an initial  cluster of d-vectors given by 
the matrix $D_{0,0}=-I$ can be used to  
produce a 
complete set of matrices on a 6-point stencil on $\Z^2$, namely $D_{0,-1},
D_{0,0}, D_{0,1},D_{0,2},D_{1,0},D_{1,1}$. Fixing the values of $D_{m,n}$ 
on these 6 points completely   
specifies an initial value problem for \eqref{dlattice}, the matrix version of 
the tropical lattice equations. 
The linear ordinary difference equation       \eqref{order6T} has the characteristic roots 
$1,1,1,-1,e^{\sfrac{2\pi\ri}{3}},   e^{\sfrac{-2\pi\ri}{3}}$, 
so in order to find the explicit formula for $D_{m,n}$, we can begin by writing down a general 
solution of this linear  equation   %   
in the form 
\beq\label{Dmnsoln} 
D_{m,n} = \mathrm{A}_0 \, n^2 +   \mathrm{B}_0 \, n + \mathrm{C}_0  + \mathrm{D}_0 \, (-1)^n + \mathrm{E}_0 \,e^{\sfrac{2\pi\ri n}{3}}  + \mathrm{E}_0^* \,  e^{-\sfrac{2\pi\ri n}{3}}, 
\eeq 
where \textit{a priori} the coefficient matrices $\mathrm{A}_0,\mathrm{B}_0$, etc.\ are all functions of $m$. Then, upon applying the first mixed relation 
in \eqref{mixedST} to the above formula, we find 
$$ 
({\cal S}-1)\, \Big( \mathrm{A}_0 (6n+9) + 3\mathrm{B}_0 - 2\mathrm{D}_0 (-1)^n \Big) =0 
$$ 
for all $n$, which implies that the coefficients $\mathrm{A}_0,\mathrm{B}_0$ and $\mathrm{D}_0$ are all constants (independent of $m$). 
Similarly, applying the second mixed relation in \eqref{mixedST} to the general formula \eqref{Dmnsoln} for $D_{m,n}$ then implies that 
$$ 
({\cal S}^2-1)\,  \mathrm{E}_0 = 0  =({\cal S}^2-1)\,  \mathrm{E}_0^*, 
$$ 
so these last two coefficients must both be period 2 functions of $m$, and we may write 
$$ 
  \mathrm{E}_0 =\mathrm{F}+(-1)^m \mathrm{G}
$$ 
for some constant matrices $\mathrm{F}$, $ \mathrm{G}$, and $\mathrm{E}_0^*$ is given by the same formula with $\mathrm{F}$, $ \mathrm{G}$ 
replaced by their complex conjugates (since the d-vectors are all real). This only leaves the $m$-dependence of the coefficient $ \mathrm{C}_0$ undetermined, but 
then the pure  linear relation in $m$, namely  \eqref{order4S}, requires that all the coefficients in \eqref{Dmnsoln} must lie in the kernel of 
the operator  ${\cal L}$, so in particular ${\cal L}\,  \mathrm{C}_0=0$, 
hence  $\mathrm{C}_0$ must have the general form 
$$ 
\mathrm{C}_0 = \mathrm{C}_1\, m^2 + \mathrm{C}_2\, m + \mathrm{C}_3 + \mathrm{C}_4 \, (-1)^m. 
$$
It remains to generate a sufficient number of matrices $D_{m,n}$ (for small $m,n$) via the system \eqref{dlattice}, in order to fix the exact values of the 
constant coeffients $\mathrm{A}_0,\mathrm{B}_0, \mathrm{C}_1,\mathrm{C}_2, \mathrm{C}_3,\mathrm{C}_4,\mathrm{D}_0$, as well as $\mathrm{F},  \mathrm{G}$ 
and  their complex conjugates, which reduces the problem to solving systems of linear equations with computer algebra. 
\end{prf} 
\begin{remark} The fact that the coefficients of the quadratic terms  $m^2$ and $n^2$ are non-zero, with no $mn$ terms in the formula (\ref{degd}),  
can be used to show that the maps 
$\hat\psi$ and $\hat{\varphi}$ correspond to independent translations along each fibre of the pencil  of curves (\ref{biqk1}). 
This means that  the  Mordell-Weil  group of the associated rational elliptic surface has minimal rank  2 (for generic parameters $c,d$). 
More detailed arguments that
show why the rank should be exactly 2 are relegated to the first appendix.  
\end{remark} 

\section{Integrable deformation of the  $\rB_2$ cluster map} 
In this section, we consider deformations of the periodic cluster map  constructed from the cluster algebra of type $\rB_{2}$.

\subsection{Deformed $\rB_{2}$ map}
\setcounter{equation}{0}

The Cartan matrix for the  $\rB_2$ root system is 
$$ 
C=\left(\begin{array}{cc} 2 & -1  \\ 
-2 & 2 
\end{array}\right), 
$$ 
and this is the companion to the exchange matrix $B=(b_{ij})$ given by 
\beq\label{B2B} 
B=\left(\begin{array}{ccc} 0 & 1  \\ 
-2 &  0 
\end{array}\right).  
\eeq 
which is obtained from $C$ by removing the diagonal terms
and adjusting the signs of the off-diagonal terms appropriately 
(with the requirement that if $b_{ij}\neq 0$, then $b_{ji}$ 
should have the opposite sign). The latter matrix is skew-symmetrizable, since for $D=\mathrm{diag}(1,2)$ we have that 
$$\Omega =BD=(\om_{ij})$$ is the skew-symmetric matrix 
%Skew-symmetrizability of $B$ is seen from the fact that $\Om=BD$ is skew-symmetric,  
%where $D=\mathrm{diag}(1,2)$ gives %we have
\beq\label{B2Om} 
\Om = \left(\begin{array}{ccc} 0 & 2  \\ 
-2 &  0 
\end{array}\right).  
\eeq 

Starting from an initial cluster 
${\bf x}=(x_1,x_2)$, 
we consider a pair of deformed mutations, of the form  % according to 
\beq \label{B2maps}\begin{array}{lcl}
\mu_1: \quad  (x_1,x_2)\mapsto (x_1',x_2), \qquad x_1'x_1 & = & a_1x_2^2+b_1, \\
\mu_2: \quad  (x_1',x_2)\mapsto (x_1',x_2'), \qquad x_2'x_2 & = & a_2x_1'+b_2. 
\end{array}
\eeq 
One can confirm that, after applying the corresponding pair of matrix mutations, namely $\mu_1$ followed by $\mu_2$, according to the rule (\ref{matmut}),  the exchange matrix \eqref{B2B} is mutation periodic under this composition of mutations, % above, 
that is to say 
$$ 
\mu_2\mu_1(B)=B.
$$ 
So,  similarly to the situation for the skew form (\ref{omA3}) constructed from the exchange matrix of type $\rA_{3}$,  by a minor variation on Theorem 1.3 in \cite{hk}, 
adjusting the presymplectic structure to the skew-symmetrizable setting 
(see \cite{inouenakanishi}, for instance),  %implies that 
the map  
${\varphi} = \mu_{2}\mu_{1}$ composed from the pair of deformed cluster 
mutations (\ref{B2maps}) preserves the %following 
log-canonical two-form 
\begin{equation}\label{symB2}
    \omega =  \sum_{i<j} \om_{ij} \dd \log x_i \wedge \dd \log x_j = \frac{2}{x_1x_2}\dd x_1 \wedge \dd x_2. 
\end{equation}
The latter is the skew form built from the 
%where the 
coefficients  of   the matrix  $\Omega$ in \eqref{B2Om},  
obtained from skew-symmetrization of 
\eqref{B2B}. In other words, $\varphi^*(\om)=\om$, and since the two-form (\ref{symB2}) is non-degenerate in this case, the deformed $\rB_2$ cluster map $\varphi$ 
is symplectic, for arbitrary values of the parameters $a_i,b_i$. However, note that when these parameters take generic values, the composition of transformations 
in \eqref{B2maps} is not a cluster map, because it does not generate Laurent polynomials in $x_1,x_2$. 

The original undeformed mutations obtained from the exchange matrix (\ref{B2B}), which generate the cluster algebra of type $\rB_2$, are recovered by 
setting all the parameters $a_i,b_i$ in \eqref{B2maps} to 1. Since the Coxeter number of $\rB_2$ is 4, Zamolodchikov periodicity implies that 
the undeformed cluster map $\varphi=\mu_2\mu_1$ 
has period $3=\frac{1}{2}\left(4+2\right)$, so that 
$$\varphi=\mu_2\mu_1 \quad (%\tilde{\varphi} \  
\text{with}\ a_{1}=b_1 =a_{2} = b_2=1 )\implies \varphi^3(\bx)=\bx.$$
Therefore, due to the periodicity, for any function $f:\,\C^2\to\C$, the associated symmetric function given by the product over an orbit, that is  
$$K_f({\bf x})=\prod_{j=0}^2(\varphi^*)^j(f)({\bf x})=\prod_{j=0}^2f\Big((\varphi^*)^j({\bf x})\Big),$$   
is invariant under  the cluster map %$\varphi$ admits
$\varphi$. Here we consider 
\begin{equation}\label{firstintB2}
     K =\prod^{2}_{j=0} (\varphi^{*})^{j}(x_{2}) =  x_2 + \frac{2}{x_2} + \frac{x_1}{x_2}  + \frac{x_2}{x_1} + \frac{1}{x_1x_2} \\
\end{equation}

Before proceeding further with the general deformed case, for arbitrary non-zero parameters $a_i,b_i$, 
we can apply rescaling $x_{i} \to \lambda_{i} x_{i}$ to each cluster variable,  with a suitable choice of parameters $(\lambda_{1},\lambda_2)\in(\C^*)^2$,  in order to arrange it so that $a_{1} = 1 = a_{2}$, 
which simplifies the calculations.  With the remaining parameters $b_1,b_2$ fixed, the iteration of the deformed map ${\varphi}$ is given by a system of recurrences
\beq \label{dB2maps}\begin{array}{lcl}
 x_{1,n+1}x_{1,n} & = & x_{2,n}^2+b_1, \\
x_{2,n+1}x_{2,n} & = & x_{1,n+1}+b_2. 
\end{array}
\eeq
An invariant function for this deformed map %${\varphi}$ 
can be constructed by 
the same % following the 
procedure as was used in \cite{hk}, and described for type $\rA_3$ above, %in which 
whereby we modify each Laurent monomial in \eqref{firstintB2} by inserting arbitrary coefficients $\kappa_{i}$ in front of each monomial, similarly to %the formula 
(\ref{Klincomb}), so that (after fixing the leading coefficient to be 1) we have  
\begin{equation}\label{dfirstintB2}
    \tilde{K}=  x_2 + \frac{\kappa_{1}}{x_2} + \frac{\kappa_{2}x_1}{x_2}  + \frac{\kappa_{3}x_2}{x_1} + \frac{\kappa_{4}}{x_1x_2} . \\
\end{equation}
Next, we proceed to impose the condition of invariance on this Laurent polynomial,  
that is ${\varphi}^{*}(\tilde{K}) = \tilde{K}$, which puts  constraints on the coefficients $\kappa_{i}$ and $b_{i}$. 
This gives rise to a necessary and sufficient condition for the deformed map \eqref{dB2maps} to be Liouville integrable, leading to the following result.

\begin{thm} The necessary and sufficient condition for a rational function of the form \eqref{dfirstintB2} to be a first integral for 
the map defined by (\ref{dB2maps}) is that 
\begin{equation}\label{condsB2}
    b_{1} = b_{2} = \beta, 
\end{equation}
in which case $\tilde{K}$ is  given by 
\begin{equation}\label{dfirstB2}
    \begin{split}
        & \tilde{K}=  x_2 + \frac{1+\beta}{x_2}  + \frac{x_1}{x_2}+ \frac{ x_2}{x_1} + \frac{\beta}{x_1x_2} \\
    \end{split}
.
\end{equation}
Hence the deformed symplectic map ${\varphi}$  given by 
\beq\label{phitilmapB2} 
 \begin{array}{lcrcl}
    {\varphi}&:&\C^2&\to&\C^2\\
    & &\bx=(x_1,x_2)&\mapsto& \bx '=(x_1',x_2') = 
\left(\frac{(x_2)^2+\beta}{x_1},\frac{x_1'+\beta}{x_2}\right) \;.  
  \end{array}
\eeq   
is Liouville integrable whenever the condition \eqref{condsB2} holds. 
\end{thm}
\begin{prf} 
The proof follows from an %of the above result can be carried out by 
explicit calculation, which is best achieved using  a computer algebra package such 
as MAPLE: 
assuming that a first integral of the form \eqref{dfirstintB2} exists, the equation $\varphi^*(\tilde{K})=\tilde{K}$ can be rewritten as an 
identity between two polynomials in $x_1,x_2$, and then comparing coefficients at each degree yields a set of linear equations in the 
coefficients $\kappa_i$; this linear system has a solution if and only if \eqref{condsB2} holds.
\end{prf} 
\begin{remark}
The level sets of the first integral \eqref{dfirstB2} are biquadratic curves,  hence we can construct a QRT map given by the composition of two involutions, of the same 
form as in (\ref{invol}). However, comparison with the formula (\ref{phitilmapB2}) shows that in this case, 
on the biquadratic pencil 
$$  
x_1(x_2)^2 + (1+\beta){x_1}  + (x_1)^2+  (x_2)^2 + \beta=\tilde{\kappa}{x_1x_2} 
$$ 
corresponding to the level sets $\tilde{K}=\tilde{\kappa}$, 
the transformation $\mu_1$ is the horizontal switch, and 
$\mu_2$ is the vertical switch (see \cite{duistermaat}), hence the %deformed 
map ${\varphi}=\mu_2\cdot \mu_1$ coincides with the QRT map. % is identical to the deformed map $\tilde{\varphi}$
\end{remark}

We have seen that, when the parameters satisfy the constraints \eqref{condsB2}, the symplectic map given by (\ref{dB2maps}) is Liouville integrable.  
However, as already mentioned above, the general deformed cluster map ${\varphi}$ is not itself a cluster map, and this continues to be true for the constrained version  
%the integrable map 
(\ref{phitilmapB2}), since for $\beta\neq 1$ the sequence of pairs of 
coordinates 
$\tilde{\varphi}^n(\bx)$ generated by the latter map do not belong to the ring of  Laurent polynomials in $x_1,x_2$. In an attempt to resolve this issue, we must go a step further,  and try to apply  Laurentification, analogously to what was carried out in \cite{hk} for deformed maps of type $\rA$ in low dimension (and described for deformed $\rA_3$ in the previous section). 

\subsection{Laurentification of deformed type $\rB_{2}$ map}
As we saw for the case of the deformed $\rA_3$ map in the previous section, the word 
Laurentification refers to a transformation that lifts a given birational map to a map acting on a new set of coordinates in higher dimensions, 
where the lifted map possesses the Laurent property. 
Several methods have been used to achieve Laurentification, such as the recursive factorization approach taken in \cite{hhkq} 
(see also \cite{viallet1, viallet} for some related results and observations). 
Here we consider another method, introduced in \cite{hkq}, which involves the singularity pattern of the iterates of the deformed map ${\varphi}$. 
Instead of performing a general analysis of singularities, we apply an empirical version of $p$-adic analysis, 
which is done by inspecting the prime factorization of the terms given by the iteration. 
To see the procedure, we consider the rational orbit of the map (\ref{phitilmapB2}) 
obtained by setting the value of the initial cluster to be $(x_{1},x_{2}) =(x_{1,0},x_{2,0}) = (1,1)$, with parameters $b_{1}=2=b_{2}$, and find 
the prime factorizations of the numerators and denominators of successive terms, as 
in the table below: 
{\renewcommand{\arraystretch}{3}
\begin{center}
\begin{tabular}{ |c || c | c | c | c | c |c|c | c | } 
  \hline
  $n$ & $1$ & $2$ & $3$ & $4$ & $5$ & $6$ & $7$ & $8$ \\ 
  \hline
  $x_{1,n}$ & $3$ & $3^2$ & $\frac{19}{5^2}$ & $\frac{569}{11^2}$ & $\frac{17^2\cdot 107}{3^2\cdot 23^2}$ & $\frac{139 \cdot 3299}{811^2}$ & $\frac{457737691}{8089^2}$ & $\frac{3 \cdot 457 \cdot 81689827}{7^2 \cdot 23039^2}$ \\
  \hline 
  $x_{2,n}$ & $5$ & $\frac{11}{5}$ & $\frac{3 \cdot 23}{5 \cdot 11}$ & $\frac{5 \cdot 811}{3 \cdot 11 \cdot 23}$ & $\frac{11 \cdot 8089}{3 \cdot 23 \cdot 811}$ & $\frac{3 \cdot 7 \cdot 23 \cdot 23039}{811 \cdot 8089}$ & $\frac{13\cdot 173 \cdot 811 \cdot 3793}{7 \cdot 23039 \cdot 8089}$ & $\frac{5^3 \cdot 41 \cdot 39461 \cdot 8089}{7 \cdot 13 \cdot 173 \cdot 23039 \cdot 3793}$ \\ 
  \hline
\end{tabular}
\end{center}
}
We can see that for each of the primes $p=5, 11 ,23, 811, 8089$ (for instance), the $p$-adic norms of $x_{1,n}$ and $x_{2,n}$ exhibit the patterns  
\begin{equation}
\begin{array}{rlllll} 
  \abs{x_{1,n}}_{p}: &  1, &  1, &  p^{2}, &  1 , & 1 \\
    \abs{x_{2,n}}_{p}: &  p^{-1}, &  p, &  p, &  p^{-1} & 1
\end{array}
\end{equation}
Furthermore,  other primes, such as  $p=19,569,107, 139, 3299$, appear successively as factors  in the numerator of  $x_{1,n}$, but not in $x_{2,n}$. This suggests that there should be the 
following singularity patterns:  
\begin{align*}
&\text{Pattern 1:} \quad \dots,(R,0),(R,\infty),(\infty^{2},\infty),(R,0),\dots\\
&\text{Pattern 2:} \quad \dots,(0,R),\dots, 
\end{align*}
where $R$ denotes a regular  (non-zero) finite value. Then we introduce two tau-functions $\tau_{n}$, $\sigma_n$, associated with Pattern 1 and Pattern 2, respectively, in order to arrange 
it so that, for isolated values of $n$, $\tau_n\equiv 0 (\bmod p)$ for the first set of primes,  and $\sigma_{n} \equiv 0 (\bmod p)$ for the second set. Then, to recover the two different singularity patterns, we define a monomial rational map 
$\pi: \, \C^5\rightarrow \C^2$, which is specified by the following 
transformation  of dependent variables:  
\begin{equation}\label{tauB2}
  \pi: \qquad   x_{1,n} = \frac{\sigma_n}{\tau_{n+1}^2}, \quad x_{2,n} = \frac{\tau_n \tau_{n+3}}{\tau_{n+1}\tau_{n+2}}. 
\end{equation}
If  the two expressions (\ref{tauB2}) are substituted directly into the components \eqref{dB2maps} of $\varphi$, with the  parameters constrained so that $b_1 =\beta = b_2$, then one obtains %following 
the system of recurrence relations 
\begin{equation}\label{systmB2}
\begin{split}
    &\sigma_n \sigma_{n+1} = \beta \tau_{n+1}^{2} \tau_{n+2}^2 + \tau_{n}^2\tau_{n+3}^2 , \\
    &\tau_{n}\tau_{n+4} = \beta \tau_{n+2}^2 + \sigma_{n+1} . 
    \end{split}
\end{equation}
If we iterate the latter pair of equations with initial values $(\sigma_0,\tau_0,\tau_1,\tau_2,\tau_3)=(1,1,1,1,1)$ and $\beta=2$, then we obtain a pair of integer sequences, with the first few 
terms presented in the following table: 
 {\renewcommand{\arraystretch}{3}
\begin{center}
\begin{tabular}{ |c || c | c | c | c | c |c|c | c | } 
  \hline
  $n$ & $0$ & $1$ & $2$ & $3$ & $4$ & $5$ & $6$ & $7$  \\ 
  \hline
  $\sigma_{n}$ & $3$ & $9$ & ${19}$ & ${569}$ & ${30923}$ & $458561$ & $457737691$ & $111996752817$ \\
  \hline 
  $\tau_{n+4}$ & $5$ & ${11}$ & ${69}$ & ${811}$ & ${8089}$ & $161273$ & $8530457$ & $202237625$ \\ 
  \hline
\end{tabular}
\end{center}
}
\noindent 
Observe that the primes appearing separately as isolated factors 
in each of these integer sequences are the same ones that were identified as factors of the numerators and denominators in the 
preceding table. 

The system of recurrence relations  %latter
\eqref{systmB2} can be interpreted as  
iteration of a  birational map $\psi: \, \C^5 \rightarrow \C^5$ which is intertwined with $\varphi$ via $\pi$, that is  
$$ 
\psi: \quad (\sigma_0,\tau_0,\tau_1,\tau_2,\tau_3) \mapsto  (\sigma_1,\tau_1,\tau_2,\tau_3,\tau_4) , 
\qquad   \varphi \cdot \pi = \pi \cdot \psi.  
$$ 
Then we would like to identify 
%beginning  
the initial data for the map $\psi$ as an initial cluster in a  seed for a cluster algebra of rank 5, so that 
$\tilde{\vb{x}}= (\tilde{x}_{1},\tilde{x}_{2}, \tilde{x}_{3},\tilde{x}_{4},\tilde{x}_{5})=  (\sigma_0,\tau_0,\tau_1,\tau_2,\tau_3)$. 
To verify that the Laurent property holds when the deformed map $\varphi$ is lifted to the map $\psi$ on the space of tau functions, we need to find a cluster algebra structure defined by an 
initial seed $(\tilde{\vb{x}},\tilde{B})$, 
for a suitable exchange matrix $\tilde{B}\in \mathrm{Mat}_5(\Z)$. We will then proceed to show that this extends to a seed $(\hat{\vb{x}},\hat{B})$, where the initial cluster 
$\hat{\vb{x}}=(\tilde{\vb{x}},\beta)$ includes the parameter $\beta$ as a frozen variable, and $\hat{B}$ is an extended $6\times 5$ exchange matrix (with an additional row to 
incorporate the frozen variable). 

To start with, we calculate the pullback of the symplectic form \eqref{symB2} by the rational map $\pi$, 
to obtain the presymplectic form  
$$
\tilde{\omega}=\pi^{*}\omega = \sum_{i<j} \tilde{\om}_{ij}, 
$$ 
which gives rise to a new % $\tilde{\omega}$, which defines a 
skew-symmetric matrix, 
\begin{equation}
    \tilde{\Omega} = (\tilde{\omega}_{ij}) = \mqty(0 & -2 & 2 & 2 & -2 \\ 2 & 0 & -4 & 0 & 0 \\ -2 & 4 & 0 & -4 & 4 \\ -2 & 0 & 4 & 0 & 0 \\ 2 & 0 & -4 & 0 & 0 )
\end{equation}
Similar to the matrix in \eqref{B2Om}, the $\tilde{\Omega}$ can be expressed as a product $\tilde{\Omega} = \tilde{B}\tilde{D}$ of skew-symmetrizable matrix $\tilde{B}$ and 
diagonal matrix $\tilde{D}$. By post-multiplying by the diagonal matrix $\tilde{D}^{-1} = \text{diag}(1,\sfrac{1}{2},\sfrac{1}{2},\sfrac{1}{2},\sfrac{1}{2})$, 
this gives the  $5\times 5$ exchange matrix 
\begin{equation}\label{exB}
    \tilde{B}= \tilde{\Omega} \tilde{D}^{-1} = (\tilde{B}_{ij}) = \mqty(0 & -1 & 1 & 1 & -1 \\ 2 & 0 & -2 & 0 & 0 \\ -2 & 2 & 0 & -2 & 2 \\ -2 & 0 & 2 & 0 & 0 \\ 2 & 0 & -2 & 0 & 0 )
\end{equation}
\\

Now observe that if we apply the composition %of mutations 
$\tmu_{2}\tmu_{1}$ for the latter exchange matrix, applying the mutation $\tmu_1$  
associated with index 1 followed by the mutation $\tmu_2$ associated with index 2, then the initial cluster 
$\tilde{\vb{x}}=(\sigma_0,\tau_0,\tau_1,\tau_2,\tau_3)$ gets transformed to 
$\tmu_2\tmu_1(\tilde{\vb{x}})=(\tx_1',\tx_2',\tx_3,\tx_4,\tx_5)=(\sigma_1,\tau_4,\tau_1,\tau_2,\tau_3)$, 
where the new cluster variables $\sigma_1, \tau_4$ are obtained from a single iteration of each of the 
recurrences in \eqref{systmB2}, setting  $n=0$ and $\beta=1$ therein. 
To generate the general sequence of mutations for tau functions that corresponds to \eqref{systmB2} with arbitrary $\beta$, it is necessary to extend the initial cluster 
to $\hat{\vb{x}}=(\tilde{\vb{x}},\beta)$ by inserting the frozen variable $\beta$,  and then a further calculation shows that we can define the  extended exchange matrix  
\begin{equation}\label{eexB}
    \hat{B} = \mqty(0 & -1 & 1 & 1 & -1 \\ 2 & 0 & -2 & 0 & 0 \\ -2 & 2 & 0 & -2 & 2 \\ -2 & 0 & 2 & 0 & 0 \\ 2 & 0 & -2& 0 & 0 \\ -1 & 0 & 0 & 0 & 1) , 
\end{equation}
which is obtained by inserting an extra row at the bottom of \eqref{exB}. The form of the recurrence system \eqref{systmB2} also requires that we permute the cluster variables after applying the two mutations $\tmu_1$ and $\tmu_2$. 

\begin{thm} \label{B2clustermap} 
% $\hat{\vb{x}}=(\sigma_{0},\tau_{0},\tau_{1},\tau_{2},\tau_{3},\beta)$
Let  ${\rho}$ be the permutation $(2543)$. 
Then $\psi = {\rho}^{-1}\tmu_{2}\tmu_{1} $ is a cluster map that fixes the extended exchange matrix 
$\hat{B}$. Iteration of $\psi$ generates two sequences of tau functions  $(\sigma_{n})$, $(\tau_{n})$ satisfying the system  
\eqref{systmB2}. 
The tau functions are elements of %the Laurent polynomial ring 
$\mathbb{Z}_{>0}[\beta,\sigma_{0}^{\pm1},\tau_{0}^{\pm1},\tau_{1}^{\pm1},\tau_{2}^{\pm1},\tau_{3}^{\pm 1} ]. $
\end{thm}
\begin{prf}
    Consider the cluster algebra with initial cluster $\hat{\vb{x}}=(\sigma_{0},\tau_{0},\tau_{1},\tau_{2},\tau_{3},\beta)$ and extended exchange matrix $\hat{B}$. 
One can see that, by applying cluster mutation to $(\sigma_0,\tau_0,\tau_1,\tau_2,\tau_3,\beta)= (\tilde{x}_{1},\tilde{x}_{2},\tilde{x}_{3},\tilde{x}_{4},\tilde{x}_{5},\tx_6)$ 
in direction 1, mutation $\tilde{\mu}_{1}$ gives the exchange relation
\begin{align*}
    \sigma_{1}\sigma_{0} =  \beta \tau_{1}^2\tau_{2}^2 +  \tau_{0}^2\tau_{3}^2 ,
\end{align*}
producing the new cluster $\tmu_1(\hat{\vb{x}})=(\sigma_1,\tau_0,\tau_1,\tau_2,\tau_3,\beta)$
and the mutated exchange matrix $\hat{B}_{1}=\tmu_1(\hat{B})$ given by 
\begin{align*}
    \hat{B}_{1}=\mqty(0 & 1 & -1 & -1 & 1 \\ -2 & 0 & 0 & 2 & 0 \\ 2 & 0 & 0 & -2 & 0 \\ 2 & -2 & 2 & 0 & -2 \\ -2 & 0 & 0 & 2 & 0 \\ 1& -1 & 0 & 0 & 0 ) . 
\end{align*}
Following this up with a mutation in direction 2, applying $\tilde{\mu}_{2}$ gives the new cluster variable $\tau_{4}$ defined by the following relation: 
\begin{align*}
     \tau_{4}\tau_{0} = \beta \tau_{2}^2 + \sigma_{1} . 
\end{align*}
The new cluster is then $\tmu_2\tmu_1(\hat{\vb{x}})=(\sigma_1,\tau_4,\tau_1,\tau_2,\tau_3,\beta)$
Therefore applying the composition of  mutations $\tmu_{2}$ and $\tmu_{1}$ %subsequently 
generates this pair of exchange relations, which corresponds to a single iteration of the map $\psi$, but requires 
an additional cyclic permutation of the middle 4 variables to obtain 
$\psi(\hat{\vb{x}})=(\sigma_1,\tau_1,\tau_2,\tau_3,\tau_4,\beta)$. % in \eqref{systmB2}. 
Furthermore, we see that the combination of two matrix mutations is equivalent to a permutation of order 4 acting on the corresponding 4 non-frozen labels, namely $\rho = (2543)$, i.e. 
\begin{align*}
\tmu_{2}\tmu_{1}(\hat{B}) = P_{1}\hat{B}P_{2} = \rho  (\hat{B})
\end{align*}
where $P_{1}$ and $P_{2}$ are row and column permutation matrices 
\begin{equation}
    P_{1}=\mqty(1 & 0 & 0 & 0 & 0 & 0  \\ 0 & 0 & 0 & 0 & 1& 0 \\ 0 & 1 & 0 & 0 & 0 &  0 \\ 0 & 0 & 1 & 0 & 0 & 0 \\ 0 & 0 & 0 & 1 & 0 & 0 \\0 & 0 & 0 & 0 & 0 & 1), \quad P_{2} =  \mqty(1 & 0 & 0 & 0 & 0 \\ 0 & 0 & 1 & 0 & 0 \\ 0 & 0 & 0 & 1 &  0 \\ 0 & 0 & 0 & 0 & 1 \\ 0 & 1 & 0 & 0 & 0)
\end{equation}
Thus we have shown that the extended exchange matrix $\hat{B}$ given by \eqref{eexB} is cluster mutation periodic 
in the generalized sense defined in \cite{nakanishi}, so that 
$\psi(\hat{B})=\hat{B}$, where the cluster map $\psi = \rho^{-1}\tmu_{2}\tmu_{1}$ generates two sequences of tau functions satisfying 
the coupled system \eqref{systmB2}. Hence, by the Laurent phenomenon in the cluster algebra, %we can state 
it follows that iteration of the %deformed 
map $\psi$ on the space of tau functions produces Laurent polynomials % ring 
that are elements of $\mathbb{Z}_{>0}[\beta,\sigma_{0}^{\pm 1},\tau_{0}^{\pm 1},\tau_{1}^{\pm 1},\tau_{2}^{\pm 1},\tau_{3}^{\pm 1} ]$ (where each monomial that appears has a positive integer coefficient, due to positivity \cite{ls, ghkk}).  
\end{prf}

\subsection{Connection with Somos-5 and a special Somos-7 recurrence }
Analogously to the observation made for the case of $\rA_3$ in the previous section, we note that the formula for $x_{2,n}$ in \eqref{tauB2} 
corresponds to the substitution used for Somos-5 in \cite{hones5}. This allows us to establish 
a connection between the system \eqref{systmB2} and a suitable Somos-5 recurrence relation.

\begin{thm}\label{thmforsomos5}
The sequence of tau functions  $(\tau_{n})$ generated by iteration of  \eqref{systmB2} satisfies a  Somos-5 relation with coefficients that are constant along each orbit, given by 
\begin{equation}\label{somos5}
 \tau_{n}\tau_{n+5} = \zeta\,\tau_{n+1}\tau_{n+4} + \theta\, \tau_{n+2}\tau_{n+3}
 \end{equation}
 where the coefficients are % given by 
 \begin{equation}\label{coeffB2}
     \zeta= 1-\beta, \quad  \theta= \beta\tilde{K} , 
% \frac{\beta \qty((\beta \tau_1^2 + \sigma_0)\tau_2^2 + \tau_0^2\tau_3^2)(\tau_1^2 + \sigma_0)}{\sigma_0\tau_0\tau_1\tau_2\tau_3}
 \end{equation}
with $\tilde{K}$ being the value of the first integral \eqref{dfirstB2}. 
Hence $u_n=x_{2,n}$ satisfies the Somos-5 QRT map \eqref{somos5QRTB2} with 
coefficients $\tilde{\al}=\zeta$, $\tilde{\be}=\theta$ as in  \eqref{coeffB2} 
along any orbit of the deformed $\rB_2$ map \eqref{phitilmapB2}. 
\end{thm}

\begin{prf}
The first three iterations of the Somos-5 sequence can be represented in  matrix form as %the 
\begin{equation}
\underbrace{\mqty(\tau_{0}\tau_{5} & \tau_{1}\tau_{4} & \tau_{2}\tau_{3} \\ \tau_{1}\tau_{6} &\tau_{2}\tau_{5} &  \tau_{3}\tau_{4} \\   \tau_{2}\tau_{7} & \tau_{3}\tau_{6} &  \tau_{4}\tau_{5})}_{M} \mqty(1\\-\zeta\\-\theta) = 0.
\end{equation}
As the vector ${\bf v} = \mqty(1,-\eta,-\theta)^{T}$ is non-zero, $\det(M) = 0 $ is a necessary condition  for the tau functions $\tau_{n}$ obtained from \eqref{systmB2} to satisfy \eqref{somos5}. 
With the help of MAPLE software, we can easily confirm that the relation holds. The coefficients $\zeta$ and $\theta$ can be found by computing the kernel %the vector in the null space 
of $M$, which turns out to be independent under shifting the indices of each tau function ($n\to n+1$): to be precise, $\zeta = 1-\beta$ is just a constant (independent of tau functions), while
$$ 
\theta= \frac{\beta \qty((\beta \tau_1^2 + \sigma_0)\tau_2^2 + \tau_0^2\tau_3^2)(\tau_1^2 + \sigma_0)}{\sigma_0\tau_0\tau_1\tau_2\tau_3}, 
$$
but this is just $\beta$ times the first integral \eqref{dfirstB2} lifted to the space of tau functions. Hence the vector  ${\bf v}$ is constant along each orbit, and 
remains in the kernel of the matrix $M$ when the replacement $\tau_n\to \tau_{n+1}$ is made for each tau function appearing therein.
\end{prf}
We have seen that, subject to the condition $b_{1} = \beta = b_{2}$,
the variable $u_n=x_{2,n}$ satisfying one half of the system \eqref{dB2maps}, 
also  satisfies the Somos-5 QRT map \eqref{somos5QRTB2} with appropriate coefficients $\tilde{\al},\tilde{\be}$. 
This suggests that each invariant curve for the deformed map $\tilde{\varphi}$, given by a level set of %described by \eqref{firstintB2} 
\eqref{dfirstB2},
is birationally equivalent to a corresponding elliptic curve associated with a level set of the  Somos-5 QRT map. According to \cite{hone2020ecm},  
each such curve is also isomorphic to a curve that corresponds  to a level set of the Lyness map 
\begin{equation}\label{lynessB2}
    w_{n+1}w_{n-1} = \tilde{\zeta}w_{n} +\tilde{\theta}
\end{equation}
(for suitable $\tilde{\zeta}$ and $\tilde{\theta}$), 
which is %a birational map of the plane corresponding 
the integrable deformation of the periodic map of type $\rA_{2}$. 
Applying the results from \cite{hone2020ecm}, it can be shown that 
%This indicates that 
the iterates $x_{2,n}$ of the deformed $\rB_2$ map $\tilde{\varphi}$ are also associated with the Lyness map, 
via the transformation 
\begin{align*}
    w_{n} = x_{2,n} + \frac{\theta}{\zeta} = \frac{1}{\zeta} \frac{\tau_{n-1}\tau_{n+4}}{\tau_{n+1}\tau_{n+2}}. 
\end{align*}
The above substitution is consistent %compatible 
with the fact that $\tau_n$ satisfies the %following 
bilinear recurrence 
\begin{equation} \label{specs7}
     \tau_{n+7}\tau_{n} = \tilde{\zeta} \tau_{n+6}\tau_{n+1} + \tilde{\theta}\tau_{n+3}\tau_{n+4}.
\end{equation}
which a special type of Somos-7 recurrence, namely the same as (\ref{lynesstau}) associated with the Lyness map (\ref{lynessgen}).  
This is another type of Somos sequence generated by a sequence of mutations in a cluster algebra of rank 7 (for further detail see \cite{fordymarsh} and \cite{Fordy_2013}). 
The same Somos-7 relation will also be seen to appear in the next section.

\begin{remark}
Another way to see that the existence of the special Somos-7 relation (\ref{specs7}) follows from Theorem \ref{thmforsomos5}, is to apply a result from 
\cite{hones5} (see also \cite{poorten}), which says that every Somos-5 sequence also satisfies a Somos-$k$ relation of odd order, for each odd integer $k\geq 7$: 
so every Somos-5 is also a Somos-7. The converse is not quite true, however: every Somos-7 does satisfy a relation of Somos-5 type, but generically it has one 
coefficient that is periodic with period 3, rather than having two constant coefficients  $\zeta,\theta$ as in \eqref{somos5}. This result is proved in Appendix B, along 
with a number of results about the Somos-7 recurrence \eqref{specs7} that have not been collected elsewhere. 
\end{remark} 

\subsection{Tropicalization and degree growth for deformed $\rB_{2}$ map} 

Given an initial cluster 
$\hat{\vb{x}}=(\sigma_{0},\tau_{0},\tau_{1},\tau_{2},\tau_{3},\beta)$ for the deformed $\rB_{2}$  cluster map, as in Theorem \ref{B2clustermap}, 
we can associate a tropical  cluster of d-vectors in $\Z^5$, encapsulated in the matrix 
\beq\label{tropB2inits}
\Big( {\bf d}_0\,\, {\bf e}_0\,\, {\bf e}_1\,\, {\bf e}_2\,\, {\bf e}_3\Big) = -I, 
\eeq 
where $I$ denotes the $5\times 5$ idenitity matrix. The Laurent property implies that the sequences of tau functions $\sigma_n,\tau_n$ 
generated by  the system \eqref{systmB2} take the form 
$$ \sigma_n = \frac{\rN_n^{(1)}(\hat{\vb{x}})}{\hat{\vb{x}}^{{\bf d}_n}}, \qquad \tau_n = \frac{\rN_n^{(2)}(\hat{\vb{x}})}{\hat{\vb{x}}^{{\bf e}_n}}, $$  
with the numerators $\rN_n^{(1)},\rN_n^{(2)}$ being polynomials in $\Z[\hat{\vb{x}}]$ that are not divisible by any of the initial cluster variables, 
while the basic results on d-vectors in \cite{fz4} imply that the sequences ${\bf d}_n,{\bf e}_n$ appearing as exponents in the denominators satisfy the 
$(\max,+)$   version of this system, given by 
\beq\label{tropB2} 
\begin{split}
    &{\bf d}_{n+1} + {\bf d}_{n} = 2 \max \big( {\bf e}_{n+2}+ {\bf e}_{n+1},  {\bf e}_{n+3}+{\bf e}_{n}\big), \\
    &{\bf e}_{n+4}+ {\bf e}_{n} =\max\big( 2 {\bf e}_{n+2}, {\bf d}_{n+1} \big) . 
    \end{split}
\eeq 
(Note that, as for the tropical version of deformed $\rA_3$ in the previous section, we do not count degrees with respect to  frozen variables, so 
there are no terms corresponding to the parameter  $\beta$ this in the system.)  

We now consider the tropical analogue of the map $\pi$ defined in \eqref{tauB2}, which leads us to introduce the quantities 
\beq\label{troptauB2}
{\bf X}_{1,n}={\bf d}_n-2{\bf e}_{n+1}, \qquad {\bf X}_{2,n}={\bf e}_{n+3}- {\bf e}_{n+2}-{\bf e}_{n+1}+{\bf e}_{n}
\eeq 
The dynamics of these combinations of d-vectors holds the key to their degree growth. 

\begin{lem}\label{B2per3} Whenever ${\bf d}_n, {\bf e}_n$ satisfy the system (\ref{tropB2}),  each component of the quantities 
\eqref{troptauB2} is a solution of the tropical (or ultradiscrete) QRT map $\varphi_{trop}$ defined by 
\beq\label{tropQRTB2} 
\begin{array}{rcl}
X_{1,n+1}+X_{1,n}& = & 2[X_{2,n}]_+, \\ 
X_{2,n+1}+X_{2,n}& = & [X_{1,n+1}]_+. 
\end{array}
\eeq  
Given arbitrary initial data $\big( X_{1,0} , X_{2,0} \big)\in\R^2$, the orbit of the latter map is periodic with period 3. 
\end{lem}  
\begin{prf} The map $\varphi_{trop}$ presented in \eqref{tropQRTB2} is the $(\max,+)$ analogue of the QRT map \eqref{phitilmapB2}, which 
follows immediately from the system  (\ref{tropB2}) by by rearranging and rewriting it in terms of the components of the quantities defined in 
\eqref{troptauB2}. The fact that every orbit is period 3 can be checked by a direct case-by-case analysis for initial data in different sectors of the plane. 
It can also be deduced from Nobe's general results on periods of 
ultradiscrete QRT maps \cite{nobe}.
\end{prf} 

\begin{remark}\label{per3rema} 
For the main case of interest, the initial tropical  cluster in \eqref{tropB2inits} produces the pair of vectors of initial values
$$  
{\bf X}_{1,0}= (-1,0,2,0,0)^T, \qquad {\bf X}_{2,0}=(0,-1,1,1,-1)^T
$$ 
for \eqref{tropQRTB2}, and this initial pair   
produces the subsequent pairs of terms 
$$  
{\bf X}_{1,1}= (1,0,0,2,0)^T, \qquad {\bf X}_{2,1}=(1,1,-1,1,1)^T
\qquad \mathrm{and} \qquad  
{\bf X}_{1,2}= (1,2,0,0,2)^T, \qquad {\bf X}_{2,2}=(0,1,1,-1,1)^T
$$ 
under iteration, with the iterates repeating thereafter. 
The components of these vectors 
are all built from two particular orbits of  
the scalar map \eqref{tropQRTB2}, namely 
\beq\label{B2roots}
(-1,0)\to (1,1)\to (1,0), \qquad  (0,-1)\to (0,1)\to (2,1),
\eeq  
which correspond precisely to the sequence of pairs of d-vectors arising from 
the Zamolodchikov periodicity  of the orbit of the original undeformed $\rB_2$ map, defined by \eqref{B2maps} with all parameters $a_i,b_i$ set to 1. 
Indeed, these two orbits can be read off from the denominators in the corresponding sequence of $\rB_2$ cluster variables, viz.
$$ 
(x_1,x_2) \longrightarrow 
\left(\frac{x_2^2+1}{x_1}, \frac{x_2^2+x_1+1}{x_1x_2}\right) \longrightarrow 
 \left(\frac{x_2^2+(x_1+1)^2}{x_1x_2^2}, \frac{x_1+1}{x_2}\right).
$$
For finite type cluster algebras, 
it is known that there is a direct correspondence between cluster variables and a specific subset of the roots in the associated 
root system, whereby the coefficients of linear combinations of simple roots determine the d-vectors (see e.g.\ Theorem 4.10 in \cite{fr}).    
In this case, the pairs of d-vectors corresponding to the exponents of $x_1,x_2$ in the monomials above are given by the sequence of 
$2\times 2$ matrices 
$$ 
\left(\begin{array}{cc} -1 & 0 \\ 0 & -1 \end{array}\right) \longrightarrow  
\left(\begin{array}{cc} 1 & 1 \\ 0 & 1 \end{array}\right) \longrightarrow  
\left(\begin{array}{cc} 1 & 0 \\ 2 & 1 \end{array}\right) , 
$$  
whose first and second rows, respectively, yield the two orbits 
\eqref{B2roots} above. 
%and correspond to the sequence of pairs of roots $(-\alpha_1,-\alpha_2)\to (\alpha_1,\alpha_1+\alpha_2) \to (\alpha_1+2\alpha_2,\alpha_2)$ in the $\rB_2$ root system. 
% Looks like our labelling CONVENTION IS C_2 (isomorphic) rather than B_2 !!! 
To determine the degree growth of the tau functions generated by  \eqref{systmB2}, we only need to make use of the fact that these particular 
sequences have period 3, rather than the more general result of Lemma \ref{B2per3}. 
\end{remark} 

\begin{thm}\label{B2degrees} 
All sequences of d-vectors ${\bf d}_n,{\bf e}_n$ that are solutions of the $(\max,+)$ system \eqref{tropB2} lie in the kernel of the 6th 
order linear difference operator 
$$
\hat{\cal L}= ({\cal T}^3-1) ({\cal T}^2-1)({\cal T}-1). 
$$
Hence the particular d-vectors corresponding to  the sequences of tau functions $\sigma_n,\tau_n$ produced by  \eqref{systmB2} 
give the leading order degree growth
\beq\label{B2leading} 
{\bf d}_n = \frac{n^2}{6}\, (1,1,1,1,1)^T +O(n), \qquad  {\bf e}_n = \frac{n^2}{12}\, (1,1,1,1,1)^T +O(n). 
\eeq 
\end{thm}
\begin{prf} 
The second relation in \eqref{troptauB2} can be written as 
$ ({\cal T}^2-1)({\cal T}-1)\,{\bf e}_n = {\bf X}_{2,n}$,  
and then by Lemma \ref{B2per3} we have 
$$ 
\hat{\cal L}\, {\bf e}_n = ({\cal T}^3-1) ({\cal T}^2-1)({\cal T}-1)\, {\bf e}_n =({\cal T}^3-1)\, {\bf X}_{2,n}=0, 
$$ 
which shows that ${\bf e}_n$ is annihilated by the linear operator  $\hat{\cal L}$ 
(which is the same operator that appeared in \eqref{order6T} in the deformed $\rA_3$ case). 
Then the    first relation in \eqref{troptauB2} implies that 
$$ 
\hat{\cal L}\, {\bf d}_n = \hat{\cal L}\, ( {\bf Y}_{1,n} -{\bf e}_{n+1} ) = ({\cal T}^2-1)({\cal T}-1) ({\bf Y}_{1,n+3}-{\bf Y}_{1,n}) -\hat{\cal L}\,{\bf e}_{n+1}= 0, 
$$ 
as required, where we used the fact that ${\bf Y}_{1,n}$ has period 3, from the same lemma.  Now for the particular  tropical 
seed \eqref{tropB2inits}, which corresponds to the d-vectors of the initial set of tau functions in \eqref{systmB2}, we can use $(\max,+)$ relations \eqref{tropB2} to calculate more terms in the sequence of d-vectors. We can also use the corresponding 
3-periodic sequence of solutions of the tropical QRT map, as presented in Remark \ref{per3rema},  to generate additional terms the sequence $({\bf e}_n)$ from the formula 
$$ 
{\bf e}_{n+3}= {\bf e}_{n+2}+{\bf e}_{n+1}-{\bf e}_{n}+{\bf X}_{2,n},
$$ 
which produces 
$$ 
{\bf e}_4=(1,1,0,0,0)^T, \qquad {\bf e}_5 =(1,2,1,0,0)^T,
$$ 
so that ${\bf e}_j$, $0\leq j\leq 5$ provide enough initial values to generate the 6 vector constants that specify the explicit solution of the linear 
difference equation  $\hat{\cal L}\, {\bf e}_n = 0$ in terms of its characteristic roots. In fact, for this choice of initial values it is straightforward 
to show by induction that the components 
of  ${\bf e}_n$ take the form 
$$ 
{\bf e}_n = (\tilde{e}_n, e_{n+3},e_{n+2},e_{n+1},e_n)
,
$$ so these vectors are completely specified in terms of the following two scalar recurrence sequences:   
$$ \begin{array}{rcl}
(\tilde{e}_n): &\quad & 0,0,0,0, 1,1,2,3,4,5,7,8,10,12,14,16,19,21,\ldots; \\  
(e_n): & \quad &  0,0,0,-1,0,0,0,1,2,2,4,5,6,  8,  10,11,14,16,\ldots .   
\end{array} 
$$ 
The sequences $\tilde{e}_n,e_n$ determine, respectively,  
the exponents of $\sigma_0$, and the exponents of the other $\tau_j$ for $j=0,1,2,3$, that appear in the denominators of $\tau_n$.
We omit the full details of the explicit formula for the sequence of vectors 
${\bf e}_n$, 
and merely note that we can  determine
the leading order behaviour ${\bf e}_n ={\bf a} \, n^2\big(1+o(n)\big) $  by  calculating the constant vector 
$$({\cal T}^3-1) ({\cal T}^2-1)\, {\bf e}_n = 12{\bf a}=(1,1,1,1,1)^T,$$ which fixes the value of ${\bf a}$. Similarly, 
a simple inductive argument can be used to show that ${\bf d}_n$ has components of the form 
$$ 
{\bf d}_n = (\tilde{d}_n, d_{n+3},d_{n+2},d_{n+1},d_n)
,
$$ 
specified by the scalars $\tilde{d}_n,d_n$. The latter two scalar sequences 
determine, respectively,  
the exponents of $\sigma_0$, and the exponents of the other $\tau_j$ for $j=0,1,2,3$, that appear in the denominators of $\sigma_n$; their  
initial terms are as follows: 
$$ \begin{array}{rcl}
(\tilde{d}_n): & \quad & -1,1,1,1,3,5,5,9,11,13,17,21,23,29,33,37,43,49,\ldots;   \\
(d_n): & \quad &        \,\,\,\,   0,0,0,0, 0,2,2,4,6,  8, 10,14,16,20,24,28,32,38,\ldots
.
\end{array} 
$$
We omit the complete details of explicit formula for   ${\bf d}_n$, 
but rather make use of the   first relation in \eqref{troptauB2} 
once again, to see that the leading order behaviour of the sequence 
is found from 
$$
{\bf d}_n = 2{\bf e}_{n+1} + {\bf X}_{1,n} = 2{\bf a}\, n^2 \big(1+o(n)\big),  
$$ 
with the same constant vector ${\bf a}$, since  ${\bf X}_{1,n}$ has period 3. 
Together, 
this yields the required expression for the leading order quadratic growth of 
${\bf d}_n,{\bf e}_n$ as in \eqref{B2leading}, with an $O(n)$ correction in each case. 
\end{prf}

\section{Integrable deformations of the  $\rB_3$ cluster map} 
\setcounter{equation}{0}

In this section, we consider the deformation of a 3D periodic cluster map which arises from mutations in the
cluster algebra of type $\rB_3$. The original cluster map in 3 dimensions has period $4 = \frac{1}{2}(6 + 2)$, which is 
$ \frac{1}{2}
\times ( \text{Coxeter number} + 2 )$, and as before our aim is to construct parameter-families of deformations of this map that result in aperiodic dynamics that is Liouville
integrable. However, in contrast to all the examples previously considered here and in \cite{hk}, for the $\rB_3$ we find that there is more than one distinct family of deformations 
that is integrable (in fact, precisely two distinct 1-parameter families, up to obvious equivalence via scaling transformations). 

\subsection{Deformed map $\rB_{3}$} 

%with parameters}
For the $\rB_3$ root system, the Cartan matrix is 
\begin{equation*}
C= \left( \begin{array}{ccc} 2 & -2 & 0  \\
-1 & 2 & -1 \\
0 & -1 & 2  \\
\end{array} \right),
\end{equation*}
which is the companion of the skew-symmetrizable exchange matrix,
\begin{equation*}
B= \left( \begin{array}{ccc} 0 & 2 & 0  \\
-1 & 0 & 1 \\
0 & -1 & 0  \\
\end{array} \right),
\end{equation*}
Skew-symmetrizability of $B$ is seen from the fact that $\Om=BD=(\om_{ij})$ is skew-symmetric,  where $D=\mathrm{diag}(2,1,1)$ and  %we have
\begin{equation*}
\Omega= \left( \begin{array}{ccc} 0 & 2 & 0  \\
-2 & 0 & 1 \\
0 & -1 & 0  \\
\end{array} \right).
\end{equation*}
We now consider the sequence of deformed mutations 
\beq \label{B3dmaps}\begin{array}{rcl}
\mu_1: \quad (x_1,x_2,x_3)\mapsto (x_1',x_2,x_3), \qquad x_1'x_1 & = & b_1+a_1 x_2, \\
 \mu_2: \quad (x_1',x_2,x_3)\mapsto (x_1',x_2',x_3), \qquad x_2'x_2 & = & b_2+a_2( x_1')^2x_3, \\ 
\mu_3: \quad (x_1',x_2',x_3)\mapsto (x_1',x_2',x_3'), \qquad x_3'x_3 & = & b_3+a_3 x_2',
\end{array}
\eeq 
where $a_j,b_j$ are arbitrary parameters. With a  generic choice of these parameters, the Laurent property no longer holds for these mutations, so the 
map ${\varphi}=\mu_3\mu_2\mu_1$ does not have the Laurent property; moreover, it is no longer completely periodic with period 4. 
%(generically, it defines an automorphism of infinite order of the field of rational functions on $\C^3$). 
However, similarly to the $\rB_2$ case,  by a minor variation on Theorem 1.3 in \cite{hk}, generalizing it % from the case of  skew-symmetric $B$ 
to the skew-symmetrizable case, it is 
not hard to see that the deformed map ${\varphi}$ given by the composition of transformations \eqref{B3dmaps} 
preserves the same presymplectic form $\om=\sum_{i<j}\om_{ij} \rd \log x_i \wedge \rd \log x_j$ as in the undeformed case.

Before considering the deformed case (\ref{B3dmaps}) further, there are two ways to simplify the calculations. Firstly, 
assuming the case of generic parameter values $a_ib_i\neq 0$ for all $i$, 
we apply the scaling action of the three-dimensional algebraic torus $(\C^*)^3$, given by $x_i\to \la_i\, x_i$, $\la_i\neq0$, and use this to remove three parameters, so that we 
may set 
$$ 
a_i\to 1, \qquad i=1,2,3,
$$ 
without loss of generality, but keep $b_i$ arbitrary for $ i=1,2,3$.  
Having simplified the space of parameters, the map $\varphi$ is equivalent to the iteration of the 
system of recurrences 
\beq\label{A3recs} 
\begin{array}{rcl}
x_{1,n+1}x_{1,n} & = & x_{2,n}+b_1, \\  
x_{2,n+1}x_{2,n} & = & x_{1,n+1}^2x_{3,n}+b_2, \\  
x_{3,n+1}x_{3,n} & = & x_{2,n+1}+b_3.   
\end{array} 
\eeq 
Secondly, because we are in an odd-dimensional situation where necessarily $\det(\Omega)=0$  
and 
$\om$ is degenerate,  we can use 
$$ 
\mathrm{ker}\,\Omega=<(1,0,2)^T>, \qquad 
\mathrm{im}\, \Omega=  (\mathrm{ker}\,\Omega)^\perp = <(0,1,0)^T, (-2,0,1)^T>
$$ 
to generate the one-parameter scaling group $(x_1,x_2,x_3)\to (\la x_1,x_2,\la^2 x_3)$, $\la\in\C^*$ (obtained from the null vector 
field $x_1\partial_{x_1}+2x_3\partial_{x_3}$ by exponentiation), 
and the projection $\pi:\, \C^3\rightarrow\C^2$ onto its monomial invariants,  
$$ 
\pi: \qquad 
y_1=x_2, \qquad y_2=\frac{x_3}{x_1^2}.  
$$ 
On the $(y_1,y_2)$-plane, $\varphi$ induces the 
reduced map $\hat{\varphi}$, such that $\pi\cdot\varphi = \hat{\varphi}\cdot \pi$, where 
\beq\label{phihat} 
\hat{\varphi}: \qquad
\left(\begin{array}{c} y_1\\ y_2 \end{array}\right) \mapsto 
\left(\begin{array}{c} y_1^{-1} \big((y_1+b_1)^2y_2+b_2\big)\\ 
y_1^{-1}
\Big(1+\frac{b_3y_1+b_2}{y_2(y_1+b_1)^2}\Big)
\end{array}\right).
\eeq 
The reduced map is symplectic, that is to say  $\hat{\varphi}^*(\hat{\om}) =\hat{\om}$, where the non-degenerate two-form  preserved by $\hat{\varphi}$ is 
\beq\label{omhat} 
\hat{\om}=\rd \log y_1 \wedge \rd \log y_2, \qquad \pi^*\hat{\om}=\om. 
\eeq 

In the original case where all parameters are 1, the reduced map (\ref{phihat}) with $b_1=b_2=b_3=1$ has period 4, because 
$x_{2,n+4}=x_{2,n}$ and $x_{3,n+4}/x_{1,n+4}^2=x_{3,n}/x_{1,n}^2$ for all $n$. In that case we can construct two 
functionally independent first integrals in the plane, $K^{(i)}, K^{(ii)}$ say.
Here we will just focus on one of these, namely 
\beq\label{B3int1} \begin{array}{rcl}
K^{(i)}&:=&\sum_{i=0}^3 (\hat{\varphi}^*)^i(y_1) 
\\
&=&y_1y_2+y_1+3y_2 +3\frac{y_2}{y_1}+\frac{y_2}{y_1^2}+\frac{5}{y_1}+\frac{1}{y_2} 
+\frac{2}{y_1^2}+\frac{2}{y_1y_2}+\frac{1}{y_1^2y_2},
\end{array} 
\eeq  
which satisfies $\hat{\varphi}^*(K^{(i)})=K^{(i)}$ when $b_1=b_2=b_3=1$.

Next, we modify $K^{(i)}$  by inserting constant coefficients in front of each of the Laurent monomials  in $y_1,y_2$ that appear, 
fixing the coefficient of the first term to be 1 without loss of generality, to obtain 
\beq\label{Khat}
\hat{K}=
y_1y_2+c_1y_1+c_2y_2 +c_3\frac{y_2}{y_1}+c_4\frac{y_2}{y_1^2}+\frac{c_5}{y_1}+\frac{c_6}{y_2} 
+\frac{c_7}{y_1^2}+\frac{c_8}{y_1y_2}+\frac{c_9}{y_1^2y_2}.
\eeq 
If we assume that these modified first integrals are preserved by the deformed map $\hat{\varphi}$ given by 
(\ref{phihat}), then this puts a finite number of constraints on the coefficients $c_i$ and the parameters $b_i$, which leads to finding necessary and sufficient conditions for the deformed symplectic map to be Liouville integrable.  Thus we obtain the following result.

\begin{thm} \label{imap}
For the deformed symplectic map (\ref{phihat}) to admit a first integral of the form (\ref{Khat}), 
it is 
necessary and sufficient 
that the parameters $b_i$ should satisfy either 
\beq\label{const1} 
b_1=b_2, \quad b_3=1, 
\eeq 
or 
\beq\label{const2} 
b_2=b_3=b_1^2.
\eeq 
If we fix $b_1=\be$, then in the case that the constraint (\ref{const1}) holds, the first integral takes the form 
\small
\beq\label{k1st} 
\hat{K}_1=
y_1y_2+y_1+(2\be +1)y_2 +\be(\be+2)\frac{y_2}{y_1}+\be^2\frac{y_2}{y_1^2}+\frac{3\be+2}{y_1}+\frac{1}{y_2} 
+\frac{2\be}{y_1^2}+\frac{2}{y_1y_2}+\frac{1}{y_1^2y_2},
\eeq 
\normalsize
while in the case that (\ref{const2}) holds, the first integral is 
\small
\beq\label{k2nd} 
\hat{K}_2=
y_1y_2+y_1+(2\be +1)y_2 +\be(\be+2)\frac{y_2}{y_1}+\be^2\frac{y_2}{y_1^2}+\frac{2\be^2+2\be+1}{y_1}+\frac{1}{y_2} 
+\frac{2\be^2}{y_1^2}+\frac{\be^2+1}{y_1y_2}+\frac{\be^2}{y_1^2y_2}.
\eeq 
\normalsize
Hence the map $\hat{\varphi}$ given by  (\ref{phihat}) is Liouville integrable whenever either condition (\ref{const1}) or (\ref{const2}) holds. 
\end{thm}

Thus we arrive at two 1-parameter families of integrable maps of the plane associated with the deformation of the $\rB_3$ cluster map, 
namely 
\beq\label{map1} 
\hat{\varphi}_{1}: \qquad
\left(\begin{array}{c} y_1\\ y_2 \end{array}\right) \mapsto 
\left(\begin{array}{c} y_1^{-1} \big((y_1+\be)^2y_2+\be\big)\\ 
y_1^{-1}
\Big(1+\frac{1}{y_2(y_1+\be)}\Big)
\end{array}\right),
\eeq 
which has the first integral $\hat{K}_1$ given by  (\ref{k1st}), 
and 
\beq\label{map2} 
\hat{\varphi}_{2}: \qquad
\left(\begin{array}{c} y_1\\ y_2 \end{array}\right) \mapsto 
\left(\begin{array}{c} y_1^{-1} \big((y_1+\be)^2y_2+\be^2\big)\\ 
y_1^{-1}
\Big(1+\frac{\be^2(y_1+1)}{y_2(y_1+\be)^2}\Big)
\end{array}\right),
\eeq 
with the first integral $\hat{K}_2$, as in (\ref{k2nd}). So each map has an invariant pencil of genus 1 curves of degree 5 and bidegree $(3,2)$; that 
is too high for a QRT map, where the bidegree is $(2,2)$ \cite{duistermaat}. Clearly 
the maps coincide for $\be=1$ when the map is completely periodic with period 4 (corresponding to a pencil of elliptic curves with 4-torsion). 
However, it seems that the maps cannot be birationally conjugate to one another for other values of $\be$; one way to see this is to look at the 
j-invariants of the curves in each pencil, which are rational functions of $\be$ and the value of the invariant $\hat{K}_j=\kappa$ (for $j=1,2$, respectively): the factorizations 
of the two different j-invariants have polynomial factors in their denominators that appear with quite different degrees, and this could  be used to show that 
there is no automorphism of $\C(\be,\kappa)$ which transforms one elliptic fibration into the other; or perhaps there is a geometrical way to see this 
more easily. It is possible to see that the two maps cannot be conjugate to one another more directly, by considering the fixed points: for generic $\be$, in the affine plane $\C^2$ the map (\ref{map1}) has three fixed points outside the line $y_1=0$ where it is singular, whereas the 
map (\ref{map2}) only has one fixed point outside this line.

\subsection{The deformed map $\hat{\varphi}_{1}$ for $\rB_3$} % $(b_{1}=b_{2}=\beta,\quad b_{3} =1 )$}
Let us consider the deformed map \eqref{map1}, which can be rewritten as the pair of recurrence relations 
\beq\label{B31recs} 
\begin{array}{rcl}
y_{1,n+1}y_{1,n} & = & (y_{1,n} + \beta)^{2}y_{2,n} + \beta \\  
y_{2,n+1}y_{2,n}y_{1,n}(y_{1,n} + \beta) & = & (y_{1,n} + \beta)y_{2,n} + 1 \\  
\end{array} 
\eeq 
Following the same process as in the previous section, we study the singularity structures of the deformed map \eqref{map1} by observing the $p$-adic properties of iterates defined over $\mathbb{Q}$. Then we find that there are two singularity patterns for $y_{1,n}$ and $y_{2,n}$, namely 
\begin{equation}\label{singB3i}
\begin{split}
    &\text{Pattern 1 :} \ (y_{1,n},y_{2,n}) = \dots,(R,\infty^{1}), (\infty^{1},R),(\infty^{1},0^{1}),(R,0^{2}),(-\beta,\infty^{2}),\dots\\[0.5em]  
    &\text{Pattern 2 :} \  (y_{1,n},y_{2,n}) =\dots, (0,R),\dots .
\end{split}
\end{equation}
This indicates that the $y_{1,n}$ and $y_{2,n}$ can be written in terms of tau functions $\tau_{n}$ and $\eta_{n}$ as  
\begin{equation}\label{B31tau}
    y_{1,n} = \frac{\eta_{n}}{\tau_{n+2}\tau_{n+3}}, \quad y_{2,n} = \rho_{n}\, \frac{\tau_{n+1}^{2}\tau_{n+2}}{\tau_{n}^{2}\tau_{n+4}}
\end{equation}
where the quantity $\rho_{n}$ is an additional prefactor. However, substituting these expressions directly into \eqref{B31recs} gives rise to relations between 
tau functions that are not in the form of cluster exchange relations. So in addition to $p$-adic analysis, we proceed to consider the singularity patterns more closely %which is determined 
via explicit analysis with the   introduction of a small quantity $\epsilon$.

A discrete dynamical system defined by a birational map can have 
two types of singularities:  
the points in phase space at which the map is undefined, and the points where the Jacobian of the map vanishes. 
From \eqref{map1}, one can see that the deformed map $\hat{\varphi}_{1}$ possesses a singularity at $y_{1} = -\beta$. Performing singularity analysis by setting $y_{1,n} = -\beta + \epsilon$, we find the  
confined singularity pattern
\begin{equation}\label{patternB3i}
        \mqty(-\beta \\ C) \to \mqty(-1 \\ \infty^{1}) \to \mqty(\infty^1 \\ -1) \to \mqty(\infty^1 \\ 0^1) \to \mqty(-1 \\ 0^2) \to \mqty(-\beta \\ \infty^2) \to \mqty(C' \\ -1/\beta)  , 
    \end{equation}
where $C,C'$ are regular values, and when $\epsilon\to 0$ the subsequent terms are not indeterminate (they are generic, regular values). 
By comparing \eqref{patternB3i} with \eqref{singB3i}, it is clear that  \eqref{patternB3i} corresponds to Pattern 1, but with more detail revealed. 
The detailed form of the singularity pattern suggests another way to  relate $y_{1,n}$ to the tau function $\tau_n$, after shifting by the parameter $\beta$,  
%$y_{1,n}$ can be 
expressing it as 
\begin{equation}\label{lyntauB3i} % {B31tau}
    y_{1,n} = -\beta + \vartheta_n \,  \frac{\tau_{n+5}\tau_{n}}{\tau_{n+3}\tau_{n+2}},  
\end{equation}
where $\vartheta_n$ is another prefactor. %
Defining a new variable $w_{n} = y_{1} +\beta$ leads to a system of three recurrence relations, expressed in terms of $w_{n}, y_{1,n}$ and $y_{2,n}$. Furthermore, subtracting the first relation in \eqref{B31recs} from $w_n$ times the second and removing a common factor of $y_{1,n}$ results in simplifying the recurrence for $y_{2,n}$, yielding the three equations   
\beq\label{B31recs1} 
\begin{array}{rcl}
w_{n} & = & y_{1,n}+\beta \\
y_{1,n+1}y_{1,n} & = & w_{n}^{2}y_{2,n} + \beta \\  
y_{2,n+1}y_{2,n}w_{n}^2 & = & y_{1,n+1} + 1 \\  
\end{array} 
\eeq 
The way that the tau function $\tau_n$ appears in \eqref{lyntauB3i} suggests that there is a close connection between the iteration of $w_{n}$ in \eqref{B31recs1} and the Lyness map. 
\begin{thm}\label{somos7thm}
    Under the iteration of \eqref{B31recs1}, the quantity $w_{n}$ satisfies the Lyness recurrence  
     \begin{equation}\label{lynessw} 
        w_{n+1}w_{n-1} = \tilde{\alpha} w_{n} + \tilde{\beta}, 
    \end{equation}
where the coefficients along each orbit of the map $\hat{\varphi}_1$ are  $\tilde{\alpha} = 1 - \beta$ and $\tilde{\beta} = \beta \hat{K}_{1} + 2 \beta^2 + \beta + 1$.
\end{thm}
\begin{prf}
By following the same approach as used in the proof of Theorem \ref{thmforsomos5}, after setting the prefactor $\vartheta_n\to 1$ in \eqref{lyntauB3i},  
one can show that $\tau_n$ satisfies a special Somos-7 relation of the same form as 
\eqref{specs7}, namely 
\begin{equation}\label{somos7B3i}
        \tau_{n+7}\tau_{n} = \tilde{\alpha} \tau_{n+6}\tau_{n+1} + \tilde{\beta}\tau_{n+3}\tau_{n+4}.   
\end{equation}  
where the $\tilde{\alpha} = 1 - \beta$, and the coefficient $\tilde{\beta}$ is given as above in terms of $\be$ and the conserved quantity 
$\hat{K}_1$. Then from \eqref{lyntauB3i} and the first relation in \eqref{B31recs1}, it is clear that $w_n$ is given in terms of the tau function $\tau_n$ by   
\beq\label{ws7gauge}  %{somos7thm}
w_n = 
\frac{\tau_{n+5}\tau_{n}}{\tau_{n+3}\tau_{n+2}},  
\eeq 
and from this it is an immediate consequence that $w_{n}$ satisfies the Lyness recurrence \eqref{lynessw} with these coefficients, which are constant along each orbit.
\end{prf}
%
%Therefore, given that $w_{n}= y_{1}+\beta = \tau_{n+2}\tau_{n-3} /(\tau_{n}\tau_{n-1})$, the tau-function  $\tau_{n}$ holds for Somos-7 recurrence relation,
%
As noted in the introduction, the special Somos-7 recurrence \eqref{somos7B3i} corresponds to a cluster algebra of rank 7, where $(\tau_n)$ is a sequence of cluster variables and 
the coefficients $\tilde{\alpha},\tilde{\be}$ are regarded as frozen variables; and in that setting, the associated exchange matrix has rank 2 
(for further details, see \cite{Fordy_2013}).  
In the discussion of the deformed $\rB_2$ map, we already noted that there is a close connection between this special Somos-7 relation and Somos-5. 
This leads to a related result for the map   defined by \eqref{B31recs}. 
\begin{thm}\label{somos5thm}
    Under iteration of \eqref{B31recs}, the quantity $v_{n} = y_{1,n} + 1$ satisfies the Somos-5 QRT map, in the form  
    \begin{equation}\label{s5vrec} 
         v_{n+1}v_{n}v_{n-1} = \hat{\alpha} v_{n} + \hat{\beta}
    \end{equation}
where the coefficients along each orbit of the map 
$\hat{\varphi}_1$ are given by $\hat{\alpha} = \tilde{K}_{1} + \beta + 3$ and $\hat{\beta} = (\beta-1)\hat{\alpha} $. 
    \end{thm}
\begin{prf}
   % The coefficients $\hat{\alpha}$ and $\hat{\beta}$ can be determined by taking the same steps as in the proof of theorem \ref{thmforsomos5}. Alternatively, one can find 
This result, including the above formulae for the  coefficients $\hat{\alpha},\hat{\be}$, is a %immediate 
consequence of  
Theorem 1 in \cite{hone2020ecm}, which states that each invariant curve of the Lyness map is birationally equivalent to an 
invariant curve corresponding to the Somos-5 QRT map, and hence there is a direct 
correspondence between the  
orbits of the two maps, whenever the parameters of the maps related to each other in a specific way. See also Proposition B.2 in the second appendix below.  
\end{prf}
Recall from the discussion around Theorem  \ref{thmforsomos5} that a 
substitution of the form  %v_{n} = y_{1,n} + 1$ \eqref{somosQRTtau} 
\begin{equation}\label{somosQRTtau}
   v_n =   \frac{\hat{\tau}_{n+4}\hat{\tau}_{n+1}}{\hat{\tau}_{n+3}\hat{\tau}_{n+2}}
\end{equation}
relates \eqref{s5vrec} directly  to the  Somos-5 recurrence, that is   
    \begin{equation}\label{somos5B3i}
       \hat{\tau}_{n+5}\hat{\tau}_{n} = \hat{\alpha}\hat{\tau}_{n+1}\hat{\tau}_{n+4} + \hat{\beta} \hat{\tau}_{n+2}\hat{\tau}_{n+3}. 
    \end{equation}
Now the substitution \eqref{somosQRTtau} and the definition of the quantity $v_n$ implies that 
$$
y_{1,n} =-1 + \frac{\hat{\tau}_{n+4}\hat{\tau}_{n+1}}{\hat{\tau}_{n+3}\hat{\tau}_{n+2}}, 
$$
but in general this is not compatible with the substitution (\ref{ws7gauge}) that relates $w_n$ to a solution of \eqref{somos7B3i}, in the sense that 
the tau functions $\tau_n$ and $\hat{\tau}_n$ need not be the same, but rather are related by a gauge factor that depends on $n$. 
Rather, the most general way to relate $v_n$ to $\tau_n$ is to write 
\begin{equation}\label{somos5B3ii} %{somos7B3i}
  v_n=  y_{1,n} + 1 =  \xi_n\, \frac{\tau_{n+4}\tau_{n+1}}{\tau_{n+3}\tau_{n+2}}, 
\end{equation}
with another prefactor $\xi_n$ that depends on $n$. It will turn out that, with an appropriate choice of gauge, this quantity is periodic with period 3. (See Theorem \ref{nonlaurent} 
below, and Appendix B.)  
%Since the  Somos-7 can be described by cluster algebra, the Laurent phenomenon suggests that $\tau_{n}$ holds the Laurent property.

Observe that, with the extra variable $v$ added to $y_1,y_2$ and $w$, the map   defined by \eqref{B31recs} is equivalent to  iteration of a system of four equations, namely  
\beq\label{B31recs2} 
\begin{array}{rcl}
w_{n} & = & y_{1,n}+\beta \\
y_{1,n+1}y_{1,n} & = & w_{n}^{2}y_{2,n} + \beta \\
v_n & = &    y_{1,n}+1 \\
y_{2,n+1}y_{2,n}w_{n}^2 & = & v_{n+1} ,   
\end{array} 
\eeq 
and upon substituting for $y_1,y_2$ from \eqref{B31tau}, for $w$ from  \eqref{lyntauB3i}, and for $v$ from \eqref{somos5B3ii} % 
the most general set of relations between the tau functions is found to be the following: 
\begin{equation}\label{systmtauB3i} 
    \begin{split}
    &\vartheta_n\, \tau_{n+5}\tau_{n} =  \beta\, \tau_{n+3}\tau_{n+2} + \eta_{n}  \\ 
    & \eta_{n+1}\eta_{n} = \rho_{n} (\vartheta_n)^2\, \tau_{n+5}^2\tau_{n+1}^2 + \beta\, \tau_{n+4}\tau_{n+3}^{2}\tau_{n+2} 
    \\
    &\xi_{n}\tau_{n+4}\tau_{n+1} = \tau_{n+3}\tau_{n+2} + \eta_{n} \\
    &\rho_{n+1}\rho_{n}(\vartheta_n)^2=\xi_{n+1} . 
    \end{split}
\end{equation}
\begin{thm} \label{nonlaurent} 
There is a choice of gauge which fixes 
$\vartheta_n\to 1$ in the system \eqref{systmtauB3i}, and implies that $\xi_{n+3}=\xi_n$ and $\rho_{n+6}=\rho_n$ for all $n$, 
with 
\beq\label{perxirho}
 \prod_{i=0}^{5}\rho_{i} = \prod_{j=1}^3 \xi_j =\hat{K}_{1} + \beta + 3
. 
\eeq  
In that case, the system 
corresponds to a lift of the deformed $\rB_3$ map $\hat{\varphi}_1$ to a birational map on an extended space of tau functions, that is 
\beq\label{bigphi} 
\Phi: \,\, (\tau_0,   \tau_1, \tau_2, \tau_3, \tau_4, \eta_0, \rho_0, \be) \mapsto 
(  \tau_1, \tau_2, \tau_3, \tau_4, \tau_5, \eta_1, \rho_1, \be), 
\eeq 
where the sequences $(\tau_n)$, $(\eta_n)$ possess the Laurent property, but the periodic coefficients $\rho_n$ do not. 
\end{thm} 
\begin{prf} 
By definition, in the context of the tau function formulae \eqref{B31tau}, a gauge transformation is any transformation of the 
tau functions which leaves the variables $y_{1,n},y_{2,n}$ invariant. If we make the replacement $\tau_n \to g_n \, \tau_n$, where 
the dependence of $g_n$ on $n$ is arbitrary, then clearly replacing $\eta_n\to g_{n+2}g_{n+3}\, \eta_n$ leaves $y_{1,n}$ the same, 
while replacing $\rho_n \to g_n^2 g_{n+4}g_{n+1}^{-2} g_{n+2}^{-1}\,\rho_n$   leaves $y_{2,n}$ unchanged. 
Now in \eqref{lyntauB3i}, regardless of what non-zero prefactor $\vartheta_n$ appears to begin with, we can always make the replacement 
$\vartheta_n\to  g_{n+2}g_{n+3} g_{n}^{-1}g_{n+5}^{-1}\vartheta_n =1$; to be precise, this is achieved by specifying any solution of a linear difference equation of 
order 5 for $\log g_n$. With that choice of gauge, the variable $w_n$  is given in terms of $\tau_n$ by \eqref{ws7gauge}, and the sequence 
$(\tau_n)$  satisfies the special Somos-7 recurrence \eqref{somos7B3i}, as in the proof of  Theorem \ref{somos7thm}. 
However, as already mentioned above, in general the prefactor $\xi_n$
appearing in \eqref{somos5B3ii}  cannot be simultaneously fixed to be 1 (rather, fixing $\xi_n\to 1$, so that $\tau_n$ satisfies 
the Somos-5 relation \eqref{somos5B3i}, 
is a \textit{different} gauge choice). Thus, in the ``Somos-7 gauge'', where $\theta_n=1$, the system of recurrences %system 
\eqref{systmtauB3i} becomes 
\beq\label{B31multil} 
\begin{array}{rcl}
 \tau_{n+5}\tau_{n} &  = &  \beta\, \tau_{n+3}\tau_{n+2} + \eta_{n}  \\ 
 \eta_{n+1}\eta_{n} & = &  \rho_{n} \, \tau_{n+5}^2\tau_{n+1}^2 + \beta\, \tau_{n+4}\tau_{n+3}^{2}\tau_{n+2} 
    \\
 \xi_{n+1}\, \tau_{n+5}\tau_{n+2} &=&  \tau_{n+4}\tau_{n+3} + \eta_{n+1} \\
 \rho_{n+1}\rho_{n}&=& \xi_{n+1} 
\end{array}
\eeq 
(having shifted $n\to n+1$ in the third relation). 
In the above, an extended ``cluster'' of initial data, including the fixed parameter (``frozen variable'') $\beta$,  is given by 
$(\tau_0,   \tau_1, \tau_2, \tau_3, \tau_4, \eta_0, \rho_0, \be)$, and via \eqref{B31tau} this fixes initial data ${\bf y}_0=(y_{1,0},y_{2,0})$ for the map $\hat{\varphi}_1$. 
Now iterating each of the equations \eqref{B31multil} one by one, in order, starting from $n=0$, produces in turn 
$\tau_5,\eta_1, \xi_1, \rho_1$, giving the image of the lifted map $\Phi$ as in \eqref{bigphi}. 
Notice that the intermediate step of finding $\xi_1$ can be skipped: for each $n$,  by combining the last two relations, we 
have 
$$ 
\rho_{n+1}\rho_n = \frac{\tau_{n+4}\tau_{n+3} + \eta_{n+1} }{\tau_{n+5}\tau_{n+2}}. 
$$ 
Hence the first two relations in  \eqref{B31multil} appear like a pair of cluster exchange relations, with one 
of them having a coefficient $\rho_n$ that is non-autonomous (dependent on $n$). Upon iteration of the map $\Phi$, 
we obtain the three sequences $(\tau_n)$, $(\eta_n)$,  $(\rho_n)$, which together specify the orbit 
${\bf y}_n = \hat{\varphi}_1^n({\bf y}_0)$, 
as well as the  sequence  $(\xi_n)$ of  intermediate values, which appear  in the formula \eqref{somos5B3ii} for the quantities $v_n$. 
Now consider the ring of Laurent polynomials 
$$ 
{\cal R} = \Z [\be, \tau_{0}^{\pm 1},\tau_{1}^{\pm 1},\tau_{2}^{\pm 1},\tau_{3}^{\pm 1},\tau_{4}^{\pm 1},\eta_{0}^{\pm 1},\rho_{0}^{\pm 1}]. 
$$ 
Direct calculation of three steps of $\Phi$   shows by inspection that 
$\xi_1,\xi_2\in{\cal R}$, and 
$$ 
\xi_3 = \frac{\tau_{3}\tau_{2} + \eta_{0} }{\tau_{4}\tau_{1}}=\xi_0\in{\cal R}, 
$$ 
hence the sequence $(\xi_n)$ has period 3, or in other words $({\cal T}^3-1)\, \xi_n =0$ (where ${\cal T}$ denotes the shift operator that sends $n\to n+1$). 
Then, upon taking logarithms on both sides of the fourth relation in \eqref{B31multil}, we have 
$$ 
({\cal T}+1) \, \log \rho_n = \log \xi_{n+1} \implies   ({\cal T}^3-1)({\cal T}+1) \, \log \rho_n = 0 
\implies ({\cal T}^6-1) \, \log \rho_n =({\cal T}^3-1)({\cal T}^3+1) \, \log \rho_n = 0, 
$$ 
hence the sequence $(\rho_n)$ has period 6, as required. However, while $\rho_0,\rho_1,\rho_5\in{\cal R}$, 
we find that $\rho_2,\rho_3,\rho_4\not\in{\cal R}$: the latter three terms have non-monomial factors appearing in their denominators, 
so they cannot be cluster variables.   A direct calculation shows that the product of three adjacent $\xi_n$ is 
$$ 
\xi_1\xi_2\xi_3  =\hat{K}_{1} + \beta + 3 \in {\cal R},  
$$
where here $\hat{K}_{1}$ is used to denote the value of the invariant along an orbit of the lifted map $\Phi$, considered 
as a function of $\tau_0,   \tau_1, \tau_2, \tau_3, \tau_4, \eta_0, \rho_0, \be$; hence
 $\hat{K}_{1}\in{\cal R}$, and using the fourth relation in \eqref{B31multil} once more, we see that the latter product is 
equal to $\rho_0\rho_1\rho_2\rho_3\rho_4\rho_5$, so \eqref{perxirho} holds, as required. 
%$\tau_5,\tau_6,\tau_7\in{\cal R}$,  $\eta_1,\tau_6,\tau_7\in{\cal R}$, 
Next, we claim that $\tau_n,\eta_n\in{\cal R}$. To see this, we just need to show that $\tau_n\in{\cal R}$ for all $n$, since 
if this holds then the first relation in \eqref{B31multil} implies immediately that 
$\eta_n =   \tau_{n+5}\tau_{n} -  \beta\, \tau_{n+3}\tau_{n+2}\in{\cal R}$. So we consider the aforementioned fact that, 
due to the gauge choice, 
$\tau_n$  satisfies the special Somos-7 recurrence \eqref{somos7B3i}, which has coefficients $\tilde{\al}, \tilde{\be}$, and 
there is an associated Lyness invariant quantity (see \cite{hone2020ecm}, for instance), which we denote by $\tilde{K}$. 
Then, by a minor modification of Theorem 3.7 in \cite{swahon} and its proof, it follows that the Somos-7 recurrence 
has the strong Laurent property, in the sense that 
$\tau_n\in\tilde{\cal R}$ for all $n\geq 0$, where 
$$ 
\tilde{\cal R}=\Z[\tilde{\al}, \tilde{\be},\tilde{K},\tau_0^{\pm 1},\tau_1^{\pm 1},\tau_2^{\pm 1},\tau_3^{\pm 1},\tau_4,\tau_5,\tau_6 ] . 
$$ 
(For further details, see Theorem B.5 in the second appendix below.)  
By inspection of the first two iterates of $\Phi$, we can verify directly that $\tau_5,\tau_6\in{\cal R}$, while from Theorem \ref{somos7thm} 
we have $\tilde{\alpha} = 1 - \beta$, $\tilde{\beta} = \beta \hat{K}_{1} + 2 \beta^2 + \beta + 1$ and a short explicit calculation with 
computer algebra 
%using the result of Theorem 1 in \cite{hone2020ecm} 
shows that $\tilde{K}=\hat{K}_{1} + 2 \beta+2$. Since, as already noted, $\hat{K}_{1}\in{\cal R}$ on an orbit of $\Phi$, 
it follows that $\tilde{\al}, \tilde{\be},\tilde{K}\in{\cal R}$, hence $\tilde{\cal R}$ is a subring of ${\cal R}$. Thus we 
see that $\tau_n\in{\cal R}$ for $n\geq 0$, 
and an analogous argument extends this to $n<0$ and completes the proof of the theorem.
\end{prf} 

\begin{remark} 
The first two relations in \eqref{B31multil} resemble  exchange relations in a cluster algebra, but the third and fourth relations (which define $\rho_{n}$) do not. Thus, the sequence of tau functions cannot be produced by cluster exchange relations with frozen variables alone. Nevertheless, this can be considered an ``almost Laurentification" of the deformed map: 
the tau functions $\tau_n$ and $\eta_n$ are Laurent polynomials, while the periodic quantities $\rho_n$ only contain a finite number of non-monomial factors in their denominators, so this is an example of the extended Laurent property \cite{mase}, where only a finite extension of the ring $\cal R$ is required. The  coefficients $\rho_n$ are reminiscent 
of y-variables in a cluster algebra with coefficients, which can be used to generate non-autonomous difference equations, including those of discrete Painlev\'e type \cite{hinoue, okubo}. 
We have attempted to construct the relations \eqref{B31multil} from a suitable Y-system, by pulling back the two-form  \eqref{omhat} to derive an associated exchange matrix (cf.\ the formulae \eqref{OmB32}, \eqref{B32B} and \eqref{B32exch} for the case of the map $\hat{\varphi}_2$ below), 
but as yet we have not succeeded 
in doing this in a consistent way.  
\end{remark}

\subsection{Deformed map $\hat{\varphi}_{2}$ for $\rB_3$} % (b_{2}=b_{3}=\beta^2,\quad b_{1} =\beta )$}

%Before we look into the singularity structure, we write the 

The action of the deformed map $\hat{\varphi}_{2}$ given by \eqref{map2} is equivalent to iteration of the coupled pair of recurrences  
\beq\label{B21recs} 
\begin{array}{rcl}
y_{1,n+1}y_{1,n} & = & w_{n}^{2}y_{2,n} + \beta^2 \\  
y_{2,n+1}y_{2,n}y_{1,n}w_{n}^{2} & = & w_{n}^{2}y_{2,n} + \beta^2(y_{1,n} + 1) , \\  
\end{array} 
\eeq 
where for convenience  we made use of the same variable $w_{n} = y_{1,n}+\beta$ 
as in the previous discussion of the map $\hat{\varphi}_{1}$. 
Subtracting the first relation from the second  gives rise to a simplified relation for $y_{2,n+1}$, and thus iterates of the map are generated by the system of four relations
\beq\label{B32recs} %%% {B21recs} 
\begin{array}{rcl}
w_{n} & = & y_{1,n} + \beta \\
y_{1,n+1}y_{1,n} & = & w_{n}^{2}y_{2,n} + \beta^2 \\  
v_{n+1} & = & y_{1,n+1} + \beta^2 \\
y_{2,n+1}y_{2,n}w_{n}^{2} & = & v_{n+1},  \\  
\end{array} 
\eeq 
where, in contrast to \eqref{B31recs2}, there is a different definition for the quantity $v_n = y_{1,n}+\be^2$. 
By looking into the prime factorization of some orbits of \eqref{B32recs} defined over $\Q$, we observe the following confined singularity patterns:   
\begin{equation}\label{singB32}
\begin{split}
    &\text{Pattern 1 :} \ (y_{1,n},y_{2,n},w_{n}) = \dots,(R,0,R), (R,\infty,R),(\infty,R,\infty),(\infty,0,\infty),(R,0,R),\dots\\[0.5em] 
    &\text{Pattern 2 :} \  (y_{1,n},y_{2,n},w_{n}) =\dots,(R,R,0), (R,\infty^{2},0),\dots \\[0.5em] 
    &\text{Pattern 3 :} \  (y_{1,n},y_{2,n},w_{n}) =\dots, (0,R,R),\dots
\end{split}
\end{equation}
(The corresponding patterns for $v_n$ have been omitted.)  
We introduce tau functions $\tau_n, \sigma_n$ and $\eta_n$ which correspond to  Patterns 1,2 and  3, respectively, such that   $y_{1,n}$, $y_{2,n}$ and $w_{n}= y_{1,n}+\beta$ 
can be written as 
\begin{equation}\label{vartrans}
    y_{1,n} = \frac{\eta_{n}}{\tau_{n+2}\tau_{n+1}}, \quad y_{2,n} = \frac{\tau_{n+4}\tau_{n+1}\tau_{n}}{\sigma_{n}^{2}\tau_{n+3}}, \quad  w_{n} = \frac{\sigma_{n+1}\sigma_{n}}{\tau_{n+2}\tau_{n+1}}, 
\end{equation}
%
%Performing a more detailed singularity analysis,
and then the fourth equation in \eqref{B32recs} immediately implies that   
\begin{equation}\label{somos7B32} % {vartrans}
 v_n=   y_{1,n} +\beta^2 =  \frac{\tau_{n+4}\tau_{n-1}}{\tau_{n+2}\tau_{n+1}}. 
\end{equation}

Upon inspecting the structure of a particular singularity further by approaching it in the limit of  a small parameter $\epsilon\to 0$, 
one can see that the singularities of  $y_{1,n}$ and $y_{2,n}$ in Pattern 1 correspond to the sequence 
\begin{equation}\label{patternB3}
        \mqty(C \\ -\frac{\beta^2(C+1)}{C^2 + 2\beta C + \beta^2}) \to \mqty(-\beta^2 \\ 0^1) \to \mqty(-1 \\ \infty^{1}) \to \mqty(\infty^1 \\ -1) \to \mqty(\infty^{1} \\ 0^1) \to \mqty(-1 \\ 0^1) \to \mqty(-\beta^2 \\ C') \\ 
\end{equation}
with $C,C'$ being regular values, 
which propagates from the point  $\Big(C,-\frac{\beta^2(C+1)}{C^2 + 2\beta C + \beta^2} \Big) $ where the Jacobian of the deformed map $\hat{\varphi}_2$ is zero. 
Noting that the value $y_1=-1$ appears in the singularity pattern, we can consider another variable $u_n=y_{1,n}+1$, and find that 
\begin{equation}\label{somos5B32}
   u_n =   y_{1,n} +1 =\xi_n\,  \frac{\tau_{n+3}\tau_{n}}{\tau_{n+2}\tau_{n+1}}, 
\end{equation}
where the prefactor $\xi_n$ cannot be removed without a change of gauge, which would modify the form of some of the 
expressions in \eqref{somos7B32}. (As explained in Appendix B, the quantity $\xi_n$ is periodic with period 3.) 

Notice that the ratios of tau functions in \eqref{somos7B32} and \eqref{somos5B32} are identical to the substitutions associated with the Lyness map and Somos-5 QRT map, respectively. 
This suggests that the quantities $v_{n} =y_{1,n} + \beta^2$ and $u_{n}=y_{1,n} + 1 $ should provide solutions of these maps under iteration, as described by the following statement.  
\begin{thm}\label{th46} 
The quantities  $v_{n}$ generated under iteration of the system of recurrences \eqref{B32recs} satisfy the Lyness map which is equivalent to the recurrence 
\begin{equation}\label{lmaps7}
    v_{n+1}v_{n-1} =\gamma v_{n} +  \delta,  
\end{equation}
where the coefficients along each orbit of the $\hat{\varphi}_2$ are specified by $\gamma= 1-\beta^{2}$ and $\delta = \beta^{2}\hat{K}_{2} + 2\beta(\beta^3 + 1)$, 
and the associated sequence of tau functions $(\tau_n)$ related via \eqref{somos7B32}
satisfies the Somos-7 recurrence 
\beq\label{B32somos7}  %%% {th46} %%% {lmaps7} 
\tau_{n+7}\tau_n = \gamma\, \tau_{n+6}\tau_{n+1} + \delta\, \tau_{n+4}\tau_{n+3} 
.
\eeq
The corresponding iterates of $u_{n}=y_{1,n}+1$ satisfy the  Somos-5 QRT map which is given by the recurrence 
\begin{equation}\label{qrtmaps5} 
    u_{n+1}u_{n}u_{n-1} = \hat{\gamma} u_{n} + \hat{\delta}, 
\end{equation}
where $\hat{\gamma} = \hat{K}_{2} + 2\beta + 2$ and $\hat{\delta} = (\beta^{2} - 1) \hat{\gamma}$.
\end{thm} 
\begin{prf}
This follows from analogous arguments to those used in proving Theorem \ref{somos7thm} and Theorem \ref{somos5thm}.  
For a more detailed explanation of the connection between the Lyness map \eqref{lmaps7} and the  Somos-5 QRT map \eqref{qrtmaps5}, 
see Proposition B.2 in the second appendix below.
\end{prf}

The tau function expressions \eqref{vartrans} and \eqref{somos7B32} can be substituted directly into \eqref{B32recs}, giving  rise to the system of equations  
\begin{equation}\label{systmtauB3ii}
\begin{split}
    &\sigma_{n+1}\sigma_{n}  = \beta \tau_{n+2}\tau_{n+1} + \eta_{n} \\ 
    &\eta_{n+1}\eta_{n} = \tau_{n+4}\tau_{n} \sigma_{n+1}^{2} + \beta^{2}\tau_{n+3}\tau_{n+2}^2\tau_{n+1}\\
    &\tau_{n+5}\tau_{n}= \beta^{2} \tau_{n+3}\tau_{n+2} + \eta_{n+1} 
\end{split}
\end{equation}
Since the three recurrences above are all of the right form for an exchange relation, it appears likely that their  iteration can be described by a sequence of 
cluster mutations in an appropriate cluster algebra. 
To verify this is the case, we set the initial cluster to be 
$$\tilde{\vb{x}} = (\tilde{x}_{1},\tilde{x}_{2},\tilde{x}_{3},\tilde{x}_{4},\tilde{x}_{5},\tilde{x}_{6},\tilde{x}_{7}) 
= ( \eta_{0}, \sigma_{0}, \tau_{0},  \tau_{1}, \tau_{2}, \tau_{3}, \tau_{4}), $$ 
and then determine a new exchange matrix via the pullback of the symplectic form \eqref{omhat} by the rational map $\tilde{\pi}:\, \C^7\to \C^2$ defined by the equations 
for $(y_{1,0},y_{2,0})\in\C^2$ given by setting $n=0$ in \eqref{vartrans}. 
As a result,  one finds a presymplectic form on the space of tau functions, written in terms of the cluster variables $\tx_j$ for $1\leq j\leq 7$ as   
 \begin{align*}
 \tilde{\om} = \tilde{\pi}^{*}\om =\sum_{ij}\tilde{\Om}_{ij} \rd \log \tilde{x}_{i} \wedge \rd \log \tilde{x}_{j} , 
 \end{align*}
 where the $7\times 7$  
matrix $\tilde{\Om}$ is given by 
 \begin{equation}\label{OmB32} 
    \tilde{\Om} = \mqty(0 & -2& 1& 1& 0 & -1 & 1 \\ 2& 0 & 0 & -2 & -2 &  0 & 0 \\ -1 & 0 & 0 & 1 & 1 & 0 & 0 \\ -1 & 2 & -1 & 0 & 1 & 1 & -1 \\ 0 & 2 & -1 & -1 & 0 & 1 & -1 \\ 1 & 0 & 0 & -1 & -1 & 0 & 0 \\ -1 & 0 & 0 & 1 & 1 & 0 & 0  ) . 
\end{equation}
Given the skew-symmetric matrix $\Om$ as above, a skew-symmetrizable exchange matrix $\tilde{B}$ such that $\tilde{B}\tilde{D}=\tilde{\Om}$ 
can be determined by post-multiplying with the diagonal matrix $\tilde{D}^{-1}=\text{diag} (1,\sfrac{1}{2}, 1, 1, 1, 1,1)$, to obtain % as shown below 
\begin{equation} \label{B32B} 
    \tilde{B}= \tilde{\Omega} \tilde{D}^{-1} = (\tilde{B}_{ij}) =  \mqty(0 & -1& 1& 1& 0 & -1 & 1 \\ 2& 0 & 0 & -2 & -2 &  0 & 0 \\ -1 & 0 & 0 & 1 & 1 & 0 & 0 \\ -1 & 1 & -1 & 0 & 1 & 1 & -1 \\ 0 & 1 & -1 & -1 & 0 & 1 & -1 \\ 1 & 0 & 0 & -1 & -1 & 0 & 0 \\ -1 & 0 & 0 & 1 & 1 & 0 & 0 )
\end{equation}

The matrix $\tilde{B}$ above generates a coefficient-free cluster algebra, but both the parameter $\be$ and its square appear in front of some of the terms in  \eqref{systmtauB3ii}.  
To incorporate this into the exchange relations,  
we extend the initial cluster by adding the frozen variable $\tx_8=\beta$ and adjoining an extra row with entries $(0,1,0,0,0,0,2)$ to the exchange matrix $\tilde{B}$. 
Then we can obtain the following statement, which constitutes the Laurentification of the deformed $\rB_3$ map $\hat{\varphi}_2$.

\begin{thm}\label{B32clusterthm} 
%[Laurentification of the deformed map] 
Let $(\hat{\vb{x}},\hat{B})$ be given as an initial seed which is composed of the extended initial cluster 
$$ \hat{\vb{x}} =(\tx_j)_{1\leq j \leq 8} = \qty(\eta_{0},\sigma_{0},\tau_{0}, \tau_{1},\tau_{2},\tau_{3},\tau_{4},\beta) $$ 
together with the associated extended exchange matrix %$\hat{B}$ given by 
 \begin{equation}\label{B32exch} 
\hat{B}= 
    \mqty(0 & -1& 1& 1& 0 & -1 & 1 \\ 2& 0 & 0 & -2 & -2 &  0 & 0 \\ -1 & 0 & 0 & 1 & 1 & 0 & 0 \\ -1 & 1 & -1 & 0 & 1 & 1 & -1 \\ 0 & 1 & -1 & -1 & 0 & 1 & -1 \\ 1 & 0 & 0 & -1 & -1 & 0 & 0 \\ -1 & 0 & 0 & 1 & 1 & 0 & 0 \\ 0 & 1 & 0 & 0 & 0 & 0 & -2 ), 
\end{equation}
and consider the permutation $\rho=(34567)$. 
Then the iteration of the  cluster map $\psi = \rho^{-1} \tilde{\mu}_{3}\tilde{\mu}_{1}\tilde{\mu}_{2}$ %for permutation $\rho = (3,4,5,6,7) $ 
is equivalent to the system of recurrences \eqref{systmtauB3ii}, 
and for all $n\in\Z$ the tau functions $\eta_n,\sigma_n,\tau_n$ are elements of the Laurent polynomial ring 
$\mathbb{Z}\qty[\beta, \eta_{0}^{\pm 1},\sigma_{0}^{\pm 1},\tau_{0}^{\pm 1}, \tau_{1}^{\pm 1},\tau_{2}^{\pm 1},\tau_{3}^{\pm 1},\tau_{4}^{\pm 1}]$, 
with positive integer coefficients. 
\end{thm}
\begin{prf} Let us consider the seed $(\hat{\vb{x}}',\hat{B}')=\tmu_{3}\tmu_{1}\tmu_{2}(\hat{\vb{x}},\hat{B}) =$ that arises from applying the sequence of mutations 
$\tmu_3\tmu_1\tmu_2$ to the given initial seed, where (as usual) we use $\tmu_j$ to denote mutations in the cluster algebra associated with Laurentification of the 
deformed map. The new cluster is $\hat{\vb{x}}'= (\tilde{x}_{1}',\tilde{x}_{2}',\tilde{x}_{3}',\tilde{x}_{4},\tilde{x}_{5},\tilde{x}_{6},\tilde{x}_{7},\tilde{x}_{8})$, 
where the new cluster variables (with primes) are obtained from the exchange relations 
\begin{equation}\label{systmexchB3ii}
    \begin{split}
        & \tilde{x}'_{2}\tilde{x}_{2}  = \tx_8 \tilde{x}_{4}\tilde{x}_{5} + \tilde{x}_{1} \\ 
        & \tilde{x}'_{1}\tilde{x}_{1}  = (\tx_8)^2 \tilde{x}_{4}(\tilde{x}_{5})^2\tilde{x}_{6} + (\tilde{x}_{2}{'})^2\tilde{x}_{3}\tilde{x}_{7} \\
        & \tilde{x}'_{3}\tilde{x}_{3}  = (\tx_8)^2 \tilde{x}_{5}\tilde{x}_{6} + \tilde{x}_{1}{'}, 
    \end{split}
\end{equation}
while  the mutated exchange matrix $\hat{B}'=\tmu_{3}\tmu_{1}\tmu_{2}(\hat{B})$ is given by 
\begin{equation}
   \mqty(0 & -1& 1& 1& 1 & 0 & -1 \\ 2& 0 & 0 & 0 & -2 &  -2 & 0 \\ -1 & 0 & 0 & 0 & 1 & 1 & 0 \\ -1 & 0 & 0 & 0 & 1 & 1 & 0\\ -1 & 1 & -1 & -1 & 0 & 1 & 1 \\  0 & 1 & -1 & -1 & -1 & 0 & 1  \\ 1 & 0 & 0 & 0 & -1 & -1 & 0 \\ 0 & 1 & -2 & 0 & 0 & 0 & 0 ) .
\end{equation}
For the new cluster variables, if we identify $\tilde{x}_{1}{'} = \eta_{1}$,  $\tilde{x}_{2}{'} = \sigma_{1} $,  $\tilde{x}_{3}{'} = \tau_{5}$ and replace all variables $\tilde{x}_{i}$ 
for $4\leq i\leq 8$ with the corresponding tau functions and frozen variable from the original cluster $\hat{\vb{x}}$, 
then we find that the exchange relations \eqref{systmexchB3ii} are equivalent to the recurrence formulae \eqref{systmtauB3ii} for $n=0$. 
As for the exchange matrix $\hat{B}{'}$, we can rewrite it in the following way: 
\begin{align*}
\tmu_{3}\tmu_{1}\tmu_{2}(\hat{B}) = P_{1}\hat{B}P_{2} = \rho  (\hat{B}). 
\end{align*}
In the above, the action of the permutation $\rho = (34567)$  is equivalent to applying the row and column permutation matrices 
\begin{equation}
    P_{1}=\mqty(1& 0& 0& 0& 0& 0& 0& 0  \\ 0& 1& 0& 0& 0& 0& 0& 0 \\ 0& 0& 0& 0& 0& 0& 1& 0 \\ 0& 0& 1& 0& 0& 0& 0& 0 \\ 0& 0& 0& 1& 0& 0& 0& 0 \\ 0& 0& 0& 0& 1& 0& 0& 0 \\0& 0& 0& 0& 0& 1& 0& 0 \\ 0& 0& 0& 0& 0& 0& 0& 1 ), \quad P_{2} =  \mqty(1 & 0 & 0 & 0 & 0 & 0 & 0 \\ 0 & 1 & 0 & 0 & 0 & 0 & 0 \\ 0 & 0 & 0 & 1 &  0 & 0 & 0 \\ 0 & 0 & 0 & 0 & 1 & 0 & 0 \\ 0 & 0 & 0 & 0 & 0 & 1 & 0 \\ 0 & 0 & 0 & 0 & 0 & 0 & 1 \\ 0& 0 & 1 & 0 & 0 & 0 & 0) . 
\end{equation}
Hence we see that the 
cluster map  defined by $\psi =\rho^{-1}\tmu_{3}\tmu_{1}\tmu_{2}$ satisfies $\psi(\hat{B})=B$, and its action on any cluster  
is equivalent to the shift $n \to n+1$ on the indices of the tau functions. 
Since they are cluster variables, these tau functions exhibit the Laurent property. Moreover, 
the coefficients of the Laurent polynomial cluster variables are positive integers, due to the positivity property \cite{ls, ghkk}. 
\end{prf}

%%%%%%%%%%%%%%%%%%%%%%%

\subsection{Tropicalization and degree growth for deformed $\rB_{3}$ cluster map}

We have seen that deformation of the periodic dynamics in type  $\rB_{3}$ 
yields two integrable symplectic maps, but only the second map 
$\hat{\varphi}_2$ given by \eqref{B21recs} 
has been shown to correspond to mutations in a cluster algebra, as in 
Theorem \ref{B32clusterthm}.  By the Laurent property, we can write
the tau functions generated by the system \eqref{systmtauB3ii} in the 
form 
$$  \eta_n = \frac{\rN_n^{(1)}(\hat{\vb{x}})}{\hat{\vb{x}}^{{\bf d}_n}}, \qquad \sigma_n  = \frac{\rN_n^{(2)}(\hat{\vb{x}})}{\hat{\vb{x}}^{{\bf e}_n}}, 
\qquad \tau_n = \frac{\rN_n^{(3)}(\hat{\vb{x}})}{\hat{\vb{x}}^{{\bf f}_n}},  $$ 
and find that the corresponding d-vectors 
${\bf d}_n,  {\bf e}_n,  {\bf f}_n\in\Z^7$ 
satisfy the $(\max,+)$ tropical relations  
\begin{equation}\label{troptauB3ii}
\begin{split}
    &{\bf e}_{n+1}+{\bf e}_{n}  
= \max(  {\bf f}_{n+2}+ {\bf f}_{n+1} , {\bf d}_{n} ) , \\ 
    &{\bf d}_{n+1}+{\bf d}_{n} =\max( {\bf f}_{n+4}+{\bf f}_{n} 
+2 {\bf e}_{n+1} , {\bf f}_{n+3}+2 {\bf f}_{n+2}+ {\bf f}_{n+1} ) , \\
    &{\bf f}_{n+5}+ {\bf f}_{n}= \max(  {\bf f}_{n+3}+ {\bf f}_{n+2} , {\bf d}_{n+1}) ,
\end{split}
\end{equation}
where, as usual, the tropical relations do not contain analogues of 
coefficient terms 
associated with the parameter $\beta$, since 
the denominators of the tau functions only depend on the non-frozen 
variables $\eta_0,\sigma_0,\tau_0,\tau_1,\tau_2,\tau_3,\tau_4$ in the initial cluster $\hat{\vb{x}}$. 

To determine the growth of the d-vectors in this case, we introduce the 
tropical analogues of the substitutions  \eqref{vartrans} and \eqref{somos7B32}, namely 
\beq\label{tropB3subs} %%% {troptauB3ii}
\begin{array}{rclcrcl}
{\bf Y}_{1,n}& =& {\bf d}_n -{\bf f}_{n+2}-{\bf f}_{n+1}, & \, &  
{\bf Y}_{2,n} & =& {\bf f}_{n+4}-{\bf f}_{n+3}+{\bf f}_{n+1}
+{\bf f}_{n}-2{\bf e}_n , \\ 
{\bf W}_{n}& =& {\bf e}_{n+1}+{\bf e}_{n}-{\bf f}_{n+2}
-{\bf f}_{n+1} , & &  
{\bf V}_{n}& =& {\bf f}_{n+4}-{\bf f}_{n+2}-{\bf f}_{n+1}
+{\bf f}_{n-1},
\end{array} 
\eeq 
which produce  periodic quantities under iteration.  
These substitutions allow us to derive the tropical version  
 of the system 
\eqref{B32recs} 
for the map $\hat{\varphi}_2$. %%%, in the form. 

\begin{lem}\label{tropB2map}
The combinations of d-vectors defined by 
\eqref{tropB3subs} satisfy the tropical analogue of the deformed $\rB_3$ map $\hat{\varphi}_2$, given by the following system of four equations:  
\beq\label{tropB32recs} %%% {B21recs} 
\begin{array}{rcl}
{\bf W}_{n} & = & [{\bf Y}_{1,n}]_+ ,  \\
{\bf Y}_{1,n+1}+{\bf Y}_{1,n} & = & [ 2{\bf W}_{n}+{\bf Y}_{2,n}]_+ , \\  
{\bf V}_{n+1} & = & [{\bf Y}_{1,n+1}]_+ ,  \\
{\bf Y}_{2,n+1}+{\bf Y}_{2,n}+ 2{\bf W}_{n} & = & {\bf V}_{n+1} .   \\  
\end{array} 
\eeq 
Given arbitrary initial values $(Y_{1,0},Y_{2,0})\in\R^2$,  every component of  this system is periodic with period 4. 
\end{lem} 
\begin{prf} The $(\max,+)$ equations relating ${\bf Y}_{1,n},{\bf Y}_{2,n},{\bf W}_n$ and ${\bf V}_n$ follow directly from the substitutions 
\eqref{tropB3subs} and the tropical analogues of the exchange relations that 
produce \eqref{troptauB3ii}.
Observe that, from the point of view of the dynamics of 
the main variables ${\bf Y}_{1,n},{\bf Y}_{2,n}$, the third equation in 
\eqref{tropB32recs} is redundant, since from the first equation it is clear that 
\beq\label{VWreln} 
{\bf V}_n = {\bf W}_n 
\eeq 
for all $n$, so we could omit the third equation and replace the fourth one by 
$$ 
{\bf Y}_{2,n+1}+{\bf Y}_{2,n}+ 2{\bf W}_{n}  =  {\bf W}_{n+1} .
$$  

For the vector system \eqref{tropB32recs}, a set of initial data 
consists of a pair of vectors ${\bf Y}_{1,0},{\bf Y}_{2,0}$, but we can 
view each component as a piecewise linear map 
$\hat{\varphi}_{2,trop}:\,\R^2\to\R^2$, and check directly that a pair  
arbitrary initial values $(Y_{1,0},Y_{2,0})\in\R^2$ 
produce a sequence of points in the plane that repeats with period 4, 
or in other words, $(\hat{\varphi}_{2,trop})^4=\mathrm{id}$. 
This can be verified by case-by-case analysis, which we leave to the reader. 
\end{prf} 

We now show how we can use Somos sequences to simplify the calculation of 
degree growth for tau functions, by using a tropical analogue  of  
Theorem \eqref{th46}. %% 

\begin{lem} \label{B32tropS7} The  d-vectors 
${\bf f}_n\in\Z^7$ that specify the denominators of 
tau functions $\tau_n$ generated by the system  \eqref{troptauB3ii} 
satisfy the tropical Somos-7 relation 
\beq\label{tropS7}
{\bf f}_{n+7}+{\bf f}_n = \max ({\bf f}_{n+6}+{\bf f}_{n+1} , {\bf c}+{\bf f}_{n+4}+{\bf f}_{n+3}),   
\eeq 
and the corresponding quantity ${\bf V}_n$ defined in \eqref{tropB3subs} 
satisfies the ultradiscrete Lyness map 
\beq\label{udlyness}
{\bf V}_{n+1} +  {\bf V}_{n-1} = \max ({\bf V}_{n}, {\bf c} ), 
\eeq 
where ${\bf c}$ is the constant vector 
$$ 
{\bf c}= (2,2,1,1,1,1,1)^T.  
$$ 
\end{lem} 
\begin{prf} Upon substituting the expression  
$\tau_n =  {\rN_n^{(3)}(\hat{\vb{x}})}/{\hat{\vb{x}}^{{\bf f}_n}}$ 
into the Somos-7 recurrence \eqref{B32somos7} and 
comparing denominators on each side, we see that the 
first term on the right has denominator 
$\hat{\vb{x}}^{{\bf f}_{n+6}+{\bf f}_{n+1}}$, since the coefficient 
$\gamma$ in front of this term is a constant that is independent of cluster variables; but the coefficient $\delta$ appearing in front of the second term is linear in   the first integral $\hat{K}_2$, as given in (\ref{k2nd}), and pulling 
this back to function of the tau functions via the map $\tilde{\pi}:\C^7\to\C^2$ 
defined by the substitutions for 
$y_{1,0},y_{2,0}$ in  \eqref{vartrans} gives a Laurent polynomial that takes the form  
$$ 
\tilde{\pi}^*(\delta) = 
\frac{{\mathrm P} (\hat{\vb{x}})} {\eta_0^2\sigma_0^2\tau_0\tau_1\tau_2\tau_3\tau_4},   
$$  
for a certain polynomial ${\mathrm P}$,  
which means that the denominator of the second term on the right-hand side 
is  $\hat{\vb{x}}^{{\bf c}+{\bf f}_{n+4}+{\bf f}_{n+3}}$, with the 
given constant vector ${\bf c}$. Hence ${\bf f}_n$ satisfies the given 
$(\max,+)$ version of Somos-7, and it follows immediately that 
${\bf V}_n$ satisfies the corresponding analogue of the Lyness map \eqref{lmaps7}, which is  
the relation (\ref{udlyness}) with the same 
constant ${\bf c}$. 
\end{prf} 

The seed  $\hat{\vb{x}}=(\eta_0,\sigma_0,\tau_0,\tau_1,\tau_2,\tau_3,\tau_4,\beta)$ gives  the  initial cluster of d-vectors specified by the matrix 
\beq\label{tropB3inits}
\Big( {\bf d}_0\,\, {\bf e}_0\,\, {\bf f}_0\,\,{\bf f}_1\,\, {\bf f}_2\,\, {\bf f}_3\,\,{\bf f}_4\,\,\Big) = -I, 
\eeq 
where $I$ here denotes the $7\times 7$ identity matrix. 
This provides  the initial data 
$$
{\bf Y}_{1,0} = (-1,0,0,1,1,0,0)^T, 
\qquad 
 {\bf Y}_{2,0} = (0,2,-1,-1,0,1,-1)^T
$$
for the vector form of the map $\hat{\varphi}_{2,trop}$, as in \eqref{tropB32recs},  
and gives 
$$ 
{\bf V}_0 = {\bf W}_0 = [{\bf Y}_{1,0}]_+ = 
 (0,0,0,1,1,0,0)^T. 
$$
As for the other root systems, we can relate these vectors, and the fact that they produce an orbit with period 4, to the Zamolodchikov periodicity of 
the original $\rB_3$ cluster map, given by \eqref{A3recs} with $b_1=b_2=b_3=1$. 
After applying a single iteration of the map $\hat{\varphi}_{2,trop}$ to these 
initial vectors, we find
$$ 
{\bf Y}_{1,1}  = [{\bf Y}_{1,1}]_+= {\bf V}_1 = {\bf W}_1  =
(1,2,0,0,1,1,0)^T, \quad 
 {\bf Y}_{2,1} = (1,0,1,-1,-1,0,1)^T. 
$$ 
Note that, since the sequences ${\bf Y}_{1,n}$ and 
${\bf Y}_{2,n}$ both have period 4, it follows that the associated sequence 
of quantities   
${\bf V}_n = [Y_{1,n}]_+$ does as well, and  this is the same as ${\bf W}_n$,  
as already noted in \eqref{VWreln}. By making further iterations of  $\hat{\varphi}_{2,trop}$, we record  that the next two values of the vector 
${\bf V}_n$ are 
$$ 
{\bf V}_2=(2,2,1,0,0,1,1)^T, \qquad 
{\bf V}_3=(1,0,1,1,0,0,1)^T, 
$$ 
with all subsequent terms determined from ${\bf V}_{n+4}={\bf V}_n$. 

Another way to see the periodicity of the quantities ${\bf V}_n$ is by noting that \eqref{udlyness}, or rather its scalar version 
\beq\label{troplyness}
{V}_{n+1} +  {V}_{n-1} = \max ({ V}_{n}, { c} ), 
\eeq 
is just one particular example of the family of  ultradiscrete QRT maps considered by Nobe. In the case we are interested in, the first two components of the vectors ${\bf V}_0$ and ${\bf V}_1$, and the first two components of the vector ${\bf c}$, correspond to two different period 4 orbits of the 
scalar relation 
(\ref{troplyness}) with parameter value $c=2$, namely  
\beq\label{troplyc2} 
 (0,1)\to (1,2)\to (2,1)\to (1,0) \quad \mathrm{and} \quad 
  (0,2)\to (2,2)\to (2,0)\to (0,0), 
\eeq 
respectively, while the other five components of these vectors correspond to the same period 4 orbit of  (\ref{troplyness}) with parameter value $c=1$, 
that is 
\beq\label{troplyc1} 
(0,0)\to (0,1) \to (1,1)\to (1,0). 
\eeq 
However, note that, from the result of Theorem 4 in \cite{nobe}, 
choosing different initial 
data and/or different values of $c$ in  (\ref{troplyness}) can produce orbits with a different (arbitrarily large) period.  

\begin{thm}\label{tropB2degree}
The d-vectors ${\bf d}_n,{\bf e}_n,{\bf f}_n$ 
satisfying the $(\max,+)$ system \eqref{troptauB3ii} all lie in the kernel of the linear difference operator 
$$ 
\tilde{{\cal L}}= ({\cal T}^2+1) ({\cal T}^2-1)^2({\cal T}^3-1).
$$
For the tau functions $\eta_n,\sigma_n,\tau_n$ generated by \eqref{systmtauB3ii}, the leading order degree growth of their 
denominators is given  by 
$$ 
{\bf d}_n = 2{\bf a}\, n^2+O(n), \quad 
 {\bf e}_n = {\bf a}\, n^2+O(n), \quad 
{\bf f}_n = {\bf a}\, n^2+O(n), \quad 
\mathrm{with}\quad  
{\bf a}= \frac{1}{24} (2,2,1,1,1,1,1)^T .
$$ 
\end{thm} 
\begin{prf}
By definition, we have 
$$ 
{\bf V}_n = ({\cal T}^5 - {\cal T}^3 - {\cal T}^2+1)\, {\bf f}_{n-1},  
$$  
but since the sequence ${\bf V}_n$ has period 4, this gives 
$$ 
({\cal T}^4-1)\, {\bf V}_{n+1} 
= ({\cal T}^4-1)({\cal T}^2-1)({\cal T}^3-1)\, {\bf f}_n = 
\tilde{{\cal L}}\, {\bf f}_n = 0. 
$$ 
Then, given that ${\bf f}_n$ lies in the kernel of 
$\tilde{{\cal L}}$, from the first formula in \eqref{tropB3subs} we find 
$$ 
\tilde{{\cal L}}\, {\bf d}_n 
= \tilde{{\cal L}}\, ({\bf f}_{n+2} + {\bf f}_{n+1}+ {\bf Y}_{1,n}) =  \tilde{{\cal L}}\, {\bf f}_{n+2} + \tilde{{\cal L}}\, {\bf f}_{n+1} 
+ ({\cal T}^2-1)({\cal T}^3-1)({\bf Y}_{1,n+4}-{\bf Y}_{1,n}) = 0, 
$$ 
where used the fact that ${\bf Y}_{1,n}$ has period 4. 
Similarly, the second formula in \eqref{tropB3subs} and the fact that 
${\bf Y}_{2,n}$ has period 4 gives 
$$ 
\tilde{{\cal L}}\, {\bf e}_n 
=\frac{1}{2} \, \tilde{{\cal L}}\,({\bf f}_{n+4} - {\bf f}_{n+3} + 
{\bf f}_{n+1} + {\bf f}_{n}-{\bf Y}_{2,n})=0. 
$$ 
Thus the first part of the statement is proved. 

For the second part of the statement, note that the value of ${\bf V}_0$ calculated above requires that 
$$ 
{\bf f}_{-1}= (0,0,0,0,0,0,1)^T, 
$$ 
and we can use the relation 
$$ 
{\bf f}_{n+4} = {\bf V}_n +{\bf f}_{n+2}+ {\bf f}_{n+1}-{\bf f}_{n-1}, 
$$ 
together with the 4-periodicity of ${\bf V}_n$, to extend the initial values given in \eqref{tropB3inits} 
and thereby produce $({\bf f}_j)_{-1\leq j\leq 7}$, 
providing the required number of terms to completely solve the 
initial value problem for the linear difference equation $\tilde{{\cal L}}\, {\bf f}_n = 0$. In passing, we 
note that this sequence of d-vectors has the particular form  
$$ 
 {\bf f}_n = (f_n^{(1)}, f_n^{(2)}, f_{n+4}^{(3)},f_{n+3}^{(3)},f_{n+2}^{(3)},f_{n+1}^{(3)},f_{n}^{(3)}), 
$$ 
given in terms of the three scalar sequences 
$$ 
\begin{array}{rcl}
(f_n^{(1)}): & \quad & 0,0,0,0,0,0,1,2,2,3,5,6,7,9,11,13,15,17,20,23,25,28,32,\ldots ,    \\
(f_n^{(2)}): & \quad &        0,0,0,0,0,0, 2,2,2,4,6,6,8,10,12,14,16,18,22,24,26,30,34,\ldots , \\ 
(f_n^{(3)}):  & \quad & 1,0,0,0,0,-1, 0,0,0,0,1,1,2,2,3,4,5,5,7,8,9,10,12,\ldots ,
\end{array} 
$$ 
which determine the degrees of $\rho_0$, $\sigma_0$ and 
$\tau_j$ for $0\leq j\leq 4$, respectively, that appear in the denominator of 
the tau functions $\tau_n$. For the leading order quadratic growth of 
${\bf f}_n$ we find 
$$ 
{\bf f}_n \sim {\bf a}\, n^2 \quad \mathrm{where} \quad 
({\cal T}+1)({\cal T}^4-1)({\cal T}^3-1)\, {\bf f}_n = 48 {\bf a} 
=(4,4,2,2,2,2,2)^T, 
$$ 
with the value of the constant ${\bf a}$ being determined above from the initial values ${\bf f}_{-1},\ldots,{\bf f}_7$. 
Then from the first two formulae in \eqref{tropB3subs}, 
together with the 4-periodicity of ${\bf Y}_{1,n}$ and ${\bf Y}_{2,n}$, we have 
$$ 
{\bf d}_n\sim {\bf f}_{n+2}+ {\bf f}_{n+1}\sim 2 {\bf f}_{n}\sim 2{\bf a}\, n^2, \qquad 
{\bf e}_n\sim\frac{1}{2} ( {\bf f}_{n+4}- {\bf f}_{n+3} 
+{\bf f}_{n+1}+ {\bf f}_{n} ) \sim {\bf f}_n \sim {\bf a}\, n^2, 
$$ 
with the same constant ${\bf a}$. This completely fixes the  quadratic  leading order behaviour of the d-vectors, and in each case the correction term is $O(n)$.
\end{prf}

\begin{remark} 
Although they do not appear to be generated by simple cluster mutations, the 
Laurent polynomials $\tau_n,\eta_n$ produced by iteration of %the map 
\eqref{bigphi}, associated with the other deformed $\rB_3$ map $\hat\varphi_1$, are also connected to a Somos-7 recurrence, namely \eqref{somos7B3i}, so it should be possible to calculate their degrees by adapting the preceding arguments suitably. 
\end{remark} 
%%%%%%%%%%%%%%%%%%%%%%%%%%%%%%%%%%%%% OLD D4 SECTION REPLACED %%%%

\section{Integrable deformations of the   $\rD_4$  cluster map} 
\setcounter{equation}{0}

In this section, we consider the deformation of a cluster map in 4D, 
which is composed of mutations in the cluster algebra of type $\rD_{4}$. We will show that there are two essentially different choices  of the deformation parameters 
that yield a discrete integrable system, each of which lifts to a cluster map in higher dimensions via Laurentification. 

\subsection{Deformed $\rD_{4}$ cluster  map}

The Cartan matrix for the $\rD_4$ root system is 
\begin{equation}
    \mqty(2 & -1 & 0 & 0 \\-1 & 2 & -1 & -1 \\ 0 & -1 & 2 & 0 \\ 0 & -1 & 0 & 2) .
\end{equation}
The corresponding exchange matrix is 
\begin{equation}
  B= \mqty(0 & -1 & 0 & 0 \\1 & 0 & -1 & -1 \\ 0 & 1 & 0 & 0 \\ 0 & 1 & 0 & 0) .
\end{equation}
The deformed mutations with parameters $a_j,b_j$ for $1\leq j\leq 4$ take the form 
\beq \label{D4dmaps}\begin{array}{rcl}
\mu_{1} : (x_1,x_2,x_3,x_4) \mapsto (x_1',x_2,x_3,x_4), \qquad x_1 x_1' &=& b_1 + a_1x_2 \\
\mu_{2} :(x_1',x_2,x_3,x_4) \mapsto (x_{1}',x_{2}',x_3,x_4), \qquad x_2 x_2' &=& b_2 + a_2 x_3 x_4 x_{1}'\\
\mu_{3} : (x_{1}',x_{2}',x_3,x_4) \mapsto (x_{1}',x_{2}',x_{3}',x_4), \qquad  x_3 x_{3}'& = & b_3 + a_3 x_2'\\
\mu_{4} : (x_1',x_2',x_3',x_4) \mapsto (x_{1}',x_{2}',x_{3}',x_{4}'), \qquad x_4 x_4' &= & b_{4} + a_4 x_{2}' 
.
\end{array}
\eeq 
The deformed map $\varphi= \mu_{4}\mu_{3}\mu_{2}\mu_{1}$   %%%%$\tilde{\varphi} = \tilde{\mu_{4}}\tilde{\mu_{3}}\tilde{\mu_{2}}\tilde{\mu_{1}}$ 
reduces to the original cluster map when we fix the parameters $a_i=1=b_i$ for all $i$ %=1,\dots,4$. 
The Coxeter number for %type 
$\rD_4$ is 6, and the periodicity for the cluster map is period $4=\frac{1}{2}(6+2)$, i.e. 
\begin{equation}
    \varphi\cdot(\vb{x},B) = (\varphi(\vb{x}),B) \quad (  
\text{with}\ a_{j}=1 = b_j )\implies \varphi^4(\vb{x}) = \vb{x} .
\end{equation}
As usual,  we can reduce the number of parameters in the problem  
by rescaling each of the cluster variables independently, $x_{i} \to \lambda_{i} x_{i}$, and choose the scalings so that the parameters $a_j$ are 
removed and the sequence of deformed mutations can be rewritten as %follows 
\beq\label{D4recs} 
\begin{array}{rcl}
x_{1,n+1} x_{1,n} &=& x_{2,n} + b_1 , \\   
x_{2,n+1} x_{2,n} &=& x_{3,n} x_{4,n} x_{1,n+1} + b_2, \\  
x_{3,n+1} x_{3,n}& = &  x_{2,n+1} + b_3 , \\ 
x_{4,n+1} x_{4,n} &= &  x_{2,n+1} + b_4
.
\end{array} 
\eeq 

Since the exchange matrix is skew-symmetric, by the result of Theorem 1.3 in \cite{hk} the deformed map preserves  
the presymplectic form $\om$ given by 
\beq\label{d4om} 
\om =  \frac{1}{x_{1}x_{2}} \dd \log x_{1} \wedge \dd \log x_{2} + \frac{1}{x_{1}x_{2}} \dd \log x_{1} \wedge \dd \log x_{2} + \frac{1}{x_{2}x_{3}} \dd \log x_{2} \wedge \dd \log x_{3} + \frac{1}{x_{2}x_{4}} \dd \log x_{2} \wedge \dd \log x_{4}. 
\eeq 
Now since $B$ is degenerate and of rank 2, 
%The rank of $B$ is 2, which suggests that 
one can reduce 
the birational map $\varphi$ from 4D to a 2-dimensional symplectic map. The null space and image of $B$ are given by 
\begin{equation}
    \ker(B) = <(1,0,0,1)^{T}, (1,0,1,0)>, \qquad \im(B)=<(0,1,0,0)^{T}, (-1,0,1,1)^{T}>.  
\end{equation}
Hence the null distribution of the presymplectic form $\omega$ is spanned by the two commuting vector fields $\bv_{1} = x_{1}\partial_{x_{1}} + x_{4}\partial_{x_{4}}$ and $\bv_{2} = x_{1} \partial_{x_{1}} + x_{3} \partial_{x_{3}}$. The space of leaves of the null foliation has local coordinates %for the foliation of the null distribution as follows, 
\begin{equation}
    y_1 = x_2,\quad y_2 = \frac{x_3x_4}{x_1}
\end{equation}
Then the  rational map defined by 
\beq\label{pimapD4} 
 \begin{array}{lcrcl}
    \pi&:&\C^4&\to&\C^2\\
    & &\bx=(x_1,x_2,x_3,x_4)&\mapsto&\by=(y_1,y_2) \; 
  \end{array}
\eeq 
reduces the cluster map $\varphi$ to the 2D symplectic map %defined on the plane, 
\beq\label{phihatmapD4} 
 \begin{array}{lcrcl}
    \hat{\varphi}&:&\C^2&\to&\C^2\\
    & &\by=(y_1,y_2)&\mapsto&\left(\dfrac{(b_1 + y_1)y_2 + b_2}{y_1}, \dfrac{(b_4 + y_2)y_1 + b_1y_2 + b_2)((b_3 + y_2)y_1 + b_1y_2 + b_2)}{y_2y_1^2(b_1 + y_1)}\right) \;, 
  \end{array}
\eeq 
which is intertwined with $\varphi$ via $\pi$, so that % satisfies the following relations, 
$$ 
\hat{\varphi}\cdot\pi = \pi\cdot\varphi, \qquad \hphi^*(\hatom)=\hatom, 
$$ 
where $\pi^*(\hatom)=\om$ is the pullback of the symplectic form 
\beq\label{symD4iy}
\hatom=\frac{1}{y_1y_2}\rd y_1\wedge \rd y_2 
\eeq 
under $\pi$. 
When all of the parameters $b_{i} = 1$,  the reduced map $\hat{\varphi}$ has period 4,  %with the same reason as the type $A_{3}$ and $B_{3}$ cases. 
and one of the first integrals associated with this map %$\hat{\varphi}$ 
takes the form 
\begin{equation}\label{K1}
    K = \sum_{i=0}^{3} (\hat{\varphi}^*)^{i}(y_1) = \frac{(1 + y_1)^3 + (2 + 5y_1 + y_1^3) y_2 + (1 + y_1)^2 y_2^2}{y_1^2y_2}
\end{equation}
By applying the same procedure as in the previous examples, %sections, 
we suppose that there is an analogous first integral that is compatible with the deformed map \eqref{phihatmapD4}, taking the form 
\begin{equation}\label{firstintD4}
    \tilde{K} = y_1 + \alpha_1 y_2 + \frac{\alpha_2 y_1}{y_2} + \frac{\alpha_3 y_2}{y_1} + \frac{\alpha_4}{y_2} + \frac{\alpha_5}{y_1} + \frac{\alpha_6 y_2}{y_1^2} + \frac{\alpha_7}{y_2y_1} + \frac{\alpha_8}{y_1^2} + \frac{\alpha_9}{y_2y_1^2}
\end{equation}
where $\alpha_{j}$ are undetermined parameters. Then imposing the requirement that $\tilde{K}$ should be preserved, so that 
$\hat{\varphi}^{*}(\tilde{K}) = \tilde{K} $, constrains these parameters and leads us to find necessary and sufficient conditions for the map $\hat{\varphi}$ to be integrable, as follows. 
\begin{thm}\label{defd4} 
For the deformed symplectic map $\hat{\varphi}$ to admit the first integral \eqref{firstintD4}, it is necessary and sufficient for the parameters $b_{i}$ to satisfy one  of the following sets of conditions: 
\begin{subequations}
\begin{equation}\label{eq:4}
     (1) \  b_2 = b_4 = %\beta = %%% DON'T NEED beta !!! %%%%%%%%%%%%%%%%
b_1 b_3 ; 
\end{equation}
\begin{equation}\label{eq:5}
    (2) \  b_1 = b_2 = 
%\beta = 
b_3 b_4 ; 
\end{equation}
\begin{equation}\label{eq:6}
    (3) \ b_2=b_3 = % \beta = 
b_1b_4 .
\end{equation}
  \end{subequations}
%where $b_{2}$ is denoted by a real parameter $\beta$. 
Hence in each of these cases, % $(1)$, $(2)$ or $(3)$, 
the deformed map $\hat{\varphi}$ given by \eqref{phihatmapD4} 
 is Liouville integrable, preserving 
the function 
\begin{equation}\label{firstintD4v1}
    \tilde{K} = y_1 + y_2 + \frac{y_1}{y_2} + \frac{(b_1 + 1)y_2}{y_1} + \frac{b_3 + b_4 + 1}{y_2} + \frac{b_1 + b_2 + b_3 + b_4 + 1}{y_1} + \frac{b_1y_2}{y_1^2} + \frac{b_3b_4 + b_3 + b_4}{y_1y_2} + \frac{2b_2}{y_1^2} + \frac{b_3b_4}{y_1^2y_2}
\end{equation}
\end{thm}
\begin{remark} Observe that the form of the original deformed mutations  (\ref{D4recs}) remain invariant under switching $x_3\leftrightarrow x_4$, $b_3\leftrightarrow b_4$, 
and similarly for the form of the reduced map  $\hat{\varphi}$ in \eqref{phihatmapD4} and the first integral \eqref{firstintD4v1} when these last two 
parameters are switched. Hence cases (1) and (3) are equivalent to one another, 
and  $(1)$ and $(2)$ are really the only two distinct cases to consider in Theorem  \ref{defd4}.
\end{remark}

For the two essentially distinct  cases of the reduced map obtained by  deformation of the $\rD_{4}$ cluster map, as identified in the preceding theorem, we have 
two 
2-parameter families of integrable maps given by  $\hat{\varphi}_{1},\hat{\varphi}_{2}$ respectively, where  
\begin{equation}\label{D41map}
\hat{\varphi}_{1}: \quad  \mqty(y_{1} \\ y_{2}) \mapsto \qty(\frac{(b_1+y_{1})y_{2} + b_1b_3}{y_{1}}, \quad  
\frac{\qty[(b_1 + y_1)y_2 + b_1b_3(y_1+1)]\cdot \qty[y_2 + b_{3}]}{y_1^2y_2})
\end{equation}
\begin{equation}\label{D42map}
    \hat{\varphi}_{2}: \quad  \mqty(y_{1} \\ y_{2}) \mapsto 
\qty(\frac{(b_3b_4+y_{1})y_{2} + b_3b_4}{y_{1}}, \quad  \frac{\qty[(b_{4} + y_{2})y_{1} + b_3b_4(y_{2} + 1)]\cdot \qty[(b_{3} + y_{2})y_{1} + 
b_3b_4 (y_{2}+1)]}{b_{4}y_{1}^2y_{2}(b_3b_4+y_{1})})
\end{equation}
where the coefficients in each map are fixed in cases (1) and (2),  respectively. The corresponding invariant functions $\tilde{K}_{1}, \tilde{K}_{2}$ are given by
\begin{equation}\label{D41firstint}
    \tilde{K}_1 = y_1 + y_2 + \frac{y_1}{y_2} + \frac{(b_1 + 1)y_2}{y_1} + \frac{b_3 + b_1b_3 + 1}{y_2} + \frac{b_1 + 2b_1b_3 + b_3 + 1}{y_1} + \frac{b_1y_2}{y_1^2} + 
\frac{b_3(b_1b_3+b_1 + 1)}{y_1y_2} + \frac{2 b_1b_3}{y_1^2} + \frac{b_1b_3^2}{y_1^2y_2} , 
\end{equation}
\begin{equation}\label{D42firstint}
     \tilde{K}_2 = y_1 + y_2 + \frac{y_1}{y_2} + \frac{(b_3b_4 + 1)y_2}{y_1} + \frac{b_3 + b_4 + 1}{y_2} 
+ \frac{2b_3b_4 + b_3 + b_4 + 1}{y_1} + \frac{b_3b_4 y_2}{y_1^2} + \frac{b_3b_4 + b_3 + b_4}{y_1y_2} + \frac{2b_3b_4}{y_1^2} + \frac{b_3b_4}{y_1^2y_2}, 
\end{equation}
which are  the particular relevant cases of the function \eqref{firstintD4v1}. 
The level sets of each of the latter  functions gives a pencil of plane curves, of which the generic member has genus 1 and hence corresponds to an elliptic curve. 
It turns out that the two functions above become equivalent to one another when $b_3=1$, upon identifying the remaining parameters $b_1$ and $b_4$ in each 
case, although the  associated maps $\hat{\varphi}_{1},\hat{\varphi}_{2}$ remain distinct from one another.  

\begin{remark} Since both sets of curves corresponding to $\tilde{K}_{1}$  and $\tilde{K}_{2}$ have bidegree  (3,2), they do not correspond to  QRT maps, which 
come from curves of bidegree (2,2) (that is, biquadratic curves).  
\end{remark}

\subsection{The deformed map $\hat{\varphi}_{1}$ for $\rD_4$} 
%with the condition $b_2 = b_4 = \beta = b_1 b_3$}

The iteration of the deformed map $\hat{\varphi}_{1}$ can be written as the following system of recurrence relations: 
\beq\label{D41recs1} % {D42recs1} 
\begin{array}{rcl}
y_{1,n+1}y_{1,n} & = &(b_1+y_{1,n})y_{2,n} + b_1b_3 \\  
y_{2,n+1}y_{2,n}y_{1,n}^2 & = & \qty((b_1 + y_{1,n})y_{2,n} + b_1b_3(y_{1,n}+1))\qty(y_{2,n} + b_{3}) . %\\  
\end{array} 
\eeq 
Using the $p$-adic method used in previous cases, we set some rational values for the parameters and initial conditions, and observe the orbit of the map $\hat{\varphi}_{1}$ 
as a sequence in $\Q^2$. Then we find three different singularity patterns, given by  
\begin{equation}\label{singD4i}
\begin{split}
    &\text{Pattern 1 :} \ (y_{1,n},y_{2,n}) = \dots,(R,0^{1}),(R,\infty^{1}) ,(\infty^{1},\infty^{1}),(\infty^{1},R),(R,0^{1}), (R,R)\dots\\[0.5em]  
    &\text{Pattern 2 :} \  (y_{1,n},y_{2,n}) =\dots, (0^{1},R),\dots \\[0.5em]  
    &\text{Pattern 3 :} \ (y_{1,n},y_{2,n}) = \dots, (R,0^{1}),(R,\infty^{1})\dots . 
\end{split}
\end{equation}
By associating a  tau function with each pattern, so that  $\tau_{n},r_n,\sigma_{n}$ correspond to Patterns 1,2,3 respectively, we are led to the change of variables % transformations   
\begin{equation}\label{vartransD4i}
      y_{1,n} = \frac{r_n}{\tau_{n-1} \tau_{n}}, \quad y_{2,n} =\frac{\sigma_{n+1}\tau_{n-2}\tau_{n+2}}{\sigma_{n}\tau_{n}\tau_{n+1}} . 
\end{equation}
If we directly substitute these variables into the recurrences \eqref{singD4i}, then we obtain the  relations  
\begin{equation}\label{y-tau}
\begin{split}
   & r_{n+1}r_{n} = \frac{(b_{1}\tau_{n-1}\tau_{n} + r_n)\sigma_{n+1}\tau_{n-2}\tau_{n+2} + b_1b_3\tau_{n-1}\tau_{n}^2\tau_{n+1}\sigma_{n}}{\sigma_{n}} \\
    &\sigma_{n+2}\tau_{n+3} = \frac{\qty[b_1b_3 \sigma_{n}\tau_{n}\tau_{n+1} + b_{1}\sigma_{n+1}\tau_{n-2}\tau_{n+2}]\cdot \qty[(b_{1}\tau_{n-1}\tau_{n} + r_{n})\sigma_{n+1}\tau_{n-2}\tau_{n+2} + b_1b_3 \sigma_{n}\tau_{n}\tau_{n+1}(\tau_{n-1}\tau_{n} + r_{n})]}{b_{1}r^{2}_{n}\tau_{n-2}\sigma_{n}} . 
\end{split}
\end{equation}

To simplify the above relations and decouple them in such a way that they represent exchange relations, 
it is helpful to observe the full singularity pattern in 4D, which emerges from applying the sequence of 
deformed mutations \eqref{D4recs} subject to the  conditions $b_2 = b_4 = \beta = b_1 b_3$. 
Using the $p$-adic method in this case suggests making a transformation on the level of the $x$-variables, defined by  
\begin{equation}\label{D4xtau}
    \begin{split}
         x_{1,n} = \rho_{n}\frac{\sigma_{n}}{\tau_{n-1}} \quad x_{2,n} = \frac{r_{n}}{\tau_{n-1}\tau_{n}} \quad  x_{3,n} = \rho_n \frac{\sigma_{n+1}}{\tau_{n}}, \quad x_{4,n} = \frac{\tau_{n-2}\tau_{n+2}}{\tau_{n-1}\tau_{n+1}} , 
    \end{split}
\end{equation}
where the extra prefactor $\rho_{n}$ corresponds to an additional singularity pattern, appearing only on this level. % of $x$-variables. 
By a short calculation, one can confirm that these new formulae are consistent with the expression for $y_{2,n}$ previously given in %fact constructed as 
\eqref{vartransD4i}, since we have 
\begin{align*}
    y_{2,n} = \frac{x_{3,n}x_{4,n}}{x_{1,n}} = \frac{\sigma_{n+1}\tau_{n-1}}{\sigma_{n}\tau_{n}}\cdot \frac{\tau_{n-2}\tau_{n+2}}{\tau_{n-1}\tau_{n+1}} = \frac{\sigma_{n+1}\tau_{n-2}\tau_{n+2}}{\sigma_{n}\tau_{n}\tau_{n+1}}, 
\end{align*}
as required. 
%Under the condition $b_2 = b_4 = \beta = b_1 b_3$, 
Thus, with the parameters constrained as in \eqref{eq:4}, the iteration of the deformed map \eqref{D4recs} is equivalent to a system of four relations, 
namely 
%on the space of tau functions is given by the following relations, 
%
\begin{equation}\label{xisystmtau} 
    \begin{split}
    &\rho_{n+1}\rho_{n}\sigma_{n+1}\sigma_{n} = b_{1} \tau_{n-1}\tau_{n} + r_{n} \\ 
    & r_{n+1}r_{n} = \rho_{n+1}\rho_{n}\sigma_{n+1}^2\tau_{n+2}\tau_{n-2} + b_1b_3\tau_{n+1}\tau_{n}^2\tau_{n-1}
    \\
    &\rho_{n+1}\rho_{n}\sigma_{n+2}\sigma_{n+1} = b_{3} \tau_{n+1}\tau_{n} + r_{n+1}\\ 
    &\tau_{n+3}\tau_{n-2} = b_1b_3 \tau_{n}\tau_{n+1} + r_{n+1} . 
    \end{split}
\end{equation}
By incorporating the above relations into \eqref{y-tau}, and eliminating $\rho_n$, we are able to decouple them into a total of three  recurrences, 
which all take the form of exchange relations, 
given by 
\begin{equation}\label{systmtauD4i}
    \begin{split}
        & \sigma_{n+2}r_{n} =b_{3}\sigma_{n}\tau_{n}\tau_{n+1} + \sigma_{n+1}\tau_{n-2}\tau_{n+2} \\ 
        &  r_{n+1}\sigma_{n} = \sigma_{n + 1}\tau_{n+2}\tau_{n-2} + b_{1}\sigma_{n+2}\tau_{n}\tau_{n-1} \\ 
        &\tau_{n+3}\tau_{n-2} = b_1b_3 \tau_{n}\tau_{n+1} + r_{n+1} . 
    \end{split}
\end{equation}

Next, in order 
to confirm that this gives a cluster map defined on a suitable space of tau functions, we need to build an appropriate exchange matrix which produces  \eqref{systmtauD4i}  
via a sequence of mutations. 
Firstly, let us combine the initial tau functions into a cluster in a seed for a coefficient-free cluster algebra, 
by setting   $$(\tilde{x}_{1},\tilde{x}_{2},\tilde{x}_{3},\tilde{x}_{4},\tilde{x}_{5}, \tilde{x}_{6},\tilde{x}_{7},\tilde{x}_{8}) = (\sigma_{0},\sigma_{1},r_{0},\tau_{-2},\tau_{-1},\tau_{0},\tau_{1},\tau_{2}),$$ 
and let $\tilde{\pi}_{1}:\, \C^8\to \C^2$ be the rational map defined by \eqref{vartransD4i}. 
Then, upon taking the pullback of the symplectic form \eqref{symD4iy} by %the rational map 
$\tilde{\pi}_{1}$,  we find 
\begin{equation}
    \tilde{\omega} = \tilde{\pi}_{1}^*(\hat{\omega}) = \sum_{1\leq i<j\leq 8}\tilde{b}^{(1)}_{ij}\dd \log \tilde{x}_{i}\wedge \dd \log \tilde{x}_{j}
\end{equation}
where the coefficients are combined into the matrix $\tilde{B}^{(1)} = (\tilde{b}^{(1)}_{ij})$ given by  
\begin{equation}\label{extmD4I}
\tilde{B}^{(1)} =   %(b^{(1)}_{ij}) = 
\mqty(0 & 0 & -1 & 0 & 1 & 1 & 0 & 0 \\ 0 & 0 & 1 & 0 & -1 & -1 & 0 & 0 \\ 1 & -1 & 0 & -1 & 0 & 1 & 1 & -1 \\ 0 & 0 & 1 & 0 & -1 & -1 & 0 & 0 \\ -1 & 1 & 0 & 1 & 0 & -1 & -1 & 1 \\-1 & 1 & -1 & 1 & 1 & 0 & -1 & 1 \\ 0 & 0 & -1 & 0 & 1 & 1 & 0 & 0 \\ 0 & 0 & 1 & 0 & -1 & -1 & 0 & 0) . 
\end{equation}
We proceed to add two extra rows, associated with the frozen variables $b_{1}$ and $b_{3}$, to the bottom of the exchange matrix \eqref{extmD4I}, which will result in the 
construction of the extended exchange matrix $\hat{B}^{(1)}$ shown in \eqref{dexchmD41} below. 
Figure \ref{quiverD4i} depicts the quiver associated with the full matrix $\hat{B}^{(1)}$. 
\begin{figure}[h]%\label{deformQD4i}
\centering
\epsfig{file={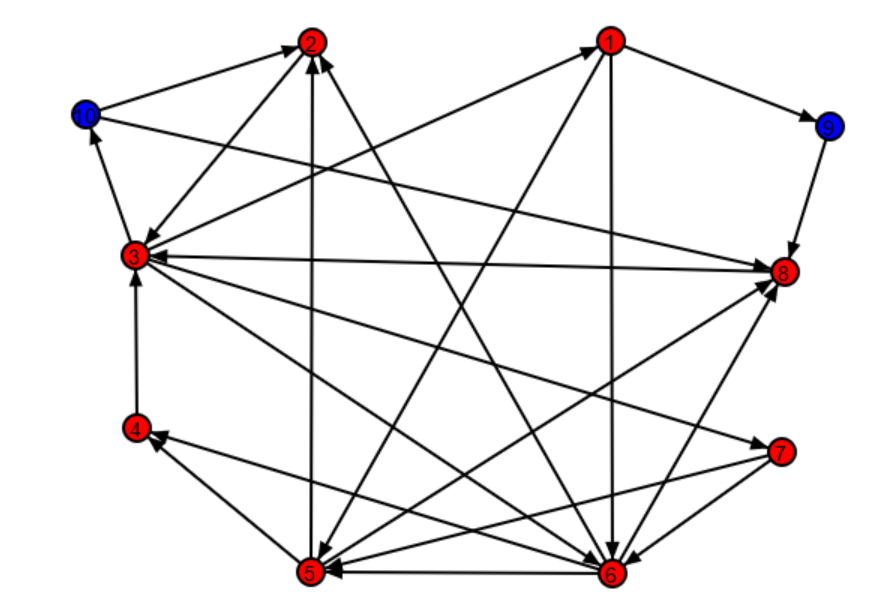}, height=3in, width=4.8in}
\caption{Extended quiver associated with the deformed $\rD_4$ cluster map $\psi_{1}$}
\label{quiverD4i}
\end{figure}
%
%Thus the matrix enables us to show the following result, 
%
\begin{thm}\label{D41clusthm} 
Given the extended initial cluster 
$$ \hat{\vb{x}} =(\tx_j)_{1\leq j \leq 10} = 
(\sigma_{0},\sigma_{1},r_{0},\tau_{-2},\tau_{-1},\tau_{0},\tau_{1},\tau_{2}, b_1, b_3),$$ 
and the  permutation $\rho_{1} = (123)(45678)$, 
the  iteration of the cluster map $\psi_{1} = \rho_{1}^{-1}\tmu_{4}\tmu_{1}\tmu_{3}$ 
defined by the extended exchange matrix  $\hat{B}^{(1)}$ in \eqref{dexchmD41}  with square submatrix  \eqref{extmD4I} is 
equivalent to the system of recurrences  \eqref{systmtauD4i}, 
%The iteration of $\psi_1$ gives 
which generates elements of  
$\mathbb{Z}_{>0}\qty[b_1,b_3,\sigma_{0}^{\pm 1},\sigma_{1}^{\pm 1},r_{0}^{\pm 1},\tau_{-2}^{\pm 1},\tau_{-1}^{\pm 1},\tau_{0}^{\pm 1},\tau_{1}^{\pm 1},\tau_{2}^{\pm 1}]$
\end{thm}
\begin{prf}
Let us consider the initial seed $(\hat{\vb{x}},\hat{B}^{(1)})$  containing the extended initial cluster 
$\hat{\vb{x}}$ as above, 
%=(\tilde{x}_{1},\tilde{x}_{2},\tilde{x}_{3},\tilde{x}_{4},\tilde{x}_{5}, \tilde{x}_{6},\tilde{x}_{7},\tilde{x}_{8},\tilde{x}_{9},\tilde{x}_{10}) = (\sigma_{0},\sigma_{1},r_{0},\tau_{-2},\tau_{-1},\tau_{0},\tau_{1},\tau_{2}, b_1, b_3)$ 
with the corresponding  extended exchange matrix given by 
\begin{equation} \label{dexchmD41}
    \hat{B}^{(1)} = \mqty(0 & 0 & -1 & 0 & 1 & 1 & 0 & 0 \\ 0 & 0 & 1 & 0 & -1 & -1 & 0 & 0 \\ 1 & -1 & 0 & -1 & 0 & 1 & 1 & -1 \\ 0 & 0 & 1 & 0 & -1 & -1 & 0 & 0 \\ -1 & 1 & 0 & 1 & 0 & -1 & -1 & 1 \\-1 & 1 & -1 & 1 & 1 & 0 & -1 & 1 \\ 0 & 0 & -1 & 0 & 1 & 1 & 0 & 0 \\ 0 & 0 & 1 & 0 & -1 & -1 & 0 & 0 \\ -1 & 0 & 0 & 0 & 0 & 0 & 0 & 1 \\ 0 & 1 & -1 & 0 & 0 & 0 & 0 & 1) .
\end{equation}
Applying cluster mutation $\tmu_3$ at node 3, followed by $\tmu_1$ and $\tmu_4$, will give 
the exchange relations 
%\begin{align*}    \sigma_{2}r_{0} = b_{3}\sigma_{0}\tau_{0}\tau_{1} + \sigma_{1}\tau_{-2} \tau_{2} \end{align*}
\beq\label{D41muts} 
 \begin{split}
        & \tilde{x}'_{3}\tilde{x}_{3}  = \tx_{10} \tilde{x}_{1}\tilde{x}_{6}\tx_7 + \tilde{x}_{2}\tx_4\tx_8 \\ 
        & \tilde{x}'_{1}\tilde{x}_{1}  = \tilde{x}_{2}\tx_4\tx_8  + \tx_9\tilde{x}_{3}{'}\tilde{x}_{5}\tilde{x}_{6} \\
        & \tilde{x}'_{4}\tilde{x}_{4}  = \tx_9\tx_{10} \tilde{x}_{6}\tilde{x}_{7} + \tilde{x}_{1}{'}, 
    \end{split}
\eeq 
which have the same form as the expressions in \eqref{systmtauD4i}. 
Under this sequence of mutations, the extended exchange matrix $\hat{B}^{(1)}$ satisfies the %following 
relation
\begin{align*}
  \mu_{4}\mu_{1}\mu_{3}(\tilde{B}^{(1)}) = \rho_{1} (\tilde{B}^{(1)}) = P_{1} \tilde{B}^{(1)} P_{2}  
\end{align*}
where the action of the permutation %of labellings 
$\rho_{1} = (123)(45678)$ on the rows/columns is  represented by the  matrices  
\begin{equation}\label{permD4I}
    P_{1}=\mqty(0& 0& 1& 0& 0& 0& 0& 0 & 0 & 0 \\ 1& 0& 0& 0& 0& 0& 0& 0 & 0 & 0 \\0& 1& 0 & 0& 0& 0& 0& 0 & 0 & 0 \\ 0& 0& 0 & 0& 0& 0& 0& 1 & 0 & 0\\ 0& 0& 0 & 1& 0& 0& 0& 0 & 0 & 0 \\ 0& 0& 0 & 0 & 1& 0& 0& 0 & 0 & 0  \\0& 0& 0 & 0 & 0& 1 & 0& 0 & 0 & 0 \\0& 0& 0 & 0 & 0& 0 & 1 & 0 & 0 & 0 \\ 0& 0& 0 & 0 & 0& 0 & 0 & 0 & 1 & 0 \\0& 0& 0 & 0 & 0& 0 & 0 & 0 & 0 & 1 ), \quad 
P_{2} =  \mqty(0 & 1 & 0 & 0 & 0 & 0 & 0 & 0 \\ 0 & 0 & 1 & 0 & 0 & 0 & 0 & 0\\1 & 0 & 0 & 0 & 0 & 0 & 0 & 0\\0 & 0 & 0 & 0 & 1 & 0 & 0 & 0\\ 0 & 0 & 0 & 0 & 0 & 1 & 0 & 0 \\ 0 & 0 & 0 & 0 & 0 & 0 & 1 & 0 \\ 0 & 0 & 0 & 0 & 0 & 0 & 0 & 1 \\ 0 & 0 & 0 & 1 & 0 & 0 & 0 & 0) . 
\end{equation}
It follows that $\hat{B}^{(1)}$ is preserved by the action of the associated cluster map, that is $\psi_1 (\hat{B}^{(1)})=\hat{B}^{(1)}$, where 
$\psi_{1} = \rho^{-1}_{1}\mu_{4}\mu_{1}\mu_{3}$, and the combination of the 
inverse permutation with the exchange relations in \eqref{D41muts} 
precisely corresponds to the shift of index $n\to n+1$ acting on 
the tau functions in each cluster, reproducing the iteration of the system \eqref{systmtauD4i}. 
Hence this cluster map is a Laurentification of  the deformed $\rD_4$ map $\hat{\varphi}_{1}$, generating Laurent polynomials in the initial cluster variables, and  positivity 
for skew-symmetric cluster algebras \cite{ls} 
implies that their coefficients are positive integers. 
%with integer ring  $$\mathbb{Z}_{>0}\qty[\sigma_{0}^{\pm},\sigma_{1}^{\pm},r_{0}^{\pm},\tau_{-2}^{\pm},\tau_{-1}^{\pm},\tau_{0}^{\pm},\tau_{1}^{\pm},\tau_{2}^{\pm}]$$
%
\end{prf}

\begin{remark}\label{remD4iQ} The subquiver in Figure \ref{quiverD4i} consisting of the 8 unfrozen nodes is mutation equivalent to another particular quiver presented by Okubo, 
which enables a  $q$-Painlev\'{e} VI equation to be constructed from an appropriate combination of coefficient mutations \cite{okubo}.
\end{remark} 

\subsection{The deformed map $\hat{\varphi}_{2}$ for $\rD_4$} 
%with the condition $b_{1} = b_{2} = \beta = b_{3}b_{4}$}
%

Let us now consider iteration of the map $\hat{\varphi}_{2}$ given by  \eqref{D42map}, which can be written as the recurrence
\beq\label{D42recs1} 
\begin{array}{rcl}
y_{1,n+1}y_{1,n} & = &(b_3b_4+y_{1,n})y_{2,n} + b_3b_4 \\  
y_{2,n+1}y_{2,n}y_{1,n}^2 b_{4} (b_3b_4 +y_{1,n}) & = & \qty((b_{4} + y_{2,n})y_{1,n} + b_3b_4(y_{2,n} + 1)) \qty((b_{3} + y_{2,n})y_{1,n} + b_3b_4 (y_{2,n}+1))\\  
\end{array} 
\eeq 
Repeating the same procedure as in the previous sections, we find three singularity patterns which arise from orbits of \eqref{D42recs1}, namely  
\begin{equation}\label{singD4ii}
\begin{split}
    &\text{Pattern 1 :} \ (y_{1,n},y_{2,n}) = \dots,(R,\infty^{1}),(\infty^{1},\infty^{1}) ,(\infty^{1},R),(R,0^{1}),(R,\infty^{1}), (R,R)\dots\\[0.5em]  
    &\text{Pattern 2 :} \  (y_{1,n},y_{2,n}) =\dots, (0^{1},R),\dots \\[0.5em]  
    &\text{Pattern 3 :} \ (y_{1,n},y_{2,n}) = \dots, (R,0^{1}),\dots 
\end{split}
\end{equation}
By relating the singularities appearing in each pattern with new variables, we define the following variable transformation, in an attempt to Laurentify the deformed map $\hat{\varphi}_{2}$:
\begin{equation}\label{var2}
    y_{1,n} = \frac{\hat{\eta}_{n}}{\hat{\tau}_{n-1}\hat{\tau}_{n}}, \quad y_{2,n} = \frac{{\rho}_{n}\hat{\tau}_{n-2}}{\hat{\tau}_{n+1}\hat{\tau}_{n}\hat{\tau}_{n-3}}
\end{equation}
By substituting these expressions into \eqref{D42recs1}, we find a rather complicated system of equations: in particular, the resulting 
expression for the product $\rho_{n+1}\rho_n$ cannot be considered as an exchange relation, as it is not immediately given as a binomial expression in the other variables. 
To resolve this problem, we look at the singularity patterns in the original 4D deformed map \eqref{D4recs} with $b_1=b_2=b_3b_4$,   
and introduce  variable transformations corresponding to these. This suggests that the  $x_{i}$ should  be expressed as  
\begin{align*}
   x_{1,n} = \frac{\htau_{n+1}\htau_{n-3}}{\htau_{n}\htau_{n-2}}, \quad x_{2,n} = \frac{\hat{\eta}_{n}}{\htau_{n}\htau_{n-1}}, \quad  x_{3,n} = \frac{\hat{r}_{n}\xi_{n}}{\htau_{n}}, \quad x_{4,n} = \frac{\hat{\sigma}_{n}}{\htau_{n}\xi_{n}}
\end{align*}
where $\xi_{n}$ satisfies $\xi_{n} \xi_{n+1} = \frac{\hat{\sigma}_{n}}{\hat{r}_{n}}$. By directly substituting these variables into \eqref{D4recs}, 
we find a system of equations written as follows:  
\begin{equation}\label{D42xeqns}
\begin{split}
    &\htau_{n+2}\htau_{n-3} = b_3b_4  \htau_{n}\htau_{n-1} + \hat{\eta}_{n} \\ 
    &\hat{\eta}_{n+1}\hat{\eta}_{n} = \hat{r}_{n}\hat{\sigma}_{n}\htau_{n-2}\htau_{n+2} + b_3b_4 \htau_{n+1}\htau_{n}^2\htau_{n-1} \\ 
    &\hat{r}_{n+1}\hat{\sigma}_{n} = b_{3}\htau_{n}\htau_{n+1} +\hat{\eta}_{n+1} \\
    & \hat{\sigma}_{n+1}\hat{r}_{n} = b_{4} \htau_{n}\htau_{n+1} + \hat{\eta}_{n+1} 
    \end{split}
\end{equation}
%
%Note the equations \eqref{D42xeqns} can be derived from both \eqref{var2} and \eqref{D42yeqns}. 
Also, by observing the singularity pattern for $y_{1,n}$ explicitly from \eqref{D42recs1}, one can see that $y_{1,n}$ should satisfy the relation %variable expression
\begin{equation} \label{d43wvar} 
    w_{1,n} := y_{1,n}+ b_3b_4 =  \frac{\htau_{n+2}\htau_{n-3}}{\htau_{n}\htau_{n-1}} , 
\end{equation}
%
%Thus this yields 
which is in agreement with what is found by combining \eqref{var2} with  
the first recurrence in \eqref{D42xeqns}. 
Furthermore, by setting $\rho_{n} = \hat{r}_{n}\hat{\sigma}_{n}$, the relation for $\rho_{n+1}\rho_n$ obtained from \eqref{D42recs1} 
follows by taking the product of the last two expressions in \eqref{D42xeqns}. 

Using the above, we can consider $(y_{1,0},y_{2,0}) =(y_1,y_2)$ and define a rational map $\tilde{\pi}_2: \, \C^8\to\C^2$ by 
$$ 
\tilde{\pi}_2: \qquad y_{1} =\frac{\hat{\eta}_0}{\htau_{-1}\htau_0}, \qquad y_2 = \frac{\hat{\sigma}_0\hat{r}_0\htau_{-2}}{\htau_1\htau_0\htau_{-3}}
$$ 
The exchange matrix describing the cluster dynamics \eqref{D42xeqns} is found by pulling back the symplectic form $\hat{\omega}$, as in \eqref{symD4iy}, via the rational map $\tilde{\pi}_2$, to obtain the presymplectic form 
\begin{align*}
    \tilde{\omega} = \tilde{\pi}_2^{*}( \hat{\omega}) = \sum_{i<j} \frac{\tilde{b}^{(2)}_{ij}}{\tilde{x}_{i}\tilde{x}_{j}} \dd \tilde{x}_{i} \wedge \dd \tilde{x}_{j} . 
\end{align*}
Now if we choose to order the  coordinates and  identify them  
with variables in a coefficient-free cluster algebra as 
$$
\qty(\tilde{x}_{1},\tilde{x}_{2},\tilde{x}_{3},\tilde{x}_{4},\tilde{x}_{5},\tilde{x}_{6},\tilde{x}_{7},\tilde{x}_{8}) = 
(\htau_{-3},\hat{r}_{0},\hat{\eta}_{0},\hat{\sigma}_{0}, \htau_{-1},\htau_{0},\htau_{1},\htau_{-2}) , 
$$ 
then we see that the map  $\tilde{\pi}_2$ is equivalent to  $\tilde{\pi}_1$ defined 
by \eqref{vartransD4i} in case (1) above, so that 
$$ 
 y_{1} =\frac{\tx_3}{\tx_{5}\tx_6}, \qquad 
y_2 = \frac{\tx_2\tx_4\tx_8}{\tx_1\tx_6\tx_{7}}
$$ 
 and the exchange matrix with entries $\tilde{b}^{(2)}_{ij}$ is identical to 
the one obtained previously, that is 
\begin{equation}\label{extmD4II}
   \tilde{B}^{(2)} =\tilde{B}^{(1)} 
\end{equation}
as in \eqref{extmD4I}. 

\begin{figure}[h]
\centering
\epsfig{file={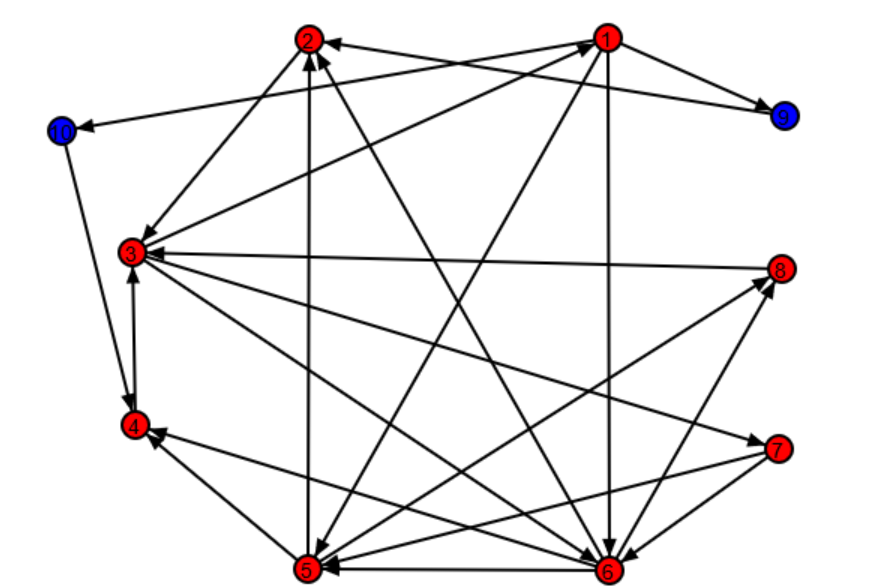}, height=3in, width=4.8in}
\caption{Extended quiver associated with the deformed $\rD_4$ cluster map $\psi_{2}$}
\label{quiverD4ii}
\end{figure}

To obtain an 
extended version of $\tilde{B}^{(2)}$ that includes $b_3,b_4$ as frozen variables and reproduces \eqref{D42xeqns} from a suitable sequence of mutations, we need to 
construct two extra rows as in \eqref{extmatD42} below. 
The result of this is represented by the quiver in Figure \ref{quiverD4ii}, with two frozen nodes.
\begin{thm}\label{D42clusthm}
Given the extended initial cluster 
$$ \hat{\vb{x}} =(\tx_j)_{1\leq j \leq 10} = 
(\htau_{-3},\hat{r}_{0},\hat{\eta}_{0},\hat{\sigma}_{0}, \htau_{-1},\htau_{0},\htau_{1},\htau_{-2},
b_3, b_4),$$ 
and the  permutation $\rho_{2} = (24)(18567)$, 
the  iteration of the cluster map $\psi_{2} = \rho_{2}^{-1}\tmu_2\tmu_{4}\tmu_{3}\tmu_{1}$ 
defined by the extended exchange matrix  $\hat{B}^{(2)}$ in \eqref{extmatD42}  with square submatrix  \eqref{extmD4I} is 
equivalent to the system of recurrences  \eqref{D42xeqns}, 
which generates elements of  
$\mathbb{Z}_{>0}\qty[b_3,b_4,\hat{r}_0^{\pm 1},\hat{\eta}_0^{\pm 1}, \hat{\sigma}_{0}^{\pm 1},\htau_{-3}^{\pm 1},\htau_{-2}^{\pm 1},\htau_{-1}^{\pm 1},\htau_{0}^{\pm 1},\htau_{1}^{\pm 1}]$. 
\end{thm}
\begin{prf}
From \eqref{extmD4II} we note that the coefficient-free cluster algebra is identical to that specified by the same $8\times 8$ exchange matrix as was found in case (1) previously, but 
we need to extend it in such a way that, once $b_3$ and $b_4$ are included as frozen variables, it is compatible with the four relations in  \eqref{D42xeqns} (whereas in case (1) there were only thee relations). 
In this way, we construct a $10\times 8$ extended exchange matrix $\hat{B}^{(2)}$ from $\tilde{B}^{(2)}=\tilde{B}^{(2)}$, given by 
\begin{equation} \label{extmatD42} 
    \hat{B}^{(2)} = 
 \mqty(0 & 0 & -1 & 0 & 1 & 1 & 0 & 0 \\ 0 & 0 & 1 & 0 & -1 & -1 & 0 & 0 \\ 1 & -1 & 0 & -1 & 0 & 1 & 1 & -1 \\ 0 & 0 & 1 & 0 & -1 & -1 & 0 & 0 \\ -1 & 1 & 0 & 1 & 0 & -1 & -1 & 1 \\-1 & 1 & -1 & 1 & 1 & 0 & -1 & 1 \\ 0 & 0 & -1 & 0 & 1 & 1 & 0 & 0 \\ 0 & 0 & 1 & 0 & -1 & -1 & 0 & 0 \\ -1 & 1 & 0 & 0 & 0 & 0 & 0 & 0  \\ -1 & 0 & 0 & 1 & 0 & 0 & 0 & 0) , 
\end{equation}
and we note that the last two rows are  different from  $\hat{B}^{(1)}$ in  \eqref{dexchmD41} 
(as can be seen by comparing Figures \ref{quiverD4i} and \ref{quiverD4ii}). 
Such an extended exchange matrix is invariant under the following 

Applying a sequence of mutations, starting with mutation $\tmu_1$ at node 1 and successively mutating at nodes 3,4 and 2, we find that the nodes are permuted by the given 
permutation $\rho_2$, so that  
\begin{align*}
    \tmu_{2}\tmu_{4}\tmu_{3}\tmu_{1}(\hat{B}_{2}) = \rho_{2}(\hat{B}_{2}), 
\end{align*}
which is a equivalent to the action of a suitable pair of row/column permutation matrices acting on $\hat{B}_{2}$. Hence the  
overall action of $\psi_2=  \rho_{2}^{-1}\tmu_2\tmu_{4}\tmu_{3}\tmu_{1}$ leaves $\hat{B}_{2}$ invariant, and it is 
straightforward to check that the corresponding combination of cluster mutations with a permutation is  equivalent to one 
iteration of the relations \eqref{D42xeqns}.  
Then as usual, because they are cluster variables,  the iterates are elements of the corresponding ring of Laurent polynomials, with 
positive integer coefficients.  
\end{prf}

\begin{remark}\label{remD4iQ} Since the subquiver with 8 unfrozen nodes in Figure \ref{quiverD4ii} is the same as that in Figure \ref{quiverD4i}, it  is also mutation equivalent to the quiver 
associated with the $q$-Painlev\'{e} VI equation in \cite{okubo}. 
\end{remark} 

\subsection{Connection with special  Somos-7 relation}

As we have already seen in examples of deformations in type $\rA$ and $\rB$, on fixed level sets the orbits of suitable tau functions satisfy a special Somos-7 relation, 
which is related to the Lyness map. There is also a corresponding Somos-5 relation, although the details of this are a bit more subtle (see Appendix B). 
We find a similar result for the  deformed maps of type $\rD_{4}$. 

\begin{thm}\label{th58} 
For each integrable case of 
the deformed $\rD_4$ map,  
the variable $w_{n} = y_{1,n} + \beta$ %, one can show that the iterates given by the deformed map $\hat{\varphi}_{1}$ 
satisfies the Lyness map in the  form 
\begin{equation}\label{lyness}
    w_{n+1}w_{n-1} = (1-\beta) w_{n} + \delta, 
\end{equation}
where in case (1) we have 
$$ \beta = b_1b_3, \qquad \delta  = \beta \tilde{K}_{1} + 2\beta^2 +b_1 + b_3,  $$ 
on each level set of the invariant function $ \tilde{K}_{1}$ given in \eqref{D41firstint}, while in case (2) the parameters are specified by 
$$ \beta = b_3b_4, \qquad \delta  = \beta \tilde{K}_{2} + 2\beta^2 +b_3 + b_4, $$
with $ \tilde{K}_{2}$ as in \eqref{D42firstint}. Furthermore, in case (1) we can express $w_n$ by the formula  
\begin{equation}\label{somos7D4}
    w_{n} = \frac{\tau_{n+2}\tau_{n-3}}{\tau_{n}\tau_{n-1}}, 
\end{equation}
where  the tau function $\tau_n$ satisfies the special  Somos-7 relation 
\begin{equation} \label{sps7again} 
    \tau_{n+7}\tau_{n} = (1-\beta)\tau_{n+6}\tau_{n+1} + \delta \tau_{n+4}\tau_{n+3}, 
\end{equation}
and for case (2) we have the same expression as \eqref{somos7D4} except that $\tau_n$ is replaced by $\htau_n$, 
where the latter satisfies the same relation \eqref{sps7again} but with the modified expression for $\beta$ and $\delta$, as above.  
Similarly, in each case the quantity 
$$ 
 \hat{w}_{n} = y_{1,n} + 1 
$$ 
satisfies the Somos-5 QRT map, in the form of the recurrence 
\begin{equation}
    \hat{w}_{n+1}\hat{w}_{n}\hat{w}_{n-1} = \zeta \hat{w}_{n} + \theta, 
\end{equation}
where, for the appropriate value of $\beta$ in each case, the coefficients are given   
by  $\theta= (\beta -1) \zeta$ with 
$$ 
\mathrm{case}\,\,\mathrm{(1):}\quad 
\zeta =  \tilde{K}_{1} + b_{1} + b_{3} +2, \qquad 
\mathrm{case}\,\,\mathrm{(2):}\quad  \zeta =  \tilde{K}_{2} + b_{3} + b_{4} +2.
$$
\end{thm} 

The proof of the preceding statements is very similar to what was done before for the other examples, so it is omitted.

\subsection{Tropicalization and degree growth for deformed $\rD_{4}$ cluster maps}

We have shown above that the two distinct integrable deformations of type $\rD_4$, given by \eqref{D41map} and \eqref{D42map}, both 
correspond to cluster maps based on the same coefficient-free cluster algebra, but with a pair of frozen variables adjoined in two different ways in each case. 
The fact that the underlying coefficient-free cluster algebra is the same means that the tropical dynamics associated with the degree growth of the cluster variables 
is almost identical for these two cases, so it is convenient to describe them simultaneously and remark briefly on the minor differences between them.  Also, 
because the analysis of the tropical dynamics is very similar to that for the other examples previously considered, we will be more sparing with the details. 

In both cases, we denote the extended initial cluster by $\hat{\vb{x}}$, as described in Theorems \ref{D41clusthm} and \ref{D42clusthm}. 
Then we can write the cluster variables in case (1) in the form 
$$ 
\sigma_n = 
\frac{\rN_n^{(1)}(\hat{\vb{x}})}{\hat{\vb{x}}^{{\bf f}_n}}, \qquad r_n  = \frac{\rN_n^{(2)}(\hat{\vb{x}})}{\hat{\vb{x}}^{{\bf e}_n}}, 
\qquad \tau_n = \frac{\rN_n^{(3)}(\hat{\vb{x}})}{\hat{\vb{x}}^{{\bf d}_n}},  
$$   
with three sets of d-vectors associated with the three different types of tau function, while in case (2) we can write the sequences of cluster variables as 
$$ 
\hat{\tau}_n = 
\frac{\hat{\rN}_n^{(1)}(\hat{\vb{x}})}{\hat{\vb{x}}^{\hat{{\bf d}}_n}}, \qquad 
\hat{r}_n  =\frac{\hat{\rN}_n^{(2)}(\hat{\vb{x}})}{\hat{\vb{x}}^{\hat{{\bf f}}_n}}, 
\qquad \hat{\eta}_n = \frac{\hat{\rN}_n^{(3)}(\hat{\vb{x}})}{\hat{\vb{x}}^{\hat{{\bf e}}_n}}, 
\qquad 
\hat{\sigma}_n = 
\frac{\hat{\rN}_n^{(4)}(\hat{\vb{x}})}{\hat{\vb{x}}^{\hat{{\bf g}}_n}}, 
$$   
where in the latter case there are four different sets of d-vectors 
with their corresponding sequences of tau functions. 
In  the first case, the d-vectors satisfy the $(\max,+)$ version of the system \eqref{systmtauD4i}, given by 
\begin{equation}\label{troptauD4i}
    \begin{split}
        & {\bf f}_{n+2}+{\bf e}_{n} =\max({\bf f}_{n}+{\bf d}_{n}+{\bf d}_{n+1} , {\bf f}_{n+1}+{\bf d}_{n-2}+{\bf d}_{n+2}) \\ 
        &  {\bf e}_{n+1}+{\bf f}_{n} = \max({\bf f}_{n + 1}+ {\bf d}_{n+2} + {\bf d}_{n-2} , {\bf f}_{n+2} + {\bf d}_{n} + {\bf d}_{n-1} )\\ 
        & {\bf d}_{n+3} + {\bf d}_{n-2} = \max( {\bf d}_{n}+ {\bf d}_{n+1} , {\bf e}_{n+1}) , 
    \end{split}
\end{equation} 
and in the second case, we find the  $(\max,+)$ version of  \eqref{D42xeqns}, % \eqref{systmtauD4i}
namely 
\begin{equation}\label{troptauD4ii}
\begin{split}
    &\hat{{\bf d}}_{n+2}+\hat{{\bf d}}_{n-3} = \max(  \hat{{\bf d}}_{n}+\hat{{\bf d}}_{n-1} , \hat{{\bf e}}_{n}) \\ 
    &\hat{{\bf e}}_{n+1}+\hat{{\bf e}}_{n} = \max( \hat{{\bf f}}_{n}+\hat{{\bf g}}_{n}+\hat{{\bf d}}_{n-2}+\hat{{\bf d}}_{n+2} , 
\hat{{\bf d}}_{n+1}+2\hat{{\bf d}}_{n}+\hat{{\bf d}}_{n-1} ) \\ 
    &\hat{{\bf f}}_{n+1}+\hat{{\bf g}}_{n} = \max(\hat{{\bf d}}_{n}+\hat{{\bf d}}_{n+1} ,\hat{{\bf e}}_{n+1} ) \\
    & \hat{{\bf g}}_{n+1}+\hat{{\bf f}}_{n} = \max(\hat{{\bf d}}_{n}+\hat{{\bf d}}_{n+1} , \hat{{\bf e}}_{n+1} ) .
    \end{split}
\end{equation}
In both cases, the initial cluster of d-vectors is specified in terms of the $8\times 8$ identity matrix, denoted $I$, with the initial cluster 
for case (1) being ordered as 
\beq\label{clinitD4i}
\Big( {\bf f}_0 \,\, {\bf f}_1 \,\,{\bf e}_0 \,\,{\bf d}_{-2} \,\,{\bf d}_{-1} \,\,{\bf d}_0 \,\,{\bf d}_1 \,\,{\bf d}_2 \Big) =-I,  
\eeq  
whereas in case (2) the ordering of the initial cluster is 
\beq\label{clinitD4ii}
\Big( \hat{{\bf d}}_{-3} \,\, \hat{{\bf f}}_{0} \,\, \hat{{\bf e}}_{0} \,\, \hat{{\bf g}}_{0} \,\, 
\hat{{\bf d}}_{-1} \,\, \hat{{\bf d}}_{0} \,\, \hat{{\bf d}}_{1} \,\, \hat{{\bf d}}_{-2} \,\,  \Big) =-I .
\eeq  

The ordering of the initial d-vectors in the initial clusters \eqref{clinitD4i} and \eqref{clinitD4ii}, respectively, corresponds in each case to the ordering 
of the initial tau functions in the set of unfrozen seed variables $(\tx_1,\tx_2,\ldots,\tx_8)$. The same ordering determines the appropriate tropical analogues 
of the symplectic coordinates $y_1,y_2$ that satisfy the maps \eqref{D41map} and \eqref{D42map}, namely 
$$ 
{\bf Y}_{1,n} = {\bf e}_n - {\bf d}_{n-1} - {\bf d}_{n} 
, \qquad 
{\bf Y}_{2,n} = {\bf f}_{n+1} - {\bf f}_{n} + {\bf d}_{n+2} - {\bf d}_{n+1} - {\bf d}_{n} + {\bf d}_{n-2} 
$$ 
in case (1), and 
 $$ 
{\bf Y}_{1,n} = \hat{{\bf e}}_n - \hat{{\bf d}}_{n-1} - \hat{{\bf d}}_{n}
, \qquad 
{\bf Y}_{2,n} =\hat{{\bf f}}_{n} + \hat{{\bf g}}_{n} - \hat{{\bf d}}_{n+1} - \hat{{\bf d}}_{n} + \hat{{\bf d}}_{n-2} - \hat{{\bf d}}_{n-3} 
$$ 
in case (2). We find that the $(\max,+)$ dynamical systems \eqref{troptauD4i} and \eqref{troptauD4ii} lead to the same ultradiscrete map 
for the vectors ${\bf Y}_{1,n},{\bf Y}_{2,n}$ in each case. 

\begin{lem}\label{tropD4map} 
The ultradiscrete analogues of \eqref{D41recs1}  and \eqref{D42recs1}, which are satisfied by the vectors ${\bf Y}_{1,n},{\bf Y}_{2,n}$, are the same 
in each case, being 
given by the following system of equations:  
\beq\label{tropD4recs} %%% {B21recs} 
\begin{array}{rcl}
{\bf W}_{n} & = & [{\bf Y}_{1,n}]_+ ,  \\
{\bf Y}_{1,n+1}+{\bf Y}_{1,n} & = & [ {\bf W}_{n}+{\bf Y}_{2,n}]_+ , \\  
{\bf Y}_{2,n+1}+{\bf Y}_{2,n}+ 2{\bf Y}_{1,n} & = & 2[{\bf Y}_{2,n}]_+ +{\bf W}_{n} .   \\  
\end{array} 
\eeq 
Given arbitrary initial values $(Y_{1,0},Y_{2,0})\in\R^2$,  every component of  this system is periodic with period 4. 
\end{lem} 
\begin{prf}
Since this is very similar to the proof of Lemma \ref{tropB2map}, it is omitted.  
\end{prf} 

The quantity ${\bf W}_n$ appearing in the system  \eqref{tropD4recs} is 
given as a combination of d-vectors by 
\beq\label{tropD41W} 
{\bf W}_n={\bf d}_{n+2}-{\bf d}_{n}-{\bf d}_{n-1}+{\bf d}_{n-3} 
\eeq 
in case (1), associated with denominators of the tau functions $\tau_n$, 
and by 
\beq\label{tropD42W} 
{\bf W}_n=\hat{{\bf d}}_{n+2}-\hat{{\bf d}}_{n}-\hat{{\bf d}}_{n-1}
+\hat{{\bf d}}_{n-3} 
\eeq 
in case (2), which is a combination of d-vectors 
associated with denominators of the tau functions $\hat{\tau}_n$. 
By the 
results of Theorem \ref{th58}, each of  these tau functions  
satisfies a Somos-7 relation of the form \eqref{sps7again}, with a 
coefficient $\delta$ that can be written  
%appearing in the for
in terms of the initial cluster variables as a Laurent polynomial of the form 
$$ 
\tilde{\pi}_j^*(\delta) = 
\frac{{\mathrm P}_j (\hat{\vb{x}})} {\tx_1\tx_2\tx_3^2\tx_4\tx_5\tx_6\tx_7\tx_8},   
$$ 
with  a suitable polynomial  ${\mathrm P}_j$ in the numerator, and the same 
form of the denominator, for $j=1,2$, respectively.  This immediately yields 
the following analogue of Lemma \ref{B32tropS7}.
\begin{lem} \label{D4tropS7} The  d-vectors 
${\bf d}_n\in\Z^8$ that specify the denominators of 
tau functions $\tau_n$ generated by the system  \eqref{troptauD4i}
satisfy the tropical Somos-7 relation 
\beq\label{tropS7}
{\bf d}_{n+7}+{\bf d}_n = \max ({\bf d}_{n+6}+{\bf d}_{n+1} , 
\hat{{\bf c}}+{\bf d}_{n+4}+{\bf d}_{n+3}),   
\eeq 
with the constant vector 
$$ 
\hat{{\bf c}}= (1,1,2,1,1,1,1,1)^T, 
$$ 
and the d-vectors $\hat{{\bf d}}_n\in\Z^8$ that specify the denominators of 
tau functions $\hat{\tau}_n$ 
generated by  \eqref{troptauD4ii} satisfy exactly the same relation. 
Also, the 
corresponding quantity ${\bf W}_n$, defined by either \eqref{tropD41W} 
or \eqref{tropD42W}, 
satisfies the ultradiscrete Lyness map 
\beq\label{udlynessW}
{\bf W}_{n+1} +  {\bf W}_{n-1} = \max ({\bf W}_{n}, \hat{{\bf c}} ), 
\eeq 
with the same  constant  $\hat{{\bf c}}$. 
\end{lem} 

Due to the way that the initial seeds for the d-vectors are combined in \eqref{clinitD4i} and \eqref{clinitD4ii}, in each case the initial tau functions 
correspond to the same initial values 
$$ 
{\bf Y}_{1,0}=(0,0,-1,0,1,1,0,0)^T, \qquad 
{\bf Y}_{2,0}=(1,-1,0,-1,0,1,1,-1)^T
$$ 
for the system \eqref{tropD4recs}. These initial values produce the period 4 orbit 
$$ 
{\bf W}_0 = (0,0,0,0,1,1,0,0)^T,
{\bf W}_1 = (1,0,1,0,0,1,1,0)^T, {\bf W}_2 = (1,1,2,1,0,0,1,1)^T,
{\bf W}_3 = (0,1,1,1,1,0,0,1)^T,
$$ 
in which the third component corresponds to an orbit  
of the scalar map \eqref{troplyness} with $c=2$, namely the first orbit 
listed in \eqref{troplyc2}, while every other component 
corresponds to the orbit \eqref{troplyc1} of the same scalar map with $c=1$. 
This allows the d-vectors for the tau functions to be completely determined in both case (1) and case (2). Since the steps of the proof of the following result are very similar to those for Theorem \ref{tropB2degree}, we leave the details for the reader. 

\begin{thm}\label{tropD4degree}
The d-vectors ${\bf d}_n,{\bf e}_n,{\bf f}_n$ 
satisfying the $(\max,+)$ system  \eqref{troptauD4i}, 
as well as the d-vectors 
$\hat{{\bf d}}_{n}$, $\hat{{\bf e}}_{n}$, $\hat{{\bf f}}_{n}$, $\hat{{\bf g}}_{n}$ 
satisfying \eqref{troptauD4ii}, 
 all lie in the kernel of the linear difference operator 
$$ 
\tilde{{\cal L}}= ({\cal T}^2+1) ({\cal T}^2-1)^2({\cal T}^3-1).
$$
For the tau functions generated by  \eqref{systmtauD4i} and 
\eqref{D42xeqns}, 
the leading order degree growth of their 
denominators is given  by 
$$ 
{\bf d}_n = \hat{{\bf a}}\, n^2+O(n), \quad 
 {\bf e}_n =2 \hat{{\bf a}}\, n^2+O(n), \quad 
{\bf f}_n = \hat{{\bf a}} \, n^2+O(n) 
$$ 
in case (1), 
and 
$$ 
\hat{{\bf d}}_n = \hat{{\bf a}}\, n^2+O(n), \quad 
 \hat{{\bf e}}_n =2 \hat{{\bf a}}\, n^2+O(n), \quad 
\hat{{\bf f}}_n = \hat{{\bf a}} \, n^2+O(n), \quad 
 \hat{{\bf g}}_n = \hat{{\bf a}} \, n^2+O(n)
$$ 
in case (2), 
with the same constant vector 
$$   
{\bf a}= \frac{1}{24} (1,1,2,1,1,1,1,1)^T 
$$ 
in each case. 
\end{thm} 
 
%%%%%%%%%%%%%%%%%%%%%%%%%%%%%%%%%%%%%%%%%%%

\section{Open problems and concluding remarks} 
\setcounter{equation}{0}

The results in this paper constitute a further proof of concept of the idea introduced by 
one of us with Kouloukas in \cite{hk}, that periodic dynamical systems arising from Zamolodchikov periodicity of cluster maps associated with finite type simple Lie algebras 
admit natural deformations to discrete dynamical  systems that are completely integrable in the Liouville sense (but no longer completely periodic), 
and further that this dynamics can be lifted to an enlarged phase space where the Laurent property holds (Laurentification).
In all of the low rank cases considered here, 
over $\C$ the level sets of first integrals are one-dimensional tori, so the spaces of initial conditions correspond to elliptic surfaces. 
This is just like the well known situation for QRT maps 
\cite{duistermaat}, which have been studied for a long time \cite{qrt}; and indeed, the 
examples of $\rA_3$ and $\rB_2$ treated above are related to particular cases of QRT maps. However, despite the prior results on Laurentification obtained in \cite{hhkq}, we do not know of any procedure to endow an arbitrary QRT map with the structure of a cluster algebra. However, if we interpret these examples over $\Q$ (rather than $\C$), 
then the Zamolodchikov periodicity of the original system can be viewed as providing a family of elliptic curves with a rational torsion point of prescribed order.

The derivation of the formula \eqref{A3dvectordeg} in the first  appendix below 
suggests that the tropical dynamics of d-vectors 
for cluster variables 
provides an efficient way to calculate the degrees of maps in projective space. 
We expect that the same approach can be applied to many 
other birational maps, shedding light on some of the unexplained observations in \cite{viallet1, viallet}.

Compared with our previous results, which were all in type $\rA$, 
the examples of $\rB_3$ and 
$\rD_4$ considered here have revealed the new feature that the same 
Zamolodchikov period dynamics can be deformed to an integrable map in more than one distinct way. Nevertheless, there seem to be very close connections between the two cases obtained for $\rB_3$; and the close connections between the case (1) and case (2) deformations for $\rD_4$ are even more apparent, given that  the underlying 
coefficient-free cluster algebra is the same for both.  

The situation for higher rank Lie algebras becomes even more interesting, since with more degrees of freedom one finds that the level sets of first integrals are 
abelian varieties of dimension greater than 1. 
In  \cite{hk} it was shown that the periodic cluster map associated with the $\rA_4$ root system has  a 2-parameter  integrable deformation, which lifts by Laurentification to a cluster map in 11 dimensions, with 2 additional frozen variables. Recently, 
 this 
construction has been generalized to 2-parameter deformations of $\rA_{2N}$ cluster maps for all $N\geq 1$, 
which lift to cluster maps in dimension $4N+3$ associated with a special family of quivers \cite{grab}. 
The complete description of this construction, and analogous results for the odd rank case ($\rA_{2N+1}$), will be the subject of future work. 
Recently we have also found extensions of the results given here to  
types $\rB$ and $\rD$ in higher rank, which are also under investigation. 

\section*{Appendix A: Elliptic surface and degree growth for deformed $\rA_3$ map} 

\renewcommand{\theequation}{A.\arabic{equation}}
%-------------------------------------------------------------------------
Here we consider the action of the maps $\hat{\varphi}$ and $\hat\psi$ on the elliptic surface defined by the first integral (\ref{biqk1}) in the 
deformed $\rA_3$ case. 
Let $(y, w)$ be a pair of inhomogeneous coordinates for $\mathbb{P}^1 \times \mathbb{P}^1$ and let $Y = \frac{1}{y}$ and $W = \frac{1}{w}$.
Then $\mathbb{P}^1 \times \mathbb{P}^1$ is covered by $4$ charts: $(y, w)$, $(Y, w)$, $(y, W)$ and $(Y, W)$.
For example, the point $(\infty, \infty) \in \mathbb{P}^1 \times \mathbb{P}^1$ corresponds to $(Y, W) = (0, 0)$.
On each chart, $K_1$ is written as
\begin{align*}
	K_1 &= \frac{(y + d w + c) ((y + 1) w + d)}{y w}  %\\	&
\qquad = \frac{(1 + d w Y + c Y) ((1 + Y) w + d Y)}{Y w} \\
	&= \frac{(y W + d + c W) ((y + 1) + d W)}{y W} %\\	&
\, = \frac{(W + d Y + c Y W) ((1 + Y) + d Y W)}{Y W}.
\end{align*}
The pencil defined by $K_1$ has $8$ base points on $\mathbb{P}^1 \times \mathbb{P}^1$:
\begin{itemize}
	\item[(1).]
	$(y, w) = \left( 0, - \frac{c}{d} \right)$,

	\item[(2).]
	$(y, w) = (0, - d)$,

	\item[(3).]
	$(y, w) = (- c, 0)$,

	\item[(4).]
	$(Y, w) = (0, 0)$,

	%\begin{itemize}
		\item[(5).]
		$\left( Y, %frac
{w}/{Y} \right) = (0, - d)$,

	%\end{itemize}

	\item[(6).]
	$(y, W) = (- 1, 0)$,

	\item[(7).]
	$(Y, W) = (0, 0)$,

	%\begin{itemize}
		\item[(8).]
		$\left( Y, %\frac
{W}/{Y} \right) = (0, - d)$.

	%\end{itemize}

\end{itemize}
We consider the \textit{generic} situation where the values of the parameters $c,d$ are such that none of these base points coincide, which means that we should assume
\beq\label{cdgeneric}
c\neq 0, 1, \qquad d\neq 0, \qquad c\neq d^2. 
\eeq 
Note that the points (5) and (8) are infinitely near (4) and (7), respectively.
Blowing up $\mathbb{P}^1 \times \mathbb{P}^1$ at these $8$ points, we obtain an elliptic surface ${\cal X}$ on which $K_1 \colon {\cal X} \to \mathbb{P}^1$ gives an elliptic fibration (see Figs~\ref{figure:blowup} and \ref{figure:surface}).
The commuting maps $\hat\varphi$ and $\hat\psi$, which were originally defined in  \eqref{dphihatmapA3} and \eqref{psihatmapA3} 
using  the affine chart $(y,w)=(u_1,u_2)$, both extend to automorphisms on ${\cal X}$.

%-------------------------------------------------------------------------
\begin{figure}
 \centering
	\begin{picture}(100, 360)
		%%%% 0
		{\thicklines
		\put(0, 30){\line(1, 0){100}}
		\put(0, 90){\line(1, 0){100}}
		\put(20, 10){\line(0, 1){100}}
		\put(80, 10){\line(0, 1){100}}
		}
		
		\put(80, 30){\circle*{5}}
		\put(85, 16){(4)}
		
		\put(80, 90){\circle*{5}}
		\put(85, 98){(7)}
		
		\put(20, 70){\circle*{5}}
		\put(0, 67){(2)}
		
		\put(20, 50){\circle*{5}}
		\put(0, 47){(1)}
		
		\put(40, 30){\circle*{5}}
		\put(32, 16){(3)}
		
		\put(50, 90){\circle*{5}}
		\put(42, 98){(6)}

		\put(50, 120){$\downarrow$}

		%%%%% 1
		{\thicklines
		\put(0, 160){\line(1, 0){50}}
		\qbezier(80, 190)(80, 175)(100, 160)
		\qbezier(50, 160)(65, 160)(80, 140)
		
		\put(20, 140){\line(0, 1){100}}
		\put(0, 220){\line(1, 0){50}}
		\qbezier(80, 190)(80, 205)(100, 220)
		\qbezier(50, 220)(65, 220)(80, 240)
		}
		
		% 1
		\put(10, 180){\line(1, 0){20}}

		% 2
		\put(10, 200){\line(1, 0){20}}

		% 3
		\put(40, 150){\line(0, 1){20}}

		% 4
		\put(60, 140){\line(1, 1){40}}

		% 6
		\put(50, 210){\line(0, 1){20}}

		% 7
		\put(100, 200){\line(-1, 1){40}}

		\put(79, 159){\circle*{5}}
		\put(67, 166){(5)}
		
		\put(79, 221){\circle*{5}}
		\put(67, 205){(8)}

		\put(50, 250){$\downarrow$}

		%%%%% 2
		{\thicklines
		\put(0, 290){\line(1, 0){50}}
		\qbezier(80, 320)(80, 305)(100, 290)
		\qbezier(50, 290)(65, 290)(80, 270)
		
		\put(20, 270){\line(0, 1){100}}
		\put(0, 350){\line(1, 0){50}}
		\qbezier(80, 320)(80, 335)(100, 350)
		\qbezier(50, 350)(65, 350)(80, 370)
		}
		
		% 1
		\put(10, 310){\line(1, 0){20}}

		% 2
		\put(10, 330){\line(1, 0){20}}

		% 3
		\put(40, 280){\line(0, 1){20}}

		% 4
		\put(60, 270){\line(1, 1){40}}

		% 5
		\put(70, 300){\line(1, -1){20}}

		% 6
		\put(50, 340){\line(0, 1){20}}

		% 7
		\put(100, 330){\line(-1, 1){40}}

		% 8
		\put(70, 340){\line(1, 1){20}}

	\end{picture}
	\caption{
		Blow-ups needed to obtain an elliptic surface from the pencil defined by $K_1$.
	}
	\label{figure:blowup}
\end{figure}
	
%-------------------------------------------------------------------------

\begin{figure}
\centering
	\begin{picture}(240, 240)
			
		{\thicklines
		\put(0, 60){\line(1, 0){100}}
		\qbezier(160, 120)(160, 90)(200, 60)
		\qbezier(100, 60)(130, 60)(160, 20)
		
		\put(40, 20){\line(0, 1){200}}
		\put(0, 180){\line(1, 0){100}}
		\qbezier(160, 120)(160, 150)(200, 180)
		\qbezier(100, 180)(130, 180)(160, 220)
		}

		\put(35, 5){$D_1$}
		\put(5, 45){$D_2$}
		\put(110, 5){$D_3$}
		\put(140, 115){$D_4$}
		\put(205, 135){$D_5$}
		\put(5, 185){$D_6$}

		% 1
		\put(20, 100){\line(1, 0){40}}
		\put(0, 95){$C_1$}

		% 2
		\put(20, 140){\line(1, 0){40}}
		\put(0, 135){$C_2$}

		% 3
		\put(80, 40){\line(0, 1){40}}
		\put(75, 25){$C_3$}

		% 4
		{\thicklines
		\put(120, 20){\line(1, 1){80}}
		}

		% 5
		\put(140, 80){\line(1, -1){40}}
		\put(185, 30){$C_5$}

		% 6
		\put(100, 160){\line(0, 1){40}}
		\put(95, 145){$C_6$}

		% 7
		{\thicklines
		\put(200, 140){\line(-1, 1){80}}
		}
			
		% 8
		\put(140, 160){\line(1, 1){40}}
		\put(185, 200){$C_8$}

	\end{picture}
	\caption{
		The elliptic surface ${\cal X}$.
		The curves $D_1$, $D_4$, $D_2$ and $D_6$ are the strict transforms of the curves $\{ y = 0 \}$, $\{ y = \infty \}$, $\{ w = 0 \}$ and $\{ w = \infty \}$ in $\mathbb{P}^1 \times \mathbb{P}^1$, respectively.
		The curves $D_1, \ldots, D_6$ have self-intersection $(- 2)$ and their intersection pattern forms the Dynkin diagram of type $A^{(1)}_5$ 
(cf.\  \cite{os}).
		The curves $C_1, C_2, C_3, C_5, C_6, C_8$ have self-intersection $(- 1)$.
	}
	\label{figure:surface}
\end{figure}

%-------------------------------------------------------------------------

Let us use intersection theory 
to calculate degree growth for $\hat\varphi$ and $\hat\psi$.
As a basis of $\Pic ({\cal X})$, it is common to use
\[
	H_y, H_w, E_1, \ldots, E_8
\]
where $H_y$ (resp.\ $H_w$) is the total transform of the class $\{ y = \text{const} \}$ (resp.\ $\{ w = \text{const} \}$) and $E_i$ is the total transform of the exceptional class of the $i$-th blowup.
For convenience, however, here we use the basis
\[
	[D_1], \ldots, [D_6], [C_2], [C_3], [C_5], [C_8]
\]
(see Figure~\ref{figure:surface}).
The matrix representations for $\hat\varphi$ and $\hat\psi$ with respect to this basis are
\[
	M_{\hat\varphi} := \begin{pmatrix}
		0 & 0 & 1 & 0 & 0 & 0 & 0 & 0 & - 1 & 0 \\
		0 & 0 & 0 & 1 & 0 & 0 & 0 & 1 & 0 & 0 \\
		0 & 0 & 0 & 0 & 1 & 0 & 0 & 1 & 1 & 0 \\
		0 & 0 & 0 & 0 & 0 & 1 & 0 & 0 & 1 & 0 \\
		1 & 0 & 0 & 0 & 0 & 0 & 0 & - 1 & 1 & 0 \\
		0 & 1 & 0 & 0 & 0 & 0 & 0 & - 1 & 0 & 0 \\
		& & & & & & 0 & 0 & - 1 & 0 \\
		& & & & & & 0 & 1 & 0 & 0 \\
		& & & & & & 0 & 1 & 1 & 1 \\
		& & & & & & 1 & - 1 & 1 & 0
	\end{pmatrix}
\]
and
\[
	M_{\hat\psi} := \begin{pmatrix}
		0 & 0 & 0 & 1 & 0 & 0 & 1 & 0 & 0 & 0 \\
		0 & 0 & 0 & 0 & 1 & 0 & 0 & 0 & 0 & - 1 \\
		0 & 0 & 0 & 0 & 0 & 1 & 0 & 0 & 1 & 0 \\
		1 & 0 & 0 & 0 & 0 & 0 & 0 & 0 & 1 & 1 \\
		0 & 1 & 0 & 0 & 0 & 0 & 1 & 0 & 1 & 2 \\
		0 & 0 & 1 & 0 & 0 & 0 & 1 & 0 & 0 & 1 \\
		& & & & & & 1 & 0 & 0 & 0 \\
		& & & & & & - 1 & 0 & - 1 & - 1 \\
		& & & & & & 0 & 0 & 1 & 0 \\
		& & & & & & 1 & 1 & 1 & 2
	\end{pmatrix},
\]
respectively.
The intersection matrix on $\Pic({\cal X})$ with respect to our basis is
\[
	A_{\cal X} := \begin{pmatrix}
		- 2 & 1 & 0 & 0 & 0 & 1 & 1 & 0 & 0 & 0 \\
		1 & - 2 & 1 & 0 & 0 & 0 & 0 & 1 & 0 & 0 \\
		0 & 1 & - 2 & 1 & 0 & 0 & 0 & 0 & 1 & 0 \\
		0 & 0 & 1 & - 2 & 1 & 0 & 0 & 0 & 0 & 0 \\
		0 & 0 & 0 & 1 & - 2 & 1 & 0 & 0 & 0 & 1 \\
		1 & 0 & 0 & 0 & 1 & - 2 & 0 & 0 & 0 & 0 \\
		1 & 0 & 0 & 0 & 0 & 0 & - 1 & 0 & 0 & 0 \\
		0 & 1 & 0 & 0 & 0 & 0 & 0 & - 1 & 0 & 0 \\
		0 & 0 & 1 & 0 & 0 & 0 & 0 & 0 & - 1 & 0 \\
		0 & 0 & 0 & 0 & 1 & 0 & 0 & 0 & 0 & - 1
	\end{pmatrix}.
\]
As in Fig~\ref{figure:p1p1vsp2}, ${\cal X}$ can be blown down to $\mathbb{P}^2$ as well.
Since the total transform of the class of lines in $\mathbb{P}^2$ is
\[
	[D_2] + 2 [D_3] + [D_4] + [C_3] + 2 [C_5],
\]
the degrees with respect to $\mathbb{P}^2$ can be calculated as
\[
	\deg \left( \hat\varphi^n \right) = \mathbf{h}^T \cdot A_{\cal X} M_{\hat\varphi}^n \mathbf{h}
\]
and
\[
	\deg \left( \hat\psi^m \right) = \mathbf{h}^T \cdot A_{\cal X} M_{\hat\psi}^m \mathbf{h},
\]
where
\[
	\mathbf{h} = (0, 1, 2, 1, 0, 0, 0, 1, 2, 0)^T.
\]
A direct calculation shows that
\[
	\lim_{n \to \infty} \frac{1}{n^2} M_{\hat\varphi}^n
	= \frac{1}{12} \begin{pmatrix}
		\mathbf{0}_{6, 6} & \mathbf{1}_{6, 4} \\
		\mathbf{0}_{4, 6} & \mathbf{0}_{4, 4}
	\end{pmatrix}
\]
and
\[
	\lim_{m \to \infty} \frac{1}{m^2} M_{\hat\psi}^m
	= \frac{1}{3} \begin{pmatrix}
		\mathbf{0}_{6, 6} & \mathbf{1}_{6, 4} \\
		\mathbf{0}_{4, 6} & \mathbf{0}_{4, 4}
	\end{pmatrix},
\]
where $\mathbf{1}_{6, 4}$ is the matrix of size $6 \times 4$ with all the entries $1$.
Hence we have
\beq\label{phintdeg}
	\deg \left( \hat\varphi^n \right) = \frac{3}{4} n^2 + O(n)
\eeq
and
\beq\label{psintdeg}
	\deg \left( \hat\psi^m \right) = \frac{9}{4} m^2 + O(m).
\eeq

The calculation of the degree growth in $\Proj^2$, as above, can also be 
derived from the tropical dynamics of the d-vectors, as determined in Theorem 
\ref{exactdegs}. Indeed, the non-frozen variables in a cluster 
associated with the deformed $\rA_3$ map $\hat\varphi$, and the QRT map $\hat\psi$ 
that  commutes with it, 
determine a point in $\Proj^2$ according to %the map 
$$ 
(\tx_1,\tx_2,\tx_3,\tx_4,\tx_5,\tx_6) \mapsto 
\Big(\tx_1\tx_4\tx_5 : (\tx_2)^2\tx_6:   \tx_2\tx_3\tx_5 \Big),
$$ 
so  the choice of  initial seed  
$$ 
(\tx_1,\tx_2,\tx_3,\tx_4,\tx_5,\tx_6) 
= (X,1,Z,1,1,Y) 
$$ 
 corresponds to an arbitrary point ${\bf X}=(X:Y:Z)\in\Proj^2$. Thus we can determine 
the action of $\hat\varphi$ and $\hat\psi$, and count the degrees of iterates of these maps, via the induced action obtained from the cluster maps 
$\tilde\varphi$ and $\tilde\psi$. For instance, by considering 
$\tilde\varphi^n(X,1,Z,1,1,Y)$, we find the sequence 
\beq\label{nonly} 
\hat{\varphi}^n({\bf X}) = 
\Big(\tau_{n-1}\tau_{n+2}\sigma_n:
(\tau_n)^2\sigma_{n+1}: \tau_n\tau_{n+1}\sigma_n\Big)\, \Big|_{\bf X}\in\Proj^2,  
\eeq 
where the subscript ${\bf X}$ on the right-hand side denotes the fact that we 
substitute the initial values $\tx_1=\tau_{-1}=X$, 
$\tx_2=\tau_0=1$, $\tx_3=\tau_1=Z$, $\tx_4=\tau_2=1$, $\tx_5=\sigma_0=1$, $\tx_6=\sigma_1=Y$ into each Laurent polynomial that appears. More generally, applying 
a combination of the two maps (associated with shifts in $m,n$ respectively), 
for any $(m,n)\in\Z^2$ we can compose $m$ steps of 
$\hat\psi$ with $n$ steps of $\hat\varphi$, to obtain the $(m,n)$ combined iterate, 
in the form  
\beq\label{projmn}
\hat{\psi}^m\hat{\varphi}^n ({\bf X}) = 
\left( 
\frac{\rN_1\rN_4\rN_5}{{\bf X}^{{\bf d}_1+{\bf d}_4+{\bf d}_5}} 
: 
\frac{(\rN_2)^2\rN_6}{{\bf X}^{2{\bf d}_2+{\bf d}_6}} 
:
\frac{\rN_2\rN_3\rN_5}{{\bf X}^{{\bf d}_2+{\bf d}_3+{\bf d}_5}} 
\right)(m,n; {\bf X}) , 
\eeq
where the dependence on $m,n$ and ${\bf X}$ indicates the 
fact that numerator polynomials $\rN_j(m,n;{\bf \tx})$ in \eqref{lpolyA3} are 
evaluated at $(\tx_1,\tx_2,\tx_3,\tx_4,\tx_5,\tx_6) 
= (X,1,Z,1,1,Y)$, and for the denominators we write
$$
{\bf X}^{{\bf d}_j}=X^{{\bf b}_1^T {\bf d}_j(m,n)} \,  
Y^{{\bf b}_6^T {\bf d}_j(m,n)} \, 
Z^{{\bf b}_3^T {\bf d}_j(m,n)},
$$ 
where ${\bf b}_k$ denotes the $k$th standard basis vector in $\R^6$. 
This leads to an exact formula for the degrees of the combined iterates of the two maps in $\Proj^2$, expressed in terms of the associated d-vectors defined on the 
$\Z^2$ lattice.

%-------------------------------------------------------------------------

\begin{figure}
\centering
	\begin{picture}(250, 360)
		%%%% 0-0
		{\thicklines
		\put(0, 30){\line(1, 0){100}}
		\put(0, 90){\line(1, 0){100}}
		\put(20, 10){\line(0, 1){100}}
		\put(80, 10){\line(0, 1){100}}
		}
		
		\put(80, 90){\circle*{5}}

		\put(30, 0){$\mathbb{P}^1 \times \mathbb{P}^1$}
		
		\put(50, 120){$\downarrow$}

		%%%% 1-0
		{\thicklines
		\put(130, 30){\line(1, 0){100}}
		\put(130, 110){\line(1, -1){100}}
		\put(150, 10){\line(0, 1){100}}
		}
		
		\put(150, 90){\circle*{5}}
		\put(210, 30){\circle*{5}}
			
		\put(180, 0){$\mathbb{P}^2$}
		
		\put(115, 120){$\searrow$}

		%%%%% 0-1
		{\thicklines
		\put(0, 160){\line(1, 0){100}}
		\put(20, 140){\line(0, 1){100}}
		\put(0, 220){\line(1, 0){50}}
		\put(80, 140){\line(0, 1){50}}
		\qbezier(80, 190)(80, 205)(100, 220)
		\qbezier(50, 220)(65, 220)(80, 240)
		}
		
		% 7
		\put(100, 200){\line(-1, 1){40}}
		
		\put(50, 250){$\downarrow$}

		%%%%% 0-2
		{\thicklines
		\put(0, 290){\line(1, 0){50}}
		\qbezier(80, 320)(80, 305)(100, 290)
		\qbezier(50, 290)(65, 290)(80, 270)
		
		\put(20, 270){\line(0, 1){100}}
		\put(0, 350){\line(1, 0){50}}
		\qbezier(80, 320)(80, 335)(100, 350)
		\qbezier(50, 350)(65, 350)(80, 370)
		}
		
		% 1
		\put(10, 310){\line(1, 0){20}}

		% 2
		\put(10, 330){\line(1, 0){20}}

		% 3
		\put(40, 280){\line(0, 1){20}}

		% 4
		\put(60, 270){\line(1, 1){40}}

		% 5
		\put(70, 300){\line(1, -1){20}}

		% 6
		\put(50, 340){\line(0, 1){20}}

		% 7
		\put(100, 330){\line(-1, 1){40}}

		% 8
		\put(70, 340){\line(1, 1){20}}

		\put(120, 320){${\cal X}$}

	\end{picture}
	\caption{
		The elliptic surface ${\cal X}$ can be obtained by blowing up $\mathbb{P}^2$ at $9$ points, instead of $\mathbb{P}^1 \times \mathbb{P}^1$ at $8$ points.
		The centers of the blow-ups on $\mathbb{P}^2$ in this figure are $[1 : 0 : 0]$ and $[0 : 1 : 0]$, where $[y : w : 1]$ are inhomogeneous coordinates on $\mathbb{P}^2$.
	}
	\label{figure:p1p1vsp2}
\end{figure}

%-------------------------------------------------------------------------

%% \begin{propn}\label{degmn}  
\noindent  
{\bf Theorem  A.1.} \textit{
For generic complex coefficients $c,d$, 
the degrees of the combined iterates of the maps $\hat\psi$ and $\hat\varphi$ 
in $\Proj^2$ are given by 
\beq\label{A3dvectordeg}  
\begin{array}{rcl}
\deg \left( \hat{\psi}^m\hat{\varphi}^n \right) 
& = & ({\bf b}_1+{\bf b}_3+{\bf b}_6)^T 
\Big({\bf d}(m,n)+{\bf d}(m,n+1) +{\bf d}(m+1,n) \Big)
+\frac{1}{2}\big(3-(-1)^n \big) \\ 
&& + \max (Y^{(1)}_{m,n},W^{(1)}_{m,n},0) + 
  \max (Y^{(3)}_{m,n},W^{(3)}_{m,n},0)+
 \max (Y^{(6)}_{m,n},W^{(6)}_{m,n},0), 
\end{array} 
\eeq
where $Y^{(k)}_{m,n},W^{(k)}_{m,n}$ respectively  
denote the $k$th component of 
each of the vectors 
defined in  \eqref{tropyw}, that is 
$$ 
Y^{(k)}_{m,n} ={\bf b}_k^T {\bf Y}_{m,n}, \qquad 
W^{(k)}_{m,n} ={\bf b}_k^T {\bf W}_{m,n} .
$$ 
}
\begin{prf} 
Each of the entries for the three homogeneous coordinates in (\ref{projmn}) 
corresponds to a product of three cluster variables with $X,Y,Z$ and 1 substituted appropriately, 
hence they are all Laurent polynomials in the homogeneous coordinates $(X:Y:Z)$ of the 
initial point in $\Proj^2$. In order to calculate the degrees of the iterates in $\Proj^2$, we need to remove the denominators, which are monomials in $X,Y,Z$, so that what remains is a coprime triple of homogeneous polynomials in these coordinates;  
%without common factors; 
then the degree $\deg ( \hat{\psi}^m\hat{\varphi}^n )$  
of the combined $(m,n)$ iterate of the maps is the common degree of 
these three polynomials. 
%Note that, upon substituting for the seed in terms of $X,Y,Z$
We can clear the denominators in (\ref{projmn}) by multiplying by 
suitable powers of the projective coordinates $X,Y,Z$. So for instance, to clear the 
powers of $X$, we must take the maximum of the exponents of $X$ that appear in  
the denominators of the three homogeneous coordinates in the formula  (\ref{projmn}), 
and multiply through by %$X$ raised to  
$$
X^{\max \Big({\bf b}_1^T({\bf d}_1 + {\bf d}_4+{\bf d}_5), 
{\bf b}_1^T(2 {\bf d}_2 + {\bf d}_6),
{\bf b}_1^T({\bf d}_2 + {\bf d}_3+{\bf d}_5)\Big)}, 
$$ 
where each  d-vector ${\bf d}_j$ is evaluated at $(m,n)\in\Z^6$, corresponding to column $j$ of the matrix $D_{m,n}$ determined in Theorem \ref{exactdegs}. If we focus on the third entry in (\ref{projmn}), then we see that the effect of rescaling by this power of $X$ is to produce an overall prefactor of 
$$
X^{\max \Big({\bf b}_1^T({\bf d}_1 -{\bf d}_2- {\bf d}_3+ {\bf d}_4), 
{\bf b}_1^T({\bf d}_2 -  {\bf d}_3- {\bf d}_5 + {\bf d}_6),
0   \Big)} = X^{\max \big(Y^{(1)}_{m,n},W^{(1)}_{m,n},0\big)}, 
$$ 
where the second equality above follows by rewriting 
each d-vector ${\bf d}_j(m,n)$ above  as an appropriate shift of a single vector ${\bf d}(m,n)$ defined 
on $\Z^2$, and then noting that  
$ {\bf Y}_{m,n}$ and $ {\bf W}_{m,n}$ are given by the relevant 
combinations of the latter vector, as in \eqref{tropyw}, with the first 
component of each, namely $Y^{(1)}_{m,n}$ and $W^{(1)}_{m,n}$,  
corresponding to the exponent of 
the homogeneous coordinate $X$. Similarly, by doing an analogous rescaling by powers of the other homogeneous coordinates $Y$ and $Z$, the third homogeneous 
coordinate in (\ref{projmn}) becomes a polynomial in $X,Y,Z$ with an overall monomial 
prefactor of 
$$ X^{\max \big(Y^{(1)}_{m,n},W^{(1)}_{m,n},0\big)}\, 
Y^{\max \big(Y^{(6)}_{m,n},W^{(6)}_{m,n},0\big)} \, 
Z^{\max \big(Y^{(3)}_{m,n},W^{(3)}_{m,n},0\big)}. 
$$ 
(The reader should hopefully not be confused by the double meaning of the letter 
$Y$ in the above formula: the unadorned letter denotes a homogeneous coordinate 
in $\Proj^2$, while  the same letter appearing in the exponents with indices and a numerical suffix denotes a tropical variable.) By construction, both the first and the second entry have now also become  polynomials in $X,Y,Z$,  also with 
monomial prefactors, so that in $\Proj^2$ we have 
\beq\label{osc}
\hat{\psi}^m\hat{\varphi}^n ({\bf X}) = \Big(\mathrm{P}_{m,n}({\bf X}):  
\mathrm{Q}_{m,n}({\bf X}) :  \mathrm{R}_{m,n}({\bf X})\Big)
\eeq 
with $\mathrm{P}_{m,n},  \mathrm{Q}_{m,n},  \mathrm{R}_{m,n}\in \C[X,Y,Z]$;  
but the monomial prefactors are such that $X$ can never divide all three of these 
polynomials simultaneously, and the same is true for $Y$ and $Z$. 
Moreover, it follows from Lemma  \ref{ywperiod23} that the prefactors oscillate 
periodically, 
repeating with period 2 in $m$ and period 3 in $n$.  

Apart from the oscillating monomial factors, the main contribution to the degree 
of the iterates comes from the product of three numerator factors $\rN_i\rN_j\rN_k$ 
appearing in  each homogeneous component of (\ref{projmn}). 
There are two essential properties of these polynomials that are inherited from the cluster algebra: firstly, due to the fact that each numerator polynomial $\rN_j$ in \eqref{lpolyA3} is not divisible by any initial cluster variable $\tx_i$, it follows that,  
when the homogeneous coordinates for $\Proj^2$ are substituted in, the resulting 
homogeneous polynomial $\rN_j({\bf X})$ is not divisible by $X$, $Y$ or $Z$; 
and secondly, for \textit{generic} values of the coefficients $c,d$, the cluster variables 
in each cluster are pairwise coprime Laurent polynomials, meaning that    
for each $(m,n)\in\Z^2$, the polynomials 
$\rN_1({\bf X}), \rN_2({\bf X}),\ldots,\rN_6({\bf X})$ appearing in 
(\ref{projmn}) are themselves pairwise coprime. (It should be noted that the latter 
assertion ceases to be valid in the non-generic cases \eqref{cdgeneric}, 
when the numerator polynomials are reducible and have certain factors in common, which cancel out and cause the degree to drop.)  This implies that the degree of 
the third component   $\mathrm{R}_{m,n}({\bf X})$, and hence the degree of the 
$(m,n)$ combined iterate of the two maps, 
is equal to the degree of the product of three numerators $\rN_2\rN_3\rN_5$ 
appearing in 
 (\ref{projmn}), plus the degree of the oscillating prefactor \eqref{osc}. 
So it remains to calculate the degree of this product of numerators, and show 
that it is equal to the first line of the formula \eqref{A3dvectordeg}. 

The products of evaluations of Laurent polynomials appearing in each entry 
on the right-hand side of  (\ref{projmn}) 
have a homogeneous degree. To start with, let us fix $m=0$, and 
consider the action of $\hat\varphi$ (shifting in $n$) on its own, 
with the sequence of products of three tau functions that appear as components in \eqref{nonly}.  
Then for the initial point in $\Proj^2$ we have the (common) homogeneous degree 
of each of the projective coordinates, namely  
$$ 
\mathrm{hdeg}(\tau_{-1}\tau_2\sigma_0 , 
(\tau_0)^2 \sigma_1 , \tau_0\tau_1\sigma_0) 
=\mathrm{hdeg}(X,Y,Z)=1, 
$$
and we can write the pattern of homogeneous degrees in the initial cluster as 
$$ \mathrm{hdeg}(\tau_{-1},\tau_{0},\tau_{1},\tau_{2},\sigma_{0},\sigma_{1}) 
= (1,0,1,0,0,1); 
$$ 
but after a single iteration of the system \eqref{tausys} we find 
$$
\mathrm{hdeg}(\tau_{0}\tau_3\sigma_1 ,
(\tau_1)^2 \sigma_2 , \tau_1\tau_2\sigma_1) 
=\mathrm{hdeg}\left(
\frac{Y\big(dY(X+Z)+cZ^2\big)}{X}, 
\frac{Z^2(dY+cZ)}{X}, YZ 
\right) =2, 
$$ 
where from 
$$ 
\tau_3 = \frac{dY(X+Z)+cZ^2}{X}, \qquad 
\sigma_2=\frac{dY+cZ}{X} 
$$
we see that the pattern of homogeneous degrees in the new cluster is  
$$ \mathrm{hdeg}(\tau_{0},\tau_{1},\tau_{2},\tau_{3},\sigma_{1},\sigma_{2}) 
= (0,1,0,1,1,0);  
$$
and thereafter, with subsequent shifts in $n$, this pattern repeats with period 2. A 
similar calculation shows that, under each shift $m\to m+1$, the pattern of 
homogeneous degrees remains the same; so overall it only depends on the 
parity of $n$. Hence, for all $(m,n)\in\Z^2$, we see that 
the homogeneous degree of the third component in (\ref{projmn})  is 
$$
\mathrm{hdeg}\left(
\frac{\rN_2\rN_3\rN_5  }
{ {\bf X}^{{\bf d}_2+{\bf d}_3+{\bf d}_5}}(m,n;{\bf X})  \right) 
=
\deg\Big(\rN_2\rN_3\rN_5 (m,n;{\bf X})\Big) - 
\deg \Big({\bf X}^{{\bf d}_2(m,n)+{\bf d}_3(m,n)+{\bf d}_5(m,n)}\Big) 
=
 \frac{1}{2}\Big(3-(-1)^n\Big);
$$  
so the rest of the formula \eqref{A3dvectordeg} then follows, by rewriting 
the sum of d-vectors in the exponent of ${\bf X}$ above as 
${\bf d}(m,n)+ {\bf d}(m,n+1)+{\bf d}(m+1,n)$ and taking the 
sum of components 1,3 and 6 of the latter vector, which correspond, respectively, to the 
powers of $X,Z$ and $Y$ in this monomial. 
\end{prf}

Observe that the leading order terms in \eqref{A3dvectordeg} are the sum of 
three components of three shifted copies of the d-vector 
${\bf d}(m,n)$, and as the growth of each component of this vector 
is given by  the entries  in (\ref{degd}), 
overall this contributes a factor of $3\times 3 =9$ times the 
 leading order of any of the entries of the matrix $D_{m,n}$, so that   
\beq\label{leaddeg} 
\deg \left( \hat{\psi}^m\hat{\varphi}^n \right)  
= \frac{9}{4}\, m^2 + \frac{3}{4}\, n^2 + O(m)+O(n), 
\eeq
which provides an independent confirmation of the results 
\eqref{phintdeg} and \eqref{psintdeg} obtained from intersection theory. 
We now explain what it tells us about the Mordell-Weil group of the elliptic 
surface ${\cal X}$. 

For all but a finite number of values of $\kappa\in \Proj^1$, the 
corresponding  fibre 
\beq\label{penka} 
(y + d w + c) \big((y + 1) w + d\big)=\kappa \, y w
\eeq 
in the pencil defining $\cal X$
is birationally equivalent to a Weierstrass  equation for a smooth cubic curve, that is   
\beq\label{weierkap}
E_\ka: \qquad \ry^2 = \rx^3 +\rA (\ka )\rx +\rB (\ka), 
\eeq 
for certain rational functions $\rA,\rB\in\C(\ka)$. 
Hence we can regard $\cal X$ as an elliptic curve $E/\C(\ka)$  
defined over the function field $\C(\ka)$, with the Mordell-Weil group being 
the group of   $\C(\ka)$-rational points. This curve has j-invariant 
$$ 
j(E_\ka) = \frac{\mathrm{f}_4(\ka)^3}{d^4\ka^2\, \mathrm{g}_4(\ka)}, 
$$ 
for certain degree 4 polynomials of the form 
$$ 
\mathrm{f}_4 (\ka) = \ka^4 -4(c+1)\,\ka^3 + \cdots +
\Big((c-1)^2 + 4d^2\Big)^2, 
\qquad 
\mathrm{g}_4 (\ka) = c\,\ka^4-(c+1)(4c-d^2)\, \ka^3 + \cdots +
(c-d^2)\Big((c-1)^2 + 4d^2\Big)^2. 
$$ 
From the latter one can read off the singular fibres, which correspond to the values of 
$\ka$ 
where the j-invariant has poles: at $0,\infty$ and the four distinct 
roots of $\mathrm{g}_4$. One can see from $j$ that there are two reducible fibres: the fibre over 0, which has 2 components (double pole in $j$ at $\ka=0$); and the fibre over $\infty$, which has 6 components (pole of order 6 at $\ka=\infty$). 

Equivalently, we can view the Mordell-Weil group as the set of sections 
$s: \, \Proj^1 \rightarrow {\cal X}$, which are rational maps satisfying 
$K_1\big(s(\ka)\big) =\ka$ for all $\ka\in\Proj^1$. To endow this with a group 
structure, we must fix a zero section $s_0$ that sends $\ka$ to ${\cal O}_\ka$, 
the identity element in each fibre. Then the maps 
$\hat\varphi$ and $\hat\psi$ correspond, respectively, to translation by 
certain points ${\cal P}_1$ and ${\cal P}_2$ in the group law of each 
(generic) fibre, which are associated with sections $s_1$ and $s_2$, say. 
In fact, in terms of the biquadratic model \eqref{penka}, viewed as a curve in 
$\Proj^1\times\Proj^1$, we find that we can 
take the points ${\cal O}=(\infty,\infty)$,  ${\cal P}_1=(\infty,0)$, 
${\cal P}_2=(-1,\infty)$ in each fibre: these are just the base points (7),(4),(6) 
 previously identified in the pencil.     

The Mordell-Weil group of $\cal X$, or equivalently, the set of 
$\C(\ka)$-rational points on the curve $E\big(\C(\ka)\big)$, has a canonical height 
function $\hat{h}: \, E\big(\C(\ka)\big)\to\R$, which is a quadratic form having 
the property that 
$\hat{h} \geq 0$, with $\hat{h}({\cal P})=0$ iff ${\cal P}$ is a  point of finite order (a torsion point), and an associated bilinear pairing 
$$ 
<{\cal P}, {\cal Q}> = \hat{h}({\cal P}+ {\cal Q}) 
-\hat{h}( {\cal P})-\hat{h}( {\cal Q}). 
$$ 
The canonical height is unique, up to an overall choice of scale. 
In the case at hand, if we fix the normalization to be consistent with the 
conventions in  \cite{ataec}, then we can use the results in Chapter III of 
the latter book, together with 
the explicit expression \eqref{leaddeg} for the degree growth 
of the maps $\hat\varphi$ and $\hat\psi$,  
to calculate the canonical heights as follows:  
$$  
\hat{h}({\cal P}_1)  =  \frac{1}{3} \lim_{n\to\infty}n^{-2}\, \deg (\hat{\varphi}^n) = \frac{1}{4}, \qquad 
\hat{h}({\cal P}_2 ) =  \frac{1}{3} \lim_{n\to\infty}n^{-2}\, \deg (\hat{\psi}^n) = \frac{3}{4},
$$ 
$$ 
\hat{h}({\cal P}_1+ {\cal P}_2 )  = \frac{1}{3} \lim_{n\to\infty}n^{-2}\, \deg \big(\hat{\psi}^n\hat{\varphi}^n\big) = 1.  
$$ 

Having calculated the canonical heights as above, we  
then find the Gram matrix with entries $<{\cal P}_i, {\cal P}_j>$ 
built from the pairing between 
these two points, given by  
$$ 
\left(\begin{array}{cc}
 <{\cal P}_1, {\cal P}_1> & <{\cal P}_1, {\cal P}_2> \\ 
<{\cal P}_1, {\cal P}_2> & <{\cal P}_2, {\cal P}_2>
\end{array}\right)  = 
\left(\begin{array}{cc}
 \sfrac{1}{2} & 0 \\ 
 0 & \sfrac{3}{2}
\end{array}\right) . 
$$ 
From this we can conclude that 
$$ 
\hat{h} ( n_1 \, {\cal P}_1+n_2\, {\cal P}_2) 
= n_1^2 \, \hat{h} (  {\cal P}_1) + n_2^2 \, \hat{h} (  {\cal P}_2)\neq 0 
$$ 
unless $n_1=n_2=0$, which implies that ${\cal P}_1$ and ${\cal P}_2$ 
are two independent generators in the Mordell-Weil group, and hence this group 
has rank at least 2. In fact, according to Theorem 3.1 in \cite{os}, since the rank of $T$, the root lattice associated with the reducible fibres, is 
calculated from the multiplicities of these fibres as 
$\mathrm{rk} \, T = (2-1)+(6-1)=6$, the rank of 
the Mordell-Weil group is 
$$
\mathrm{rk} \,E\big(\C(\ka)\big) = 8 - \mathrm{rk}\, T=8-6 =2, 
$$ 
as expected.

It it is worthwhile to note that, if we treat the frozen variables $c,d$ as parameters, 
then the j-invariant  $j(E_\ka)$ above is defined 
over $\Q(c,d,\ka)$, so we can use generic \textit{rational} values $c,d\in\Q$ and 
rational initial data for the maps to generate interesting examples of 
elliptic curves defined over $\Q$. One of the simplest cases that produces a curve 
$E(\Q)$ of rank 2 is the choice 
$$ 
c=-2, \qquad d=-1, \qquad (y,w)=(1,1)\implies \ka =-2. 
$$
In that case, the curve \eqref{penka} becomes  
$$ 
(y - w -2) \big((y + 1) w -1\big)=-2  y w,
$$ 
which has j-invariant 
$$ 
j = -\frac{1771561}{1588} 
,
$$ 
being birationally equivalent to the  minimal model 
$$ 
\ry^2+\rx\ry+\ry=\rx^3-3\rx+2, 
$$ 
where the latter has Mordell-Weil generators $(\rx,\ry)=(1,0)$ 
and $(-1,2)$.\footnote{See {\tt https://www.lmfdb.org/EllipticCurve/Q/794/a/1} for more details.}

\section*{Appendix B: 3-invariant for the special Somos-7 recurrence} 

\renewcommand{\theequation}{B.\arabic{equation}}

Here we consider the special Somos-7 recurrence relation \eqref{s7rec} associated with the Lyness map, 
and present a number of general results about its solutions. 
We begin by showing that every solution $\tau_n$ also satisfies a relation of Somos-5 type with a coefficient $\xi_n$ 
that is periodic with period 3, of the form   
\beq\label{s5per} 
\tau_{n+5}\tau_n = \xi_n\, \tau_{n+4}\tau_{n+1} - a\, \tau_{n+3}\tau_{n+2}.  
\eeq 
This result is connected to several of the examples in this paper, where such Somos-7 relations appear in various places. 
There is another, analogous result for the Somos-5 recurrence relation itself, namely that every Somos-5 sequence 
satisfies a relation of  Somos-4 type with one coefficient that varies according to the parity of the index $n$ \cite{hones5}.  
The corresponding result for the special Somos-7 recurrence can be stated as follows. 

\noindent  
{\bf Proposition B.1.} \textit{Suppose that the sequence $(\tau_n)$ satisfies the special Somos-7 relation 
\beq\label{sps7} 
\tau_{n+7}\tau_n = a\, \tau_{n+6}\tau_{n+1} + b\, \tau_{n+4}\tau_{n+3},
\eeq 
with constant coefficients $a,b$. Then the quantity 
\beq\label{3invt} 
\xi_n = \frac{\tau_{n+5}\tau_n + a \, \tau_{n+3}\tau_{n+2}}{\tau_{n+4}\tau_{n+1} } 
\eeq 
is periodic with period 3, that is $\xi_{n+3}=\xi_n$ for all $n$.} 
\begin{prf}
We begin by defining the sequence of ratios 
\beq\label{fratio} 
f_n = \frac{\tau_{n+2}\tau_n}{\tau_{n+1}^2}.  
\eeq 
In terms of $f_n$, the Somos-7 recurrence is equivalent to the relation 
\beq\label{fmap}
f_{n+5}f_{n+4}^2 f_{n+3}^3 f_{n+2}^3 f_{n+1}^2 f_{n} = a\, f_{n+4} f_{n+3}^2 f_{n+2}^2 f_{n+1} + b.   
\eeq 
Upon shifting $n\to n+1$ and subtracting (or equivalently, applying the total difference operator to the constant $b$), we 
find 
$$ 
f_{n+6}f_{n+5}^2 f_{n+4}^3 f_{n+3}^3 f_{n+2}^2 f_{n+1} -
f_{n+5}f_{n+4}^2 f_{n+3}^3 f_{n+2}^3 f_{n+1}^2 f_{n} =
 a ( f_{n+5} f_{n+4}^2 f_{n+3}^2 f_{n+2} - 
  f_{n+4} f_{n+3}^2 f_{n+2}^2 f_{n+1}), 
$$ 
and then after dividing by $f_{n+5}f_{n+4}^2 f_{n+3}^2 f_{n+2}^2 f_{n+1}$ and rearranging, 
this becomes 
$$ 
f_{n+6}f_{n+5} f_{n+4} f_{n+3} + \frac{a}{f_{n+5} f_{n+4} } = 
f_{n+3}f_{n+2} f_{n+1} f_{n} + \frac{a}{f_{n+2} f_{n+1} }.  
$$
If we substitute for $f_n$ with the formula (\ref{fratio}), then the above equality says precisely that 
$$ 
\xi_{n+3}=\xi_n, 
$$ 
which is the required result. 
\end{prf} 
The recurrence (\ref{fmap}) is equivalent to iteration of a map in 5 dimensions, that is 
$$\Psi: \quad (f_0,f_1,f_2,f_3,f_4) \mapsto (f_1,f_2,f_3,f_4,f_5), $$ 
which corresponds to a lift of the Lyness map \eqref{lynessgen} 
lying between it and Somos-7. The proof of the preceding result  
%can be rephrased as saying that  $\xi_0$ is a 3-invariant for the Somos-7 recurrence, 
shows that, when considered as a function on the 5D phase space, the quantity 
$$ 
\xi_0 = 
f_{3}f_{2} f_{1} f_{0} + \frac{a}{f_{2} f_{1} } ,
$$
is a 3-invariant for the map $\Psi$, in the sense that 
$(\Psi^*)^3 (\xi_0)=\xi_0$, and by viewing Somos-7 as the 7D map defined by   \eqref{sps7}, 
$\xi_0$ lifts via (\ref{fratio}) to a 3-invariant for this as well. 

Note that, since the 
three quantities $\xi_0,\xi_1,\xi_2$ are functionally independent, the  
three elementary symmetric functions 
\beq\label{symf} 
\xi_0+\xi_1+\xi_2,  
\qquad \xi_0\xi_1+\xi_1\xi_2+\xi_2\xi_0, \qquad  
\xi_0\xi_1\xi_2 
\eeq 
provide three invariant functions that are also 
functionally independent; so they constitute three first integrals, both for $\Psi$ and for 
Somos-7. This  is a particular case of the fact that the general Somos-7 relation, with an extra term proportional to  
$\tau_{n+5}\tau_{n+2}$ included on the right-hand side, has  three functionally independent 
invariants, as was shown by a different method in   
\cite{Fordy_2013}.   It was also remarked there that only one of these three first integrals survives reduction to the plane 
with coordinates $(u_0,u_1)$ where the Lyness map 
$$ 
\hat{\varphi}_L: \qquad 
\left(\begin{array}{c} u_0 \\ 
u_1 \end{array}\right) \mapsto \left(\begin{array}{c} u_1 \\ 
(au_1+b)/u_0 
 \end{array}\right)
$$ is defined. 
To be precise, if we define the map $\pi: \, \C^7 \rightarrow \C^2$ by 
\beq\label{pis7}  
\pi: \qquad u_0 = \frac{\tau_0\tau_5}{\tau_2\tau_3}, \qquad   u_1=\frac{\tau_1\tau_{6}}{\tau_{3}\tau_{4}},  
\eeq
then the intermediate lift corresponds to $\tilde{\pi}: \, \C^5 \rightarrow \C^2$, given by 
 $$ 
u_0 = f_0(f_1f_2)^2f_3, \qquad   u_1=f_1(f_2f_3)^2f_4,  
$$ 
and the coordinates $f_j$ make it easy to verify that the only surviving member of the  invariant set \eqref{symf} is the product 
\beq\label{prodinvt}
\xi_0\xi_1\xi_2 = a \tilde{K}+b, 
\eeq 
where 
\beq\label{firstint} 
\tilde{K} = \frac{u_0u_1(u_0+u_1) +(u_0+u_1)^2+(a^2+b)(u_0+u_1)+ab}{u_0u_1}
\eeq 
is the form of the first integral for the Lyness map $\hat{\varphi}_L $ that is given in  \cite{hone2020ecm}. 
(As usual, we are abusing notation slightly when we identify the left- and right-hand sides of 
\eqref{prodinvt}: we should really replace $\tilde{K}$ with $\pi^*( \tilde{K})$ or 
$\tilde{\pi}^*( \tilde{K})$, depending on whether we wish to regard the $\xi_j$ as 
functions of the quantities $\tau_n$ or $f_n$.) 

\noindent  
{\bf Proposition B.2.} \textit{Suppose that the sequence $(u_n)$ satisfies the Lyness map, in the form of the recurrence 
$$ 
u_{n+1}u_{n-1}=au_n+b. 
$$ 
Then the quantity $v_n=u_n+a$ gives a solution of the Somos-5 QRT map, in the form 
\beq\label{s5qrtmapv}
v_{n+1}v_nv_{n-1}=\hat{\al} v_n +\hat{\be}, \qquad \mathrm{with}\quad \hat{\al}=  a \tilde{K}+b, \quad \hat{\be}=-a\hat{\al}, 
\eeq 
where $\tilde{K}$ is the first integral associated with the Lyness map, given by \eqref{firstint}.
}
\begin{prf} In terms of a tau function $\tau_n$ satisfying  \eqref{sps7}, we may write 
$$ 
u_n = \frac{\tau_{n+5}\tau_n}{\tau_{n+3}\tau_{n+2}} \implies v_n = u_n+a = \frac{\tau_{n+5}\tau_n+a\,\tau_{n+3}\tau_{n+2}}{\tau_{n+3}\tau_{n+2}}
=\xi_n \, \frac{\tau_{n+4}\tau_{n+1}}{\tau_{n+3}\tau_{n+2}}
$$ 
from \eqref{3invt}. Hence, by multiplying out and cancelling the terms that appear in the product of three adjacent $v_n$, we find 
$$
v_{n+1}v_nv_{n-1} = \xi_{n+1}\xi_n\xi_{n-1}\, \frac{\tau_{n+5}\tau_n}{\tau_{n+3}\tau_{n+2}} = ( a \tilde{K}+b)\, u_n, 
$$ 
where we have used \eqref{prodinvt}. If we set $\hat{\al} = a \tilde{K}+b$ and replace $u_n$ by $v_n-a$, then we obtain the 
Somos-5 QRT map in the form \eqref{s5qrtmapv}, as required. 
\end{prf} 

The preceding result shows how the Somos-5 QRT map in Theorem \ref{somos5thm} 
appears as a consequence of the Lyness map in Theorem \ref{somos7thm}, and also explains the 
connection between the corresponding instances of these maps appearing in Theorems \ref{th46} and \ref{th58}. 
We now proceed to present some additional results about the solutions of the special Somos-7 recurrence. 

\noindent 
{\bf Theorem B.3.} \textit{
The general solution of the initial value problem for \eqref{sps7}, with generic (non-zero) values of the parameters $a,b$ and initial data $\tau_j$, $0\leq j\leq 6$, has the form 
\beq\label{sigmas7} 
\tau_n = \exp \big(c_1 +c_2 \, n + c_3\,  (-1)^n + c_4 \, e^{\sfrac{2n\pi\ri}{3}} + c_5 \, e^{-\sfrac{2n\pi\ri}{3}}\big) 
 \, \frac{ \sigma (z_0 + nz) }{ \sigma(z)^{n^2} },  
\eeq  
where $\sigma (\cdot) = \sigma(\cdot ; g_2,g_3)$ is the Weierstrass sigma function associated with the elliptic curve 
\beq\label{weier} E: \qquad y^2 = 4x^3 -g_2 x-g_3. \eeq 
} 
\begin{prf} 
The proof of this result goes along very similar lines to the corresponding result for Somos-5 in \cite{hones5}, so we will sketch the main details and leave it for the reader to fill in the rest. The first part of the proof 
requires verifying that the analytic formula does indeed satisfy \eqref{sps7}, viewed as a difference equation in $n$, while the second part is to show that it gives the general solution, in the sense that for 
a generic set of coefficients $a,b$ and initial values $\tau_j$, it is possible to find constants $c_j$ and $z_0,z,g_2,g_3$ that solve the corresponding initial value problem. For the first part,  note that the set of gauge transformations $\tau_n\to \rg_n \, \tau_n$, which leave invariant both \eqref{sps7} and the image $(u_0,u_1)$ of the map $\pi$ 
in \eqref{pis7}, consists of an algebraic torus $(\C^*)^5$ defined by the equation $\rg_{n}\rg_{n+5}(\rg_{n+2} \rg_{n+3})^{-1}=1$, which implies 
$$({\cal T}^5 - {\cal T}^3 -  {\cal T}^2 +1)\, \log \rg_n = 0, $$ 
and the solution of the latter equation is precisely the exponential prefactor in (\ref{sigmas7}) 
with arbitrary constants $c_j$, $1\leq j\leq 5$. Therefore, to verify the form of the analytic solution, it remains only to check the 
part involving $\sigma$. This can be done using standard results on Weierstrass functions, and in particular the three-term 
relation for the sigma function, which shows that \eqref{sigmas7} is a solution    of \eqref{sps7} if and only if the coefficients are parametrized by $z,g_2,g_3$ according to the formulae 
\beq\label{absi}a = \frac{\sigma(4z)}{\sigma(2z)\sigma(z)^{12}}, \qquad b  = - \frac{\sigma(6z)}{\sigma(3z)\sigma(2z)\sigma(z)^{23}} . \eeq
Now for the second part,  begin by observing that the total count of coefficients plus initial values is $9$, which is the same as the count of arbitrary constants $c_j$  plus parameters $z_0,z,g_2,g_3$, so it is necessary to account for how these constants/parameters are determined from the initial value problem. Ignoring  the non-gauge part of the formula, 
 the four parameters  $z_0,z,g_2,g_3$ come from the solution of the initial value problem for the iterates of the Lyness map  for given $a,b$ and initial data $(u_0,u_1)$, which has the form 
$$  (\hat{\varphi}_L)^n (u_0, u_1) = (u_n, u_{n+1}), \qquad \mathrm{with} \qquad u_n =\frac {\sigma\big(z_0+nz\big)\sigma\big(z_0+(n+5)z\big)}{\sigma\big(z_0+(n+2)z\big)\sigma\big(z_0+(n+3)z\big)\sigma(z)^{12}}.$$ 
Then the solution to the algebraic problem of reconstructing a Weierstrass cubic curve $E$ with invariants $g_2,g_3$ together with two 
points $P_0 = \Big(\wp(z_0),\wp'(z_0)\Big)$, $P = \Big(\wp(z),\wp'(z)\Big)\in E$, 
starting from an initial point $(u_0,u_1)$ on a fixed level curve of the first integral  \eqref{firstint} for the Lyness map, is presented explicitly in 
\cite{hone2020ecm}, Theorem 1, up to rescaling the $y$ coordinate by a factor of 2. (Compared with \eqref{weier}, the Weierstrass curve in  
 \cite{hone2020ecm} is written as $y^2=x^3 +Ax+B$.) For generic values of $a,b,\tilde{K}$, the affine biquadratic equation % for the corresponding
$$ u_0u_1(u_0+u_1) +(u_0+u_1)^2+(a^2+b)(u_0+u_1)+ab-\tilde{K}u_0u_1=0 $$  
defines a smooth curve of genus 1 in $\Proj^1 \times \Proj^1$, which is birationally equivalent to a Weierstrass cubic defined by 
$g_2,g_3$, with non-vanishing discriminant $g_2^3 -27g_3^2\neq 0$. Given the solution to the purely algebraic problem of determining 
the elliptic curve $E$ and the $(x,y)$ coordinates of the pair of points $P_0,P$, such that the orbit of the Lyness map corresponds to the sequence 
$P_0+nP\in E$, 
it is necessary to determine  the associated  pair of points on the Jacobian of the curve, that is the values $z_0,z\in\Jac (E) = \C/\Lambda\cong E$, which 
are found by evaluating the elliptic integrals 
$$ z_0 = \int_\infty^{P_0} \frac{\rd x}{y}, \qquad z = \int_\infty^P \frac{\rd x}{y}, $$ 
taken modulo the lattice of periods $\Lambda$. Finally, once  $z_0,z,g_2,g_3$ have been fixed,  the system of 5 linear equations 
$$ 
c_1 +c_2 \, n + c_3\,  (-1)^n + c_4 \, e^{\sfrac{2n\pi\ri}{3}} + c_5 \, e^{-\sfrac{2n\pi\ri}{3}} = \log \left(\frac{\tau_n\, \sigma(z)^{n^2}}{\sigma(z_0+nz)} \right), 
\quad n=0,1,2,3,4 
$$  
allows the values of the  constants $c_j$ appearing in the gauge factor 
to be determined from the initial values.
\end{prf} 

In order to discuss Somos-$k$ relations of higher order that are satisfied by solutions of \eqref{sps7}, 
it is convenient to introduce the sequence of division polynomials $(\ra_n)$ associated with an elliptic curve \eqref{weier}, 
whose terms correspond to the multiples $nP\in E$ (see 
\cite{hones5} and references therein). 
Given the point $P=\big(\wp (z),\wp'(z)\big)$ parametrized by $z\in \Jac (E)$, these can be defined analytically 
by the formula 
\beq\label{edsana} 
\ra_n = \frac{\sigma (nz)}{\sigma(z)^{n^2}}. 
\eeq 
From the definition, the sequence is clearly antisymmetric, in the sense that  $\ra_n=-\ra_{-n}$ for all $n$. 
This sequence is another solution of  the recurrence \eqref{sps7}, corresponding to the same value of the first 
integral $\tilde{K}$ as for the solution \eqref{sigmas7}, but with different initial conditions; we also refer to  $(\ra_n)$ as the elliptic divisibility 
sequence associated with \eqref{sigmas7}. In particular, when the parameters and initial conditions are chosen 
suitably, %}$ 
then $(\ra_n)$ consists entirely of integers  which satisfy the divisibility property $\ra_n|\ra_m$ whenever $n|m$: this is the usual meaning of the 
name elliptic divisibility sequence (EDS). 

In general, the terms of the EDS associated with a solution of \eqref{sps7} 
are determined  completely as algebraic functions of $a,b,\tilde{K}$. Note that we have $\ra_0=0, \ra_1=1$, and the first few terms are specified by 
$$ \begin{array}{l} (\ra_2)^4 = \tilde{K} +a, \quad 
(\ra_3)^3 = a\tilde{K} + b, \quad 
\ra_4 = \ra_2 a,  \quad 
\ra_5=a^2-b,  \quad 
\ra_6 = -\ra_2\ra_3 b, \quad 
\ra_7=-ab\tilde{K}-a^4+a^2b-b^2, \\ 
\ra_8 = -\ra_2 a (ab\tilde{K} +a^4 -2a^b+2b^2), \quad 
\ra_9 = -\ra_3 (a^3 b\tilde{K} +a^6-2a^4b+3a^2b^2-b^3). 
\end{array}  
$$
%The Somos-type relations  satisfied by  
The terms of the EDS satisfy a Somos-$k$ relation for each $k\geq 4$, 
which means that 
they can be considered as polynomials in $\ra_2,\ra_3,a,b,\tilde{K}$ with the appropriate divisibility property, namely 
$$ 
\ra_n \in {\cal R}^*:= \Z[\ra_2,\ra_3,a,b,\tilde{K}]/\sim , \qquad 
\mathrm{with } \quad n|m \implies \ra_n | \ra_m \quad \mathrm{in}\quad {\cal R}^* , 
$$
where $\sim$ denotes the equivalence relation defined by the algebraic identities $(\ra_2)^4 = \tilde{K} +a$, $(\ra_3)^3 = a\tilde{K} + b$. 
Now from the analytic definition \eqref{edsana} and the formula \eqref{sigmas7}, we can use the three-term relation for the sigma function to 
derive infinitely many higher Somos relations for $\tau_n$, which must be of odd order $k=2j+1$, but with the further requirement that   
$k$ cannot be a multiple of 3, since relations of this order are the only ones that are invariant under the gauge transformations $\rg_n$ described 
in the proof of Theorem B.2. In this way, we arrive at infinitely many relations that can be written concisely in the following form: 
\beq\label{highersk}
\ra_3\ra_4 \tau_{n+j+1}\tau_{n-j} = \left| \begin{array}{cc} 
\ra_{j+1} \tau_{n+4} & \ra_{j-3} \tau_{n} \\ 
\ra_{j+4} \tau_{n+1} & \ra_j \tau_{n-3}
\end{array}  \right| 
. 
\eeq 
However, the above relation can be modified by making use of the Somos-7 recurrence \eqref{sps7} to replace the product $\tau_{n+4}\tau_{n-3}$ on the 
right-hand side of \eqref{highersk}, and further simplified by using the fact that the terms $\ra_j$ of the associated EDS satisfy the same Somos-7 relation. 
Thus we arrive at

\noindent
\textbf{Corollary B.4.} \textit{For all integers $k\geq 7$ that are odd and not a multiple of 3, every solution of \eqref{sps7} satisfies a Somos-$k$ relation with constant coefficients, 
given by %\eqref
\beq\label{nhighersk} \tau_{n+j+1}\tau_{n-j} = \al_j \, \tau_{n+3}\tau_{n-2} + \be_j\, \tau_{n+1}\tau_n, 
\eeq  
with $k=2j+1$ and $j\not\equiv 1 (\bmod \,3)$, 
where 
\beq\label{albej}
\al_j =\frac{\ra_{j+1}\ra_j}{\ra_3\ra_2} , \qquad \be_j = - \frac{\ra_{j+3}\ra_{j-2}}{\ra_3\ra_2}
.
\eeq 
}

We can now conclude this appendix with a strong version of the Laurent property for \eqref{sps7}, which was an ingredient in the proof of Theorem \ref{nonlaurent}.

\noindent
\textbf{Theorem B.5.} \textit{The iterates of the special Somos-7 recurrence \eqref{sps7} possess the strong Laurent property, in the sense that  
$\tau_n\in\tilde{\cal R}$ for all $n\geq 0$, where 
$$ 
\tilde{\cal R}=\Z[a, b,\tilde{K},\tau_0^{\pm 1},\tau_1^{\pm 1},\tau_2^{\pm 1},\tau_3^{\pm 1},\tau_4,\tau_5,\tau_6 ] . 
$$ 
}
\begin{prf}
The proof follows very closely the approach used  in \cite{swahon} to show analogous results for Somos-4 and Somos-5. To begin with, it is clear that $\tau_j\in\tilde{\cal R}$ for 
$0\leq j\leq 10$, taking the 7 initial data and applying \eqref{sps7} four times, which requires division by $\tau_0,\tau_1,\tau_2,\tau_3$. 
Thereafter, the higher relations (Somos-11, Somos-13, etc.) can be used to calculate $\tau_j$ for $j\geq 11$, setting $n=j,j+1,j+2$ in \eqref{nhighersk} with $j\geq 5$ 
(for  $j\not\equiv 1 (\bmod \,3)$), 
so that only divisions by $\tau_0,\tau_1,\tau_2$ are necessary, and all $\tau_n$ appearing on the right-hand side are previously determined elements of $\tilde{\cal R}$. 
The result then follows by induction, once it has been shown that the coefficients $\al_j,\be_j\in \Z[a,b,\tilde{K}]$. 
To see this, it is enough to   verify from the recursive relations satisfied by the EDS that each $\ra_n$ is equal to a prefactor times an element of  $\Z[a,b,\tilde{K}]$, where 
the prefactor simply repeats the pattern $1,\ra_2,\ra_3,\ra_2,1,\ra_2\ra_3$ with period 6, and hence for each  $j\not\equiv 1 (\bmod \,3)$ the coefficients in 
\eqref{albej} are of the desired form. 
\end{prf}

\vspace{.2in} 
\noindent \textbf{Acknowledgments:} This research was supported by 
 grant 
 IEC\textbackslash R3\textbackslash 193024 from the Royal Society, 
and  
%This work was supported by 
a Grant-in-Aid for Scientific Research of Japan Society for the Promotion of Science, JSPS KAKENHI Grant Number 23K12996. 
We used Bernhard Keller's quiver mutation application to produce some of the figures (see {\tt https://webusers.imj-prg.fr/\~{}bernhard.keller/quivermutation/}).

\end{document}